\documentclass[structabstract]{aa}
\usepackage{graphicx}
\usepackage{txfonts}
\usepackage{color}
\usepackage{soul}
\usepackage{rotating}
\usepackage{natbib}
\usepackage{multirow}
\usepackage{amsmath}

\def\eg{\hbox{e.g.,}}
\def\ie{\hbox{i.e.,}}
\newcommand{\kms}{km\,s$^{-1}$}
\newcommand \Ha {H$\alpha$}
\newcommand \Bra {Br$\alpha$}

\newcommand \Mdot {\hbox{$\dot M$}}
\newcommand \Mdotmax {\hbox{$\dot M_{\rm max}$}}
\newcommand \Mdotth {\hbox{$\dot M_{\rm th}^{\rm Vink}$}}
\newcommand \Rstar {$R_{\ast}$}
\newcommand \Lstar {$L_{\ast}$}
\newcommand \Mstar {$M_{\ast}$}
\newcommand \Lsun{\hbox{L$_\odot$}}
\newcommand \Msun{\hbox{$M_\odot$}}
\newcommand \Rsun{\hbox{$R_\odot$}}
\newcommand \Msunyr{\hbox{$M_\odot\,$yr$^{-1}$}}
\newcommand \massloss{\hbox{$\times\,10^{-6} \Msunyr$}}
\newcommand \vinf{\hbox{$\upsilon_\infty$}}

\newcommand{\teff}{$T_{\rm eff}$}
\newcommand \tjumpcold{$T^{\rm jump2}_{\rm eff}$}
\newcommand \tjumphot{$T^{\rm jump1}_{\rm eff}$}

\newcommand \fcl{\hbox{$f_{ \rm cl}$}}
\newcommand \fclmin{\hbox{$f^{\rm min}_{\rm cl}$}}
\newcommand\clfarmax{\hbox{$f_{\rm max}^{\rm far}$}} 
\newcommand\cloutmax{\hbox{$f_{\rm max}^{\rm out}$}} 
\newcommand\clmidmax{\hbox{$f_{\rm max}^{\rm mid}$}}
\newcommand \clin{\hbox{$f^{\rm in}_{\rm cl}$}}
\newcommand \clmid{\hbox{$f^{\rm mid}_{\rm cl}$}}
\newcommand \clout{\hbox{$f^{\rm out}_{\rm cl}$}}
\newcommand \clfar{\hbox{$f^{\rm far}_{\rm cl}$}}
\newcommand \rin{\hbox{$r_{\rm in}$}}
\newcommand \rmid{\hbox{$r_{\rm mid}$}}
\newcommand \rout{\hbox{$r_{\rm out}$}}
\newcommand \rfar{\hbox{$r_{\rm far}$}}
\newcommand \rinc{\hbox{\rin = 1.05 \Rstar}}
\newcommand \rmidc{\hbox{\rmid = 2 \Rstar}}
\newcommand \routc{\hbox{\rout = 15 \Rstar}}
\newcommand \rfarc{\hbox{\rfar$>$ 50 \Rstar}}

\newcommand \micron{\hbox{$\mu$m}}

\newcommand	\xhe{\hbox{Y$_{\rm He}$}}

\defcitealias{Vink2000}{V00}
\defcitealias{Vink2001}{V01}
\defcitealias{Puls2006}{P06}
\defcitealias{Castor1975}{CAK}

\newcommand{\rowstyle}[1]{\gdef\currentrowstyle{#1}#1\ignorespaces}

\begin{document}

   \title{Upper Mass-Loss Limits and Clumping in the Intermediate and Outer Wind Regions of OB stars
  }

\author{M. M. Rubio-D\'iez \inst{1,2}
\and J. O. Sundqvist\inst{1,3}
\and F. Najarro\inst{1}
\and A. Traficante \inst{4}
\and J. Puls\inst{5}
\and L. Calzoletti \inst{4,6}
\and D. Figer \inst{7}
}

\institute{
	$^1$Centro de Astrobiolog\'ia, CSIC-INTA, Ctra de Torrej\'on a Ajalvir km 4, E-28850 Torrej\'on de Ardoz, Madrid, Spain \\ email: {\tt mmrd@cab.inta-csic.es, m.m.rubiodiez@gmail.com} \\
	$^2$ Facultad de F\'isicas, Universidad Aut\'onoma de Madrid, Campus Cantoblanco, Ctra Colmenar km 15, 28049 Madrid, Spain.\\
	$^3$ Instituut voor Sterrenkunde, KU Leuven, Celestijnenlaan 200D, 3001 Leuven, Belgium\\
	$^4$ IAPS-INAF, via Fosso del Cavaliere, 100, I-00133 Roma, Italy\\
	$^5$ Universit\"ats-Sternwarte M\"unchen, Scheinerstr. 1, 81679 M\"unchen, Germany\\
	$^6$ European Space Astronomy Centre (ESAC)/ESA, PO Box 78, 28690 Villanueva
de la Ca\~nada, Madrid, Spain\\
	$^7$ Center for Detectors, Rochester Institute of Technology, Rochester, New York 14623-5603, USA\\}

  \date{Received December 11, 2020; accepted June 24, 2021}


  \abstract
   {Mass-loss is a key parameter throughout the evolution of massive stars, and determines the feedback with the surrounding interstellar medium. The presence of inhomogeinities in stellar winds (\textit{clumping}) leads to severe discrepancies not only among different mass-loss rate diagnostics, but also between empirical estimates and theoretical predictions.}
   {We aim to probe the radial clumping stratification of OB stars in the intermediate and outer wind regions (r\,$\gtrsim$\,2\,\Rstar; radial distance to photosphere), to derive upper limits for mass-loss rates, and to compare to current mass-loss implementation. Our sample includes 13 B\,Supergiants, the largest sample of such objects in which clumping has been analysed so far.}
   {Together with archival optical to radio observations, we obtain new far-infrared continuum observations for a sample of 25 OB stars.  Our new data uniquely constrain the clumping properties of the intermediate wind region. By using density-squared diagnostics, we further derive the minimum radial stratification of the clumping factor through the stellar wind, \fclmin\,$(r)$, and the corresponding maximum mass-loss rate, \Mdotmax, normalising clumping factors to the outermost wind region (\clfar\,=\,1).}
   {We find that the clumping degree for r\,$\gtrsim$\,2\,\Rstar decreases or stays constant with increasing radius, regardless of luminosity class or spectral type, for 22 of 25 sources in our sample. However, a dependence of the clumping degree on luminosity class and spectral type at the intermediate region relative to the outer ones has been observed: O\,Supergiants (OSGs) present, on average, a factor 2 larger clumping factors than B\,Supergiants (BSGs). Interestingly, the clumping structure of roughly 1/3 of the OB Supergiants in our sample is such that the maximum clumping occurs close to the wind base (r\,$\lesssim$\,2\,\Rstar), and then decreases
monotonically. This contrasts the more frequent case where the lowermost clumping increases towards a maximum, and needs to be
addressed by theoretical models.

   In addition, we find that the estimated \Mdotmax\ for BSGs is at least one order of magnitude (before finally decreasing) lower than the values usually adopted by stellar evolution models, whereas for OSGs upper observational limits and predictions agree within errors. This implies large reductions of mass-loss rates applied in evolution-models for BSGs, independently of the actual clumping properties of these winds. However, hydrodynamical models of clumping suggest absolute clumping factors in the outermost radio-emitting wind on the order \clfar $\approx$ 4\,--\,9;
assuming these values would imply a reduction of mass-loss rates included in stellar evolution models by a factor 2\,--\,3 for OSGs (above \teff\,$\sim$\,26\,500\,K) and by factors 6\,--\,200 for BSGs below the so-called first bi-stability jump (below \teff\,$\sim$\,22\,000\,K). 
While such reductions agree well with new theoretical mass-loss calculations for OSGs, our empirical findings call for a thorough re-investigation of BSG mass-loss rates and their associated effects on stellar evolution.
}
   {}

   
\titlerunning{Mass loss and Clumping in the wind of OB stars}
 \maketitle
%

\section{Introduction}\label{intro}

Wind outflows from OB\,Supergiants are the most widely studied examples of the radiatively driven-wind physical phenomenon  (\eg\ \citealt{Friend1986}, \citealt{Pauldrach1986}, \citealt{Puls1996}). 
These radiation-driven winds were first described theoretically by \cite{Lucy1970} and \cite{Castor1975}, hereafter \citetalias{Castor1975}, assuming a stationary, homogeneous and spherically symmetric outflow. By comparing observed spectral lines with expanding Non-Local Thermodynamic Equilibrium (NLTE) model atmospheres (\eg\ {\sc cmfgen}, \citealt{Hillier1998}; {\sc fastwind}, \citealt{Santolaya-Rey1997} and \citealt{Puls2005}; {\sc powr}, \citealt{Grafener2002})\footnote{Some current atmosphere models account, among other physics (\citealt{Hillier1987}), for X-Ray emission (\citealt{Feldmeier1997}), line-blanketing effects (\citealt{Anderson1985}) and clumping using various parameterisations (\citealt{Najarro2008}, \citealt{Sundqvist2018b}).}, it is possible to derive both physical stellar and wind parameters by means of quantitative spectroscopy (\eg\ \citealt{Najarro1995}, \citealt{Puls1996}, \citealt{Herrero2002}, \citealt{Urbaneja2003}, \citealt{Puls2005}, \citealt{Berlanas2018}, \citealt{Sander2014}, \citealt{Martins2019}). However, despite the initial success of theoretical predictions based on stationary outflows (\citealt{Vink2000}, \citealt{Kudritzki2002}, \citealt{Puls2003}), it is well established nowadays that stellar winds from massive stars are time-dependent and structured in velocity and density, displaying small-scale inhomogeneities (see review by \citealt{Puls2008}, \citealt{Hamann2008}, \citealt{Sander2017b}). These inhomogeneities change the structure of the atmosphere and wind, affecting the quantitative spectroscopic-based diagnostics used to obtain crucial stellar and wind parameters of massive stars.

Hydrodynamical wind simulations have shown that the presence of strong instabilities\footnote{The line-deshadowing instability (LDI).} in the line-driven wind leads to formation of small-scale regions of very high densities (\citealt{Owocki1988}, \citealt{Feldmeier1995}). These dense `wind clumps' can be described using either the fractional volume of dense gas, the volume filling factor ($f_{\rm v}$, \citealt{Abbott1981}), or via a clumping factor (\fcl, \citealt{Owocki1988}):  
\begin{equation}
f_{\rm cl}= \dfrac{\langle\rho^2\rangle}{\langle\rho\rangle^2} \ge 1, \label{eqn.1}
\end{equation}
 where $\langle\rho^2\rangle$ and $\langle\rho\rangle$ are the mean of the squared density, and the mean density, respectively, and where under the assumption of a negligible amount of gas in between the dense clumps one has $f_{\rm cl}=f_{\rm v}^{-1}$.  
 
Time-dependent simulations further show that the clumping factor across the wind is not homogenous, but presents radial stratification, \fcl\,=\,\fcl\,$(r)$. Theoretical predictions agree on O\,Supergiants (OSGs) in that the clumping structure first increases rapidly in the wind-acceleration region, and then starts to fade again in regions far away from the star (\citealt{Runacres2002, Runacres2005}). However, depending on the approximations and physical conditions used in the simulations (\eg\ treatment of the source function, photospheric perturbations or dimensionality), the onset of the clumping is predicted further away (\citealt{Runacres2002}, \citealt{Dessart2003, Dessart2005}) or closer to the photosphere (\citealt{Sundqvist2011}, \citealt{Sundqvist2013}, \citealt{Sundqvist2018a}, \citealt{Driessen2019}), which can affect formation of critical wind diagnostics as well as the spot where maximum clumping is achieved. The actual physical conditions close to the base wind (r $\lesssim$ 2\,\Rstar) represent an open challenge in line-driven wind theory. 

It has been suggested than a turbulent photosphere (e.g. \citealt{Cantiello2009}), may be lead to the formation on clumps at the base of the wind. Of course, this process would affect quantitatively the density structure of the wind, at least close to the stellar photosphere. However, it is not clear how the influence of a turbulent photosphere could persist far out in the wind, and there is not theoretical simulations to probe. Therefore, for the purposes of this work hydrodynamical LDI theory is used, since it has been widely tested successfully.

Since the presence of a density structure affects the matter-radiation interaction across the wind, a key consequence of wind-clumping regards its impact upon the \textit{effective} opacity of the wind (see detailed overview in \citealt{Sundqvist2018b}). For processes scaling linearly with $\rho$, the mean opacity of the clumped medium is the same as for the homogeneous wind model, whereas for processes scaling with $\rho^2$, the mean opacity is enhanced by a factor $f_{\rm cl}$. Empirically, this means that clumping differently affects the spectral diagnostic used to derive wind parameters. Assuming that all clumps are optically thin for a given mass-loss rate leaves unaltered diagnostic X-ray lines, electron-scattering wings, and scattering resonance-lines ($\rho$-dependent; \eg\ C\,{\sc iv}, P\,{\sc v}), whereas it causes an opacity-enhancement of recombination lines or free-free continuum ($\rho^2$-dependent; \eg\ \Ha, Br$_\gamma$,  mid-/far-infrared and radio continua). Therefore, if clumping is not appropriately taken into account, inconsistencies in mass-loss rate estimates between different diagnostics will arise (\citealt{Fullerton2006}, \citealt{Cohen2010}, \citealt{Sundqvist2011}). In particular, it is well established now that mass-loss rates of OB-stars derived from the classical diagnostic \Ha\  line are overestimated by a factor $\sqrt f_{\rm cl}$ 
if unclumped models are used in the analysis. 

Moreover, if clumps become optically thick this leads to additional light-leakage through porous channels in between the clumps. Porosity can occur either spatially or in velocity-space due to Doppler shifts in the rapidly accelerating wind. Studies of velocity-space porosity have shown that clumps indeed very easily become optically thick in UV resonance-lines and that this effect is critical to include when analysing such P-Cygni line formation \citep{Oskinova2007, Hillier2008, Sundqvist2010, Surlan2013}. On the other hand, such velocity-space porosity does not affect the continuum diagnostics studied in this work. And as shown by \citet{Sundqvist2018b}, under typical OB-star wind conditions effects of spatial porosity upon  
the formation of mid-/far-infrared (MIR and FIR) and radio continua should be negligible. Nonetheless, also these diagnostics are subject to uncertainties related to wind-inhomogeneities, in particular the level of clumping in the inner to intermediate (MIR and FIR) and outermost (radio forming region) wind parts. These uncertainties, however, can be reduced by combining suitable diagnostics, effectively mapping the wind at different radial distances, from close to the base (V-band, \Ha), over intermediate regions (\Bra, MIR/FIR continua), to the outermost region (radio continuum). A consistent analysis would here place tight constraints on the radial clumping-stratification. Moreover, if used in combination with ionisation and velocity law diagnostics (resonance lines; \eg\  P\,{\sc v}), or X-ray emission (\citealt{Cohen2010}), then absolute clumping factors and mass-loss rates could be uniquely derived.

First efforts have been performed by \cite{Puls2006} (hereafter, \citetalias{Puls2006}) and \cite{Najarro2008, Najarro2011}, by means of \Ha, \Bra\ and IR/mm/radio-continuum emission models for a sample of Galactic hot O stars. These studies show that whereas O\,Giants seem to have similar clumping degrees throughout the wind, OSGs show, on average, a clumping factor of about four times larger in the inner wind than in the outermost one. Thus, the mass-loss rate estimates for OSGs derived from \Ha\ and smooth wind models should be scaled down by a minimum factor 2. 
The empirical clumping stratification derived by \citetalias{Puls2006} and \cite{Najarro2011} suggested that the clumping degree starts to decrease at r $\approx$\,2\,--\,6\,\Rstar. This overall agrees rather well with the predicted clumping stratification at r $\lesssim$ 20\,\Rstar\ by hydrodynamics LDI simulations of OSGs (\citealt{Sundqvist2011, Sundqvist2013, Driessen2019}). However, due to the scarcity of reliable continuum observations of the analysed OB stars at FIR wavelengths, the clumping factor in the intermediate wind region (2\,\Rstar$\lesssim$\,r\,$\lesssim$\,15\,\Rstar) remains poorly constrained.
 
To investigate the radial clumping stratification of OB stars at r\,$\gtrsim$\,2\,\Rstar\ we obtained {\sc Herschel-PACS} 70 and 100\,\micron\ fluxes (and 160\,\micron\ flux for the brighter sources) for a carefully selected sample of OB stars. The sample consists of 25 OB Galactic Supergiant, Giant and Dwarf stars, covering spectral types from O3 to B3\,-\,4. Targets were selected to cover a wide OB-star parameter space, allowing us to analyse the behaviour of wind-clumping stratifications as a function of luminosity class and spectral type. Moreover, the large number of OB\,Supergiants in the sample allows for a deeper, statistically more significant analysis than previous work. In particular, our analysis represents a first attempt to derive the average properties of the clumping structure of B\,Supergiants (BSGs) by means of a multi-wavelength analysis, carefully comparing to the corresponding OSG results. In addition, the analysis allows us to derive reliable upper limits to mass-loss rates, and to compare these with the existing mass-loss rates recipes (\citealt{Vink2000,Vink2001}, hereafter \citetalias{Vink2000} and \citetalias{Vink2001}) usually used in evolutionary tracks. 

This represents a unique opportunity to test (at least in a relative way) theoretical mass-loss predictions for BSGs across the so-called bi-stability jump. Since the quantitative mass-loss rates across this jump are critical for stellar evolution modelling (\eg\ \citealt{Vink2010}, \citealt{Keszthelyi2017}), this has a quite important impact also upon our general knowledge about the massive-star life-cycle.

The paper is organised as follows. Section\,\ref{observations} summarises the obtained {\sc Herschel-PACS} observations and flux estimates of the sample, as well as the collected observations at different wavelengths from the literature. A description of the methodology, assumptions and procedures can be found in Section\,\ref{modeling}, whereas the performed analyses and corresponding results are presented in Section\,\ref{analysis}. We discuss our findings in Section\,\ref{discussion}, and present conclusions and a summary in Section\,\ref{summary}.

\section{Observations and data reduction}\label{observations}

Although the main goal of this work is to constrain maximum mass-loss rates and degrees of clumping in the intermediate wind region of OB\,Supergiants, our initial 27 star sample also contains a few OB\,Giants and Dwarfs, including one LBV, two confirmed binaries, four early B-Hypergiants (eBHGs) and one magnetic star (see Table\,\ref{tabledata}). This sample allowed us not only to analyse clumping stratification by luminosity class and spectral type, but also to investigate clumping properties of some peculiar objects for the first time. Finally, although blue supergiant stars are known to display photometric and spectroscopic variability, suggested to be linked to stellar pulsations, our sample has been carefully selected to encompass Blue Supergiants in long stable stages (\citealt{Clark2012}), with the exception of HD\,198478, an extreme case of spectroscopic variability\footnote{Periods in the variability of the photospheric lines from 2.7 hours to 22.5 days  related to pulsation activity have been recently reported \cite{Kraus2015}.}, with an important scatter in the stellar and wind parameters for this source (see Table\,\ref{tablehd198478}). Nonetheless, we included this star in the sample in order to further analyse the uncertainty in our estimated values due to the use of certain sets of stellar and wind parameters (see discussion Appendix\,\ref{apxBSg}), and to account for the effect of possible strong variability not detected so far for the rest of stars in our sample.

In addition, we collected archival data in the literature for V-band, near-infrared (NIR), MIR, mm and radio fluxes, when available. In Table\,\ref{tabledata} we summarise the sample of stars observed with  {\sc Herschel-PACS} and the data collected for this work. PACS column refers to our FIR flux observations. A description of the data reduction and the photometry processing of {\sc Herschel-PACS} observations can be found in the following subsection, and the collected data are summarised in the successive sections. 

\begin{sidewaystable*}
\begin{center}
\caption{List of the OB stars observed with {\sc Herschel-PACS} (IP: OT1\_mrubio\_1/OT2\_mrubio\_2) and the photometric data compiled from the literature used in this work (see references below). Our sample of stars is sorted by luminosity class and spectral type for clarity. }
\begin{tabular}{l l l l l l}
\hline\hline
\rowstyle{\bfseries}%
Source 		&{\bf{Spec. Type}} 	 & {\bf V/NIR/MIR (bands)}  & 	{\bf FIR (IRAS)}	  			&	{\bf mm/Radio}			&   {\bf Ref }\\\hline
CyOB2\#7     & O3 If               	 &	V/JHK/LMNQ  	 	 &  			     	&    2, 3.5, 6, 21 cm 					&	1, 5, 22, 23, 30, 33, 39, 44\\
HD66811	    & O4 I(n)f              	 &  V/JHK/LMNQ		 & 60 \micron  	&	0.85, 1.3 mm/ 2, 3.6, 6, 20cm 	&  5, 8, 9, 10,33, 44	\\
CyOB2\#11		& O5 If+	   			 &  V/JHK/LMNQ		 &					&   2, 3.5, 6 cm						&	20, 22, 23, 32, 33, 39, 44\\
HD210839		& O6 I(n)f 	   			 &	V/JHK/LMNQZ		 & 60 \micron 	&	1.35, 7 mm/ 2, 3.5, 6, 21 cm		&	5, 23, 33, 36, 44\\
HD152408		& O8 Iafpe   			 &  V/JHK/LMNQ		&					&	1.3 mm/ 2, 6 cm					&	1, 5, 19, 20, 21, 23, 26, 30, 32, 44     \\
HD151804 		& O8 Iab         		 &	V/JHK/LMNQ	    &						&	6 cm								&	1, 3, 5, 9, 23, 33, 44 \\   
HD149404      & O9 Ia        		      & V/JHK/LMNQZ		& 60	\micron	&	3.6 cm								&	1, 5, 15, 20, 21, 23, 24, 30   \\  
HD30614		& O9.5 Ia      		 &	V/JHK/LMNQ    		&					&  2, 3.5, 6 cm						&	5, 9, 20, 33, 37, 44\\\hline
HD37128		&  B0 Ia       		    	 &	V/JHK/LMNQ		&					&	0.85, 1.2, 1.3 mm/ 2, 3.6, 6, 6.2, 20, 90cm	& 1, 4, 5, 6, 10, 16, 20, 26, 37, 44\\                                 		          
HD38771		& B0.5 Ia          		&  V/JHK/LMNQ			& 60, 100 \micron          &    6 cm								&	1, 5, 19, 20, 44\\
HD154090		& B0.7 Ia       		&	V/JHK/LMNQ		& 60, 100 \micron         &	  3.6 cm						    & 	1, 5, 8, 23, 32, 44\\
HD193237		& B1 Ia                 	&	V/JHK/LMNQZ		& 60, 100 \micron         &   	1.2 mm/ 2, 3.5, 6, 20cm		&	1, 5, 6, 9, 20, 23, 32, 37, 44  \\
HD24398		&  B1 I              	&	V/JHK/LMNQ		& 60, 100 \micron          &   6 cm								&	1, 5, 9, 20, 23, 44   \\
HD169454		&  B1.5 Ia+            	&	V/JHK/LMNQ         &      				&	2, 6 cm								&	1, 5, 9, 17, 19, 23, 32, 44\\
HD152236		&  B1.5 Ia+     	    &	V/JHK/LMNQ         &    					&	1.3 mm/ 2, 3.5, 6.2 cm			&	3, 5, 8, 9, 15, 25, 26, 28, 38, 43, 44\\
HD190603		&  B1.5Ia			    	&	V/JHK/LMNQ		&					&										&	1, 5, 37, 44				\\
HD41117		&  B2 Ia          		&	V/JHK/LMNQ		& 60, 100 \micron &	2, 3.5, 6 cm						&	1, 5, 23, 32, 37, 44\\ 
HD194279		&  B2 Ia         		&  V/JHK/LMNQ          &                 		&	2, 3.5, 6 cm     					&	1, 5, 32, 37, 44\\[4pt]
HD198478		& B2.5 Ia+        		&	V/JHK/LMNQ		& 60, 100 \micron &	3.5, 6 cm							&	1, 5, 23, 32, 37, 44\\ 
HD80077		& B2/B3 Iae       		 &	V/JHK/LMNQ		& 60, 100 \micron &	3.5, 6.2 cm							&	1, 11, 14, 16, 20, 27, 32\\
HD53138		& B3 Ia                	 &	V/JHK/LMNQ		& 60, 100 \micron &  										&	1, 5, 20, 32, 44\\
CyOB2\#12	& B3/4 Ia              	 &	V/JHK/LMNQ		& 60, 100 \micron  &	1.2, 7 mm/ 1.3, 2, 3.5, 3.6, 6, 21 cm &2, 3, 5, 6, 7, 9, 12, 13, 18, 20, 23, 25, \\
				&							 &										&							&			 &	29, 31, 32, 34, 35, 37, 39, 40, 41, 42 \\
\hline
HD24912		& O7.5 III (n)((f))  	&  V/JHK/LMNQZ		& 60  \micron		&   7 mm/ 2, 3.5, 6 cm				&	20, 21, 23, 30, 33   \\
HD36861        & O8 III((f))   		&  V/JHK/LMNQZ		&					&	2, 3.5, 6 cm						&	5, 23, 33, 44 \\
HD37043		& O9III        			&  V/JHK/LMNQ		&					&	2, 3.5, 3.6, 6 cm				&	5, 20, 24, 32, 33, 44\\\hline
HD149757      & O9.5 Vnn             &  V/JHK/LMNQZ		&					&	6 cm							&	1, 5, 9, 19, 20, 21, 23, 30, 44\\  
HD149438		& B0.2 V            	&  V/JHK/LMNQ			&					&	6 cm							&	1, 5, 9, 23, 39, 44\\     
 \hline\hline
\end{tabular}
\tablebib{[1]: 2MASS Survey;  [2]: \citealt{Abbott1981}; [3]: \citealt{Abbott1984}; [4]: \citealt{Abbott1985}; [5]: AKARI Survey; [6]\citealt{Altenhoff1994}; [7]: \citealt{Becker1985}; [8]: \citealt{Benaglia2007}; [9]: \citealt{Bieging1989}; [10]:\citealt{Blomme2003}; 
[11]: \citealt{Carpay1991}; [12]: \citealt{Chaldu1973}; [13]: \citealt{Contreras1996}; 
[14]: \citealt{Ducati2002}; 
[15]: Glimpse Survey; [16]: \citealt{Gezari2005}; [17]: \citealt{Gutermuth2014};
[18]: \citealt{Harris1978}; [9]: \citealt{Howarth1991}; 
[20]: IRAS; [21]: IRAC/SPITZER from IRS file; [22]: IRAC/SPITZER from CygOB2 survey catalog;  [23]: IRS/Spitzer; 
[24]: \citealt{Lamers1993}; [25]: \citealt{Leitherer1982}; [26]: \citealt{Leitherer1991}; [27]: \citealt{Leitherer1995}; [28]: \citealt{Lopez1984};
[29]: \citealt{Massey1991}; [30]: MIPS/Spitzer; [31]: \citealt{Morford2016}; [32]: MSCX6; 
[33]: \citealt{Puls2006} (and references therein); 
[34]: \citealt{Rieke1974}; [35]: \citealt{Rieke1985}; 
[36]: \citealt{Schnerr2007}; [37]: \citealt{Scuderi1998}; [38]: \citealt{Sterken1997};
[39]: \citealt{TorresDodgen1991};
[40]: \citealt{Waldron1998}; [41]: \citealt{Wendker1980}; [42]: \citealt{White1983}; [43]: \citealt{Whittet1980}; [44]: WISE; 
}
\label{tabledata}
\end{center}
\end{sidewaystable*}

\subsection{FIR observations}\label{datapacs}

 FIR flux observations for our sample of 27 OB stars (Table\,\ref{tabledata}) at 70 and 100\,\micron\ were taken with the Photodetector Array Camera and Spectrometer (PACS; \citealt{Poglitsch2010}) onboard the {\sc Herschel} spacecraft (\citealt{Pilbratt2010}) in photometer mode (PI: Rubio-D\'iez, ID: OT1\_mrubio\_1, OT2\_mrubio\_2). A few of the objects in our sample have additional FIR-IRAS 60 and 100\,\micron\ data available, however most of them are upper limits (see references in Table\,\ref{tabledata}). 
 
Observations spans from December 2011 to March 2013, and were done using the mini-scan map observing mode. In this mode, two scan maps are taken along the two array diagonal directions (110 and 70\,degrees) at a constant speed of 20$''$/s in parallel lines. The nominal spatial resolution for this scan velocity is 3.2$''$/pixel.  Since our targets are point sources, a scan length of 2.5\,arcmin was enough to cover the sky region of interest. The exposure times were estimated to reach a S/N $\ge$ 10, using theoretical emission flux estimates. 
Additionally, PACS operates simultaneously in the two sidebands characteristic of heterodyne receivers. Thus, for each source in our sample a total of eight scans were obtained, two mini-scan maps at 70 and 100\,\micron\ (blue and green band), and four scans at extra band 160\,\micron\ (red band). The PACS point spread function (PSF) for these photometric bands have a full-width at half-maximum (FWHM) default value of 5.2$''$, 7.7$''$ and 12$''$, respectively.   
 
\begin{figure*}[htp]
\begin{center}
\includegraphics[width=18cm]{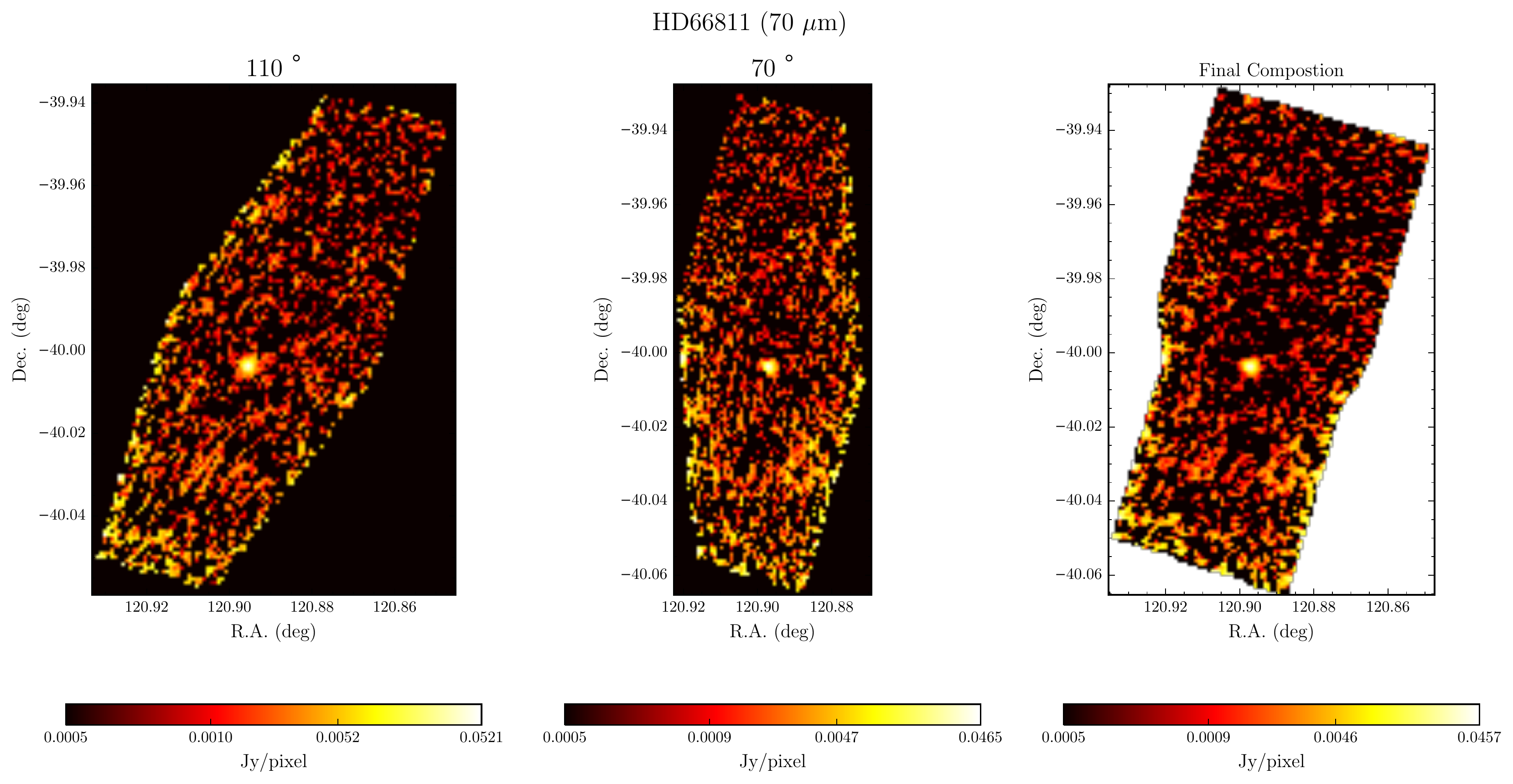}
\caption{From Left to Right, PACS 70 $\mu$m miniscan maps taken at position angles 70$^{\circ}$ and 110$^{\circ}$, and the final composed image of the O\,Supergiant star HD\,66811 ($\zeta$\,Pup).
}
\label{pacsexample}
\end{center}
\end{figure*}

\begin{table*}[!ht]
\begin{center}
\caption{Herschel/PACS 70, 100 and 160\,\micron\ fluxes for our sample stars. Sources are listed as in Table~\ref{tabledata}. The 1$\sigma$ error of the flux includes the uncertainty in flux calibration from PACS, the error in the photometric procedure and the error arising from the aperture and colour corrections (see Sec.\,\ref{observations}). The quality of each of these flux densities is designated by the quality flags (qF$_{\lambda}$) and is either of high quality (qF$_{\lambda}$ = A; S/N $\ge$ 10), of moderate quality (qF$_{\lambda}$ = B; 5 $\le$ S/N $<$ 10), or an upper limit (qF$_{\lambda}$ = U; S/N $\le$ 5 and/or confused detection).}
\begin{tabular}{l c c c c c c}
\hline\hline
Source	   		&	F$_{\rm 70 \mu m}$ $\pm$ $\sigma$  (Jy)   &  qF$_{\rm 70 \mu m}$ &	F$_{\rm 100 \mu m}$ $\pm$ $\sigma$  (Jy) 	&	qF$_{\rm 100 \mu m}$ &F$_{\rm 160 \mu m}$ $\pm$ $\sigma$  (Jy)  & qF$_{\rm 160 \mu m}$ \\\hline
CygOB2\#7   &   ---	   			   &		& ---					& 				&   ---	\\
HD66811   		&   0.32     $\pm$ 0.045 	& A		 &  0.220   $\pm$   0.025  & A		 &   0.3  $\pm$ 0.1  & B\\
CygOB2\#11  	&   $\lesssim$ 0.253  			&  U	 &  $\lesssim$0.51  				   & U     &   $\lesssim$ 2.53		   & U \\
HD210839  	&   0.07     $\pm$ 0.01  & A	 	 &  0.06    $\pm$   0.01     & A		  &   0.06 $\pm$  0.03 &	B\\
HD151804  		&   0.092   $\pm$ 0.011  & A		 &  0.076   $\pm$   0.01	   & A		 &   $\lesssim$ 0.03 			  &	U\\
HD152408  	&   0.16     $\pm$ 0.02   & A		 &  0.14    $\pm$   0.02 	   & A		 & 	---							& 	\\
HD149404  	&   0.147    $\pm$ 0.017  & A  	 &  0.18	$\pm$        & B		 &   $\lesssim$ 0.67			  & U\\
HD30614   		&   0.07    $\pm$ 0.01 	&  A	 &  0.042  $\pm$   0.005   & A		 &   ---					 & \\\hline
HD37128   		&   0.32     $\pm$ 0.04   & A	 &  0.212   $\pm$   0.025   & A		 &   $\lesssim$ 0.43            &  U \\
HD38771   		&   0.122    $\pm$ 0.015  & A     &  0.063  $\pm$   0.01  	   & A		 &  ---					 &	  \\
HD154090  	&   0.058    $\pm$ 0.008 & A	 &  0.038  $\pm$   0.005   & A        &  ---					&	\\
HD193237  	&   1.4       $\pm$ 0.1     & A     &  1.27    $\pm$   0.13	   &  A		  &  0.875 $\pm$  0.075 & A \\
HD24398   	&   0.093    $\pm$ 0.013  & A	 &  0.043  $\pm$   0.006   &  A		  &  $\lesssim$ 0.054			 & U \\
HD169454  	&   0.21      $\pm$ 0.03   & A    & $\lesssim$ 0.36    		   & U        &---					&\\
HD152236  	&   0.34     $\pm$ 0.04   & A	 &  $\lesssim$ 0.8   		   & U		  & 		---				& 	 \\
HD190603  	&	---					   	& 		 &---						   & 		  &---						& 	 \\
HD41117   		&   0.07     $\pm$ 0.01 	 & A	 &  0.045  $\pm$   0.006    &  A	  &  ---					& 	\\
HD194279		&	0.046    $\pm$0.020    & B	 & 0.0167   $\pm$  0.008    & B		  & ---				    & 	\\
HD198478  	&   0.086    $\pm$ 0.017  &  A	 &  0.043  $\pm$   0.006    & A		  &  ---					&	 \\
HD80077   	&   0.135    $\pm$ 0.017   & A	 &  0.14    $\pm$   0.07	    & B		  &   ---					&	\\
HD53138   		&   0.08     $\pm$ 0.01	 &  A	 &  0.054  $\pm$   0.007    & A		  &  $\lesssim$ 0.0345		&	U\\
CygOB2\#12   &   0.49     $\pm$ 0.06    & A	 &  0.44   $\pm$   0.08	    & A		  &  $\lesssim$ 1.98			&	U\\\hline
HD24912   		&   0.045    $\pm$ 0.007  & A	 &  0.034  $\pm$   0.004    & A  	  &   $\lesssim$ 0.041		& U  \\
HD36861   		&   0.024    $\pm$ 0.008  & A	 &  0.020  $\pm$   0.012     & B		  & ---					&		\\
HD37043   	&   0.019    $\pm$ 0.006  & B	 & ---							&		  &  ---					&	\\\hline
HD149757      &   0.079    $\pm$ 0.011   & A	 &  0.041 $\pm$   0.005		& A		  &   $\lesssim$ 0.054			&	U\\
HD149438  	&   0.053    $\pm$ 0.06  & A	 &  0.039   $\pm$   0.005	 	& A		  &   $\lesssim$ 0.067			&	U \\
\hline\hline\\
\end{tabular}
\label{tablepacsfluxes}
\end{center}
\end{table*}

The maps were processed using the map reconstruction task PhotProject with the Multi-resolution median transform (MMT) deglitching method 
implemented in HIPE (Herschel Interactive Processing Environment, \citealt{Ott2010}). A second deglitching grade was applied. Subsequently, the processed mini-scan maps were combined into a final map (Figure\,\ref{pacsexample}). We observed that whereas for the faint sources a second deglitching does not makes major changes in the final maps, for our brightest sources this step improved the accuracy of the recovered fluxes. Therefore, for the photometry in our sample stars we decided to use the second deglitching processed maps. 
Since the observing strategy focused on flux observations at the blue and green bands, the quality of the final extra red maps are generally poor and just a few of sources have been detected at this band, most of them giving upper limits. Table\,\ref{tablepacsfluxes} displays the observed sources and the bands where they were detected. Only two sources, of the initial sample, have not been detected or reliably detected, HD\,190603 and CygOB2\#7, respectively. Thus, our final sample is conformed of a total of 25 OB stars (Tables\,\ref{tabledistances} and \ref{tablestellarparams}). \\

\noindent{\textit{ \sc Herschel/PACS}} Photometry\\

To perform the point source aperture photometry we used {\textit{Hyper}} (\citealt{Traficante2015}). This routine was initially designed for FIR point source photometry in complex sky regions in the framework of the {\sc Herschel} infrared survey of the Galactic plane (Hi-Gal). Because of its modularity and versatility, this code can be easily adapted to our {\sc Herschel-PACS} observations, and it has been successfully used by different authors already (\eg\ \citealt{Benedettini2015}, \citealt{Svoboda2016}, \citealt{Li2018}, \citealt{Paulson2020}). 

{\textit{Hyper}} combines 2D multi-Gaussian fitting with aperture photometry to provide reliable photometry in regions with variable background, and in crowded fields. The 2D Gaussian fitting takes into account the beam observations and the source elongation to estimate the region over which to integrate the source flux, \ie\ it computes the PSF of the source onto the final map. The background was locally evaluated and removed for each source using polynomial fits of various orders. The code selected the background based on the lowest r.m.s of the individual residual maps. For most of the sources, a polynomial with order not larger than 2 was used. In addition, in the case of blended sources, {\textit{Hyper}} would perform a simultaneous multi-Gaussian fit of the main source and its companion(s), subtracting the modelled companion(s) from the science target. This was no the case for any of our targets. Finally, although this code allows simultaneous multi-wavelength photometry, we performed a careful and individual aperture photometry for each source, by defining a map-region around the sources of interest. This allowed to optimise the 2D-gaussian fitting procedure for the faintest sources, as well as flux estimations. Moreover, for those maps where the background varies significantly along the whole field of view, this enables a proper detection of the target, even in the case where only upper flux limits were estimated.

The final fluxes of the OB stars at the blue, green and red bands were obtained applying the aperture and colour corrections to the extracted fluxes of the sources. We estimated the uncertainty in the measurements as well as the S/N of the detection.

The error in the estimated final flux includes the uncertainty in the maps calibration from PACS (10\%), the error from the photometric procedure (from {\it{Hyper}} outputs) and from the aperture and colour corrections applied to the extracted flux. These corrections are needed to {\it{i)}} account for the missing flux due to the finite aperture; and {\it{ii)}} convert monochromatic flux densities in the PACS data products, which refer to a constant energy spectrum, to the true object Spectral Energy Density (SED) flux densities at the reference band wavelengths 70, 100 and 160\,\micron, using the tabulated aperture and colour correction factors to the extracted PACS flux as 1/$f_{\rm apc} f_{\rm cc}$\footnote{\url{https://www.cosmos.esa.int/documents/12133/996891/PACS+Photometer+Passbands+and+Colour+Correction+Factors+for+Various+Source+SEDs}}.

Finally, attending to the S/N we flagged flux values with quality flags (qF$_{\lambda}$) as: 'A' for detections with S/N\,$\ge$\,10, and 'B' for detections with 5\,$\le$\,S/N\,$<$\,10. In addition, measured fluxes with S/N\,$\le$\,5 or affected by image artifacts, are flagged as upper limits (U). 
In Table\,\ref{tablepacsfluxes} we summarise the obtained flux values and the corresponding errors at 70, 100 and 160\,\micron\ of the sample. 
 
\subsection{V/NIR and MIR observations}\label{datanir}

The V, NIR and MIR magnitudes used in this work are listed in Table\,\ref{tabledata} and were mostly collected from the literature (see references within). All sources are observed in the V and NIR bands, most of them also at MIR wavelengths. Since these observations were made with different instruments (different photometric systems) we performed an absolute flux calibration. We used Vega as a calibrator to convert the magnitudes into absolute fluxes by extrapolating the visual absolute flux calibration of Vega to a specific wavelength by model atmosphere (Kurucz models) and calculating the absolute fluxes of Vega in the different photometric systems. A detailed description of the absolute flux calibration procedure can be found in \cite{Puls2006}.

 \subsection{Millimeter and Radio Observations}\label{dataradio}
 
 Millimeter and radio observations were collected from literature (see Table\,\ref{tabledata} and references therein). Most of our sample stars were observed with the Very Large Array (VLA) at 2, 3.5 and 6\,cm or with the Australia Telescope Compact Array (ATCA) at 3.6 and 6\,cm in different epochs. In addition, several objects were also observed at 1, 3, 13, 20 and 90\,cm (J-VLA) and at 21\,cm (e-MERLIN). Only for one object, HD\,53138, no radio observations are available, whereas for 7 of the 25 targets only upper limits could be obtained. 
 
 With respect to the sub-mm and mm flux continuum measurements, we collected from the literature those obtained at 0.7, 0.85, 1.3 and 1.35\,mm, when available. Only 7 stars in our sample have millimeter observations (Submillimetre Common-User Bolometer Array, SCUBA).

\subsection{\textit{GAIA} distances}\label{gaia} 
This work uses distance determinations based on \textit{GAIA} DR2 parallaxes (\citealt{Luri2018}, \citealt{Gaia2016, Gaia2018b}).
We only used parallaxes where the associated errors are lower or similar to 15\% (8 out of 25 objects; marked with an asterisk in Table\,\ref{tabledistances}), otherwise the obtained distance becomes unreliable when compared to previous estimates (see below). The differences between \textit{GAIA} DR2 and older distance estimates ranges from 0.01\,kpc to $\approx$\,1\,kpc, with an average of 0.47\,kpc. The largest deviations correspond to HD\,169454, HD\,80077 and CyOB2\#12, whose differences with the new values are $\sim$\,0.6, 1.05\, and 0.9\,kpc, respectively. 

In particular, for CygOB2\#12 the distance derived from \textit{GAIA} DR2 parallaxes (0.85\,kpc) puts the star much closer than its association, Cygnus OB2 (1.75\,kpc). This casts some doubts about its physical properties (\citealt{naze2019}). For this source we provide the results of the analysis for both distances (two entries in Tables\,\ref{tableclfactors}, \ref{tableclbestfit} and \ref{tablebetachange}), although only the result pertaining to the larger, association distance (1.75\,kpc) is considered in the discussion\footnote{Recent \textit{GAIA} Early DR3 parallax estimates for CygOB2\#12 relocated the star within its stellar association, at $\sim$\,1.78\,kpc. For the rest of our sample DR3 parallaxes are similar to \textit{GAIA} DR2 measurements, except for HD\,210839 and HD\,36861, whose new distances are a factor 1.4 larger than implied from \textit{GAIA} DR2 (see discussion in Sect.\,\ref{modeling}).}.

The case of HD\,80077 is different. Since there were always severe doubts about this star being a membership of the Pismis\,11 cluster (located at 3.6\,kpc; \citealt{Marco2008}), and given the lower uncertainty in the parallaxes by \textit{GAIA} DR2, we keep this distance (2.55\,kpc) for the analysis. 

For the rest of the sample, initially we used photometric distances from the literature (see Tables\,\ref{tabledistances} and \ref{tablestellarparams}), whose uncertainties range between 0.1 and 0.4\,kpc\footnote{Most authors assumed absolute magnitude uncertainties between 0.3 and 0.5 magnitudes.}, and locate most of the stars in their associated cluster. For HD\,152236 ($\tau^1$\,Sco) we assumed a distance of 1.64\,kpc (\citealt{Sana2006}), used in previous studies (\citealt{Clark2012}).

In Table\,\ref{tabledistances} we present the final used distances and the derived stellar radii and extinction parameters for the sample as described in Section\,\ref{modeling}. For HD\,66811 ($\zeta$\,Pup), in order to be consistent with previous analyses by \citetalias{Puls2006} (see Table~\ref{tablestellarparams}), we used two different distances: a commonly used shorter distance (0.46\,kpc), assuming $\zeta$\,Pup is part of Vela OB2 association, and a larger one (0.73 kpc, \citealt{Sahu1993}), which corresponds to the star being runaway.  
Although we provide the results of the analysis for both distances (two entries in Tables\,\ref{tableclfactors}, \ref{tableclbestfit} and \ref{tablebetachange}), the discussion in this work refers to the shorter distance, unless otherwise specified.

\begin{table}[!ht]
\begin{center}
\caption{Final used distances and associated errors, reddening parameters and derived stellar radii. Sources are sorted as in Table\,\ref{tabledata}. Asterisked values in the distance column correspond to \textit{GAIA} DR2 distances, and the two values for HD\,66811 and CygOB2\#12 refer to different alternatives for these sources (see Sec.\,\ref{gaia}), quoted with their associated asymmetric errors. Some parameters have been modified from the initial values found in the literature (see Table\,\ref{tablestellarparams}) by the de-reddening procedure (see Sec.\,\ref{modeling}).}
\small{
\begin{tabular}{l l l l l l}
\hline\hline
Source 	&	$d$ (kpc)	& e$_{\rm d}$ (kpc)	&	E(B-V)	&	R$_{\rm V}$	& \Rstar\ (\Rsun)\\\hline
HD66811\ \ \ \ \ \ \ \       &  0.46\ \ \ \ \ \ \  	& 0.1 \ \ \ \  	&	0.04 \ \ \   	& 	3.1 \ \ \ \   &	18.6 \  \\
			&  0.73 	& 0.14	&	0.04 	& 	3.1 &	29.7\\ 
CyOB2\#11	& 	1.72* 	& $_{-0.08}^{+0.08}$	&	1.745 	& 	3.1 &	23.15 \\   
HD210839	&  0.62*   & $_{-0.04}^{+0.05}$	&	0.485	&	3.1	&	13.35 \\   
HD152408	&	1.68 	& 0.4	&	0.5 	&	2.98 &	30.0 \\ 
HD151804	&	1.8  	& 0.4	&	0.38 	&	3.13 &	36.0\\
HD149404  &	1.3		& 0.3	&	0.68 	&	3.4 &	37.0\\
HD30614   & 	0.789	& 0.1	&	0.25	&	3.5 &	20.70\\\hline 
HD37128   &	0.342	& 0.1	&	0.413	&	3.1 & 	24.68\\
HD38771   &	0.4 	& 0.1	&	0.04	&	3.1 &	23.0\\ 
HD154090	&	1.096 	& 0.25	&	0.59 	&	2.7 &	39.0\\ 
HD193237  &	1.7 		& 0.2	&	0.58 	&	3.1 & 	97.14\\
HD24398   & 0.303* & $_{-0.002}^{+0.003}$ & 0.30 	&	2.9	&	21.12\\ 
HD169454  &	2.13* 	& $_{-0.19}^{+0.23}$	&	1.13 	&	3.19 &	98.2\\  
HD152236  &	1.64 	& 0.4	&	0.7 	&	3.18 & 104.0\\
HD41117     &	1.5 		& 0.2	&	0.415	& 	3.05 &	61.0\\
HD194279  &	1.71*  	& $_{-0.07}^{+0.08}$	&	1.23	&	3.2 &	78.4\\  
HD198478  &	0.83	& 0.1	&	0.51	&	3.0 &	35.8\\
HD80077   & 	2.55*  & $_{-0.19}^{+0.22}$	&	1.5  	& 	3.2 &	167.0\\ 
HD53138    &	1.135   & 0.25	&	0.01 	&	3.2 &	65.0\\ 
CyOB2\#12 &	1.75	& 0.16	&	3.38	&	3.0 &	240.0\\
			 &	0.85*	& $_{-0.08}^{+0.1}$	&	3.38	&	3.0 &	116.7\\\hline
HD24912 	 &	0.854  & 0.1 	&	0.3  	&	3.18 &	23.4\\
HD36861    &	0.27* 	& $_{-0.03}^{+0.05}$	&	0.073	&	5.0 &	7.8 \\ 
HD37043    &	0.5		& 0.1	&	0.035	&	5.0 &	17.9 \\\hline
HD149757	 &	0.165	& 0.023 &	0.29	&	2.88 &	8.9\\
HD149438   &	0.167 	& 0.023 &	0.012 	& 2.8 &	5.65\\	 
 \hline\hline
\end{tabular}
\label{tabledistances}}
\end{center}
\end{table}

\section{Modeling}\label{modeling}

To investigate the wind clumping stratification and mass-loss rates of the 25 OB stars in our final sample, we modeled their observed spectral energy distribution (SED) from V/NIR to radio wavelengths. We used the interactive procedure developed by \citetalias{Puls2006}, which is based on continuum emission models (photospheric plus wind emission), assuming a spherically symmetric wind calibrated against full NLTE stellar atmosphere models, and accounting for optically thin clumping. Although the basic method thus neglects potential effects from porosity (see Sec.\,\ref{intro}), these should be relatively small for the diagnostics and spectral ranges considered in this work (see \citealt{Sundqvist2018b} for a discussion). Below we present a summary of the approach to model SEDs depending on the stellar and wind parameters of the objects. An in-depth description and verification of the method can be found in \citetalias{Puls2006}.

\subsection{Infrared and radio flux emission}

The infrared and radio fluxes are calculated using the approximations described by \cite{Lamers1984a}. Following \citetalias{Puls2006}, we use:\\ 

\noindent a) \textit{Wind velocity law.}
\begin{equation}
\upsilon (r)=\vinf\left(1 - \frac{b}{r} \right)^{\beta} \label{eqn.2} ,
\end{equation}

\noindent with $b = 1-(\upsilon_{\rm min}/\vinf)^{1/\beta}$ and the minimum velocity, $\upsilon_{\rm min}$, set to 10 km\,s$^{-1}$.\\

\noindent b) \textit{Electron Temperature.}\\ The electron temperature was computed using Lucy's temperature law (\citealt{Lucy1971}) with a lower temperature cut-off at 0.5\,\teff. \\

\noindent c) \textit{Ionisation equilibrium.}\\For all the objects in our sample, hydrogen is considered to be completely ionised, whereas the helium ionisation structure used in the modeling of the SED depends on the temperature of the source and the wavelength domain (for the rationale see \citetalias{Puls2006}). Specifically, for the majority of the sample (17\,000\,K\,$\lesssim$\,\teff\,$\lesssim$\,35\,000\,K) helium is considered singly ionised in the radio regime whereas for the coolest (\teff\,$\lesssim$\,17\,000 K) and hotter objects ( \teff\,$\gtrsim$\,35\,000 K), it is assumed to be neutral and fully ionised, respectively. In the NIR, helium is assumed to be fully (\teff\,$\gtrsim$\,32\,500 K) or singly (13\,000\,K\,$\lesssim$\,\teff \,$\lesssim$\,32\,000\,K) ionised. A particular treatment is required for P\,Cyg (HD\,193237) whose ionised helium structures depart from the above standard scalings with effective temperature (\citealt{Najarro1998}). As such, for P\,Cyg we assumed that helium is fully recombined in the outer parts of the wind (radio domain) and singly ionised in the inner ones (NIR domain). \\

\noindent d) \textit{Photospheric input fluxes.}\\ For $\lambda < 1\,\micron$, where the stellar photosphere dominates, the emitted flux is modeled using Kurucz's fluxes. On the other hand, for $\lambda > 1\,\micron$,  where the wind starts to dominate the resulting flux, a black body emission model is used.\\ 
 
\noindent e) \textit{Wind Clumping treatment.}\\
 Clumping is included using the following approach: all material in the wind is redistributed into clumps which are over-dense with respect to the average density, we assume optically thin clumps (see above), and further that the inter-clump medium is effectively 
void (\citealt{Abbott1981}, \citealt{Schmutz1995}). 

Under these assumptions, the spatial mean density ($\langle\rho\rangle = \Mdot/4 \pi r^2 \upsilon$) and mean squared density can be expressed as a function of the volume filling factor ($f_{\rm v}$) as:   
\begin{equation}
\langle\rho\rangle = f_{\rm v}\ \rho^+, 
\label{eqn.3}
\end{equation}
\vskip-18pt
\begin{equation}
\langle\rho^2\rangle = f_{\rm v}\ (\rho^+)^2, 
\label{eqn.4}
\end{equation}
\noindent where $\rho^+$ denotes the density inside the clump. Thus, the clumping factor (\fcl) as described by Eq.\!\,\ref{eqn.1}, 
\begin{equation} 
f_{\rm cl} = f_{\rm v}^{-1},
\label{eqn.5}
\end{equation}
\vskip-12pt
\begin{equation}
\rho^+ = f_{\rm cl}\ \langle\rho\rangle \ ,
\label{eqn.6}
\end{equation}
\noindent becomes the inverse of the volume filling factor, describing the over-density of the clumps as compared to the mean density.

Within this approach, the opacity depends only on the matter and the physical processes inside the clumps (recombination, scattering, absorption, etc.). Since the mean opacity of the processes with opacities depending on $\rho^2$ is enhanced by a factor $f_{\rm cl}$ (see Sec.\,\ref{intro}), the optical depth invariant for thermal emission, $Q'$ (\citealt{Lamers1984a}, \citetalias{Puls2006}), is modified as:
\begin{equation}
Q' =\dfrac{\dot M \sqrt f_{\rm cl}}{R_{\star}^{3/2}}.  
\label{eqn.7} 
\end{equation}

Therefore, since the opacities of all processes considered in this analysis depend on $\rho^2$ (IR/mm/radio emission), the effects of clumping can be included in the models by multiplying the opacities evaluated for the mean wind by a clumping factor. Thus, the theoretical-emitted flux (\citealt{Panagia1975}, \citealt{Wright1975}), 
\begin{equation}
F_{\nu}\propto \dfrac{\dot M^{4/3}}{d^2}, 
\label{eqn.8}
\end{equation}
for such processes might be extended according to: 
\begin{equation}
F_{\nu}\propto{\left\lgroup\dfrac{\dot M \sqrt \fcl}{R_{\star}^{3/2}}\right\rgroup}^{4/3} {\left\lgroup\dfrac{R_{\star}}{d}\right\rgroup}^{2} \equiv Q'^{\,4/3} {\left\lgroup\dfrac{R_{\star}}{d}\right\rgroup}^{2}
\label{eqn.9}
\end{equation}

Consequently, the emitted flux is sensitive to changes in clumping factors, allowing us to constrain the clumping structure of the wind by fitting emission models to flux observations for a given source. Since clumping is not constant throughout the wind but presents a radial stratification, \fcl ($r$/\Rstar), it is to be evaluated in different regions of the wind.  
   
\begin{table}[!ht]
\begin{center}
\caption{Defined wind region boundaries and the corresponding clumping factors used in this work.}	
\begin{tabular}{c|ccccc}\hline\hline
Region		&	1		&	2				&	3				&	4				&	5		\\\hline
r/\Rstar	&	1....\rin	&	\rin.....\rmid 	&	\rmid....\rout	&	\rout...\rfar	&	$>$ \rfar \\
\fcl		&	1		&	\clin			&	\clmid			&	\clout			&	\clfar	\\\hline
\end{tabular}
\label{tablewindregion}
\end{center}
\end{table} 

\begin{figure}[!ht]
\begin{center}
\includegraphics[width=8.5cm]{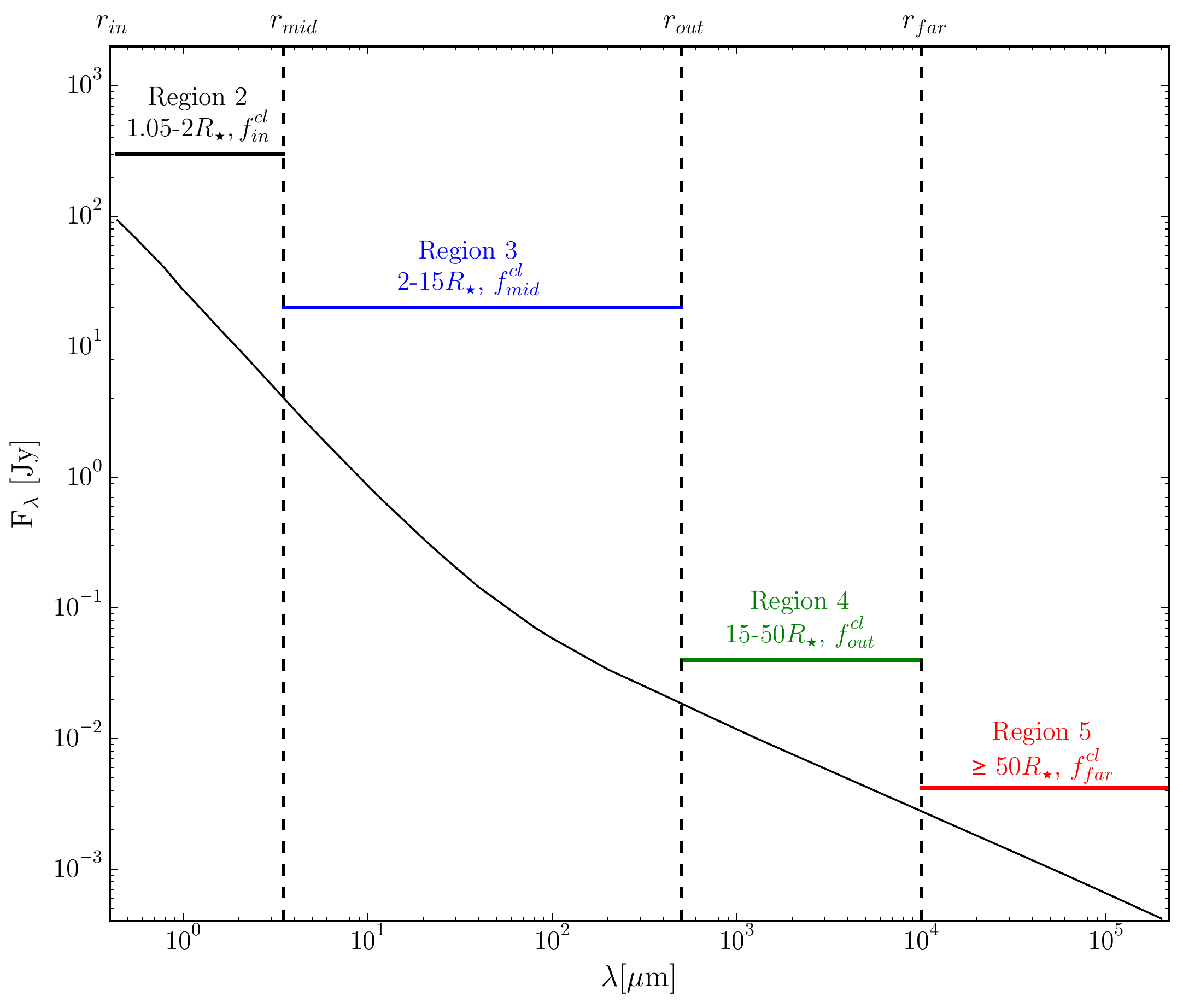}
\caption{Schematic of the wind emission: different wind regions as a function of the spectral range and their corresponding radial distance to the photosphere ($r$), as defined in Table\,\ref{tablewindregion}. The emission model corresponds to the unclumped stellar wind of a O Supergiant star (\teff\,=\,33\,kK, \Mdot\,=\,8.6\,\massloss). 
}
\label{fig.windregions}
\end{center}
\end{figure}

For that objective, the wind is divided in different regions with corresponding clumping factors as defined in Table\,\ref{tablewindregion}. Typical boundaries are \rinc, \rmidc, \routc\ and \rfarc, which roughly agree to the formation zones of \Ha\ and NIR (Region 1\,--\,2), MIR/FIR (Region 3), mm (Region 4) and radio (Region 5). Note that \textit{i)}\,the clumping at the base of the stellar wind (1\,$<$\,r/\Rstar\,$<$\,\rin) is set to 1 within the fitting procedure, and \textit{ii)}\,that the above-mentioned limits for the wind regions in Table\,\ref{tablewindregion} can be adapted within the fitting procedure when necessary (see below). Figure \ref{fig.windregions} displays a sketch of the defined wind regions as a function of the spectral energy distribution.\\

The parametrisation adopted in this work is designed to empirically constrain the clumping factor in different radial wind zones. Therefore, the derived clumping factors should be considered as average values describing the global behaviour of clumping throughout the wind. Since all used diagnostics depend on $\rho^2$, it is not possible to derive absolute values for clumping factors and mass-loss rates, but only relative ones. What we really obtain is the scaling invariant $Q' \propto \dot M \sqrt f_{\rm cl}$ for a given radius interval. To derive absolute values, a simultaneous analysis of $\rho$-dependent diagnostics (e.g. resonance lines, see Sec.\,\ref{intro}) would be required to break the degeneracy, which is out of the scope of this work. However, since all derived (minimum) clumping factors and (maximum) mass-loss rates can be scaled via $Q'$, it is still possible to carry out important comparisons with theoretical predictions as well as with previously derived empirical values.

It is worth noting that our fitting-results are independent of the individual values for distance and stellar radius, and that they can be easily scaled to different values 
as long as Q' and (\Rstar/$d$) remain conserved (see Eq. \!\,\ref{eqn.9}). Of course, changes in distances (and thus in stellar radius) affect directly the derived absolute values of mass-loss rates, but it does not affect the behaviour of the clumping structure (e.g. HD\,66811 and CygOB2\#12, see Table\,\ref{tableclfactors}). 

All comparisons with other empirical studies and theoretical predictions were performed for the same values of distance and stellar radius, for a given source. Thus, for comparisons with other empirical studies, their mass-loss rates were scaled to the distance and stellar radius used in our analysis, whenever necessary. Similarly, stellar parameters used to compute theoretical predictions for \Mdot\ (Section\,\ref{comparison}) are the same as those used to derive the empirical \Mdotmax\ estimates. Regardless of variations in absolute values of mass-loss rates with changing distance to the objects, the general behaviour found in this work will thus be unaffected, and the ratio between empirical and theoretical mass-loss rates is remains conserved. For example, we estimate that a 40\% increase in distance\footnote{40\% is the largest variation in distance between \textit{GAIA} DR2 and early DR3 measurements for our sample, see Section\,\ref{gaia}.} leads to a decrease in empirical to theoretical mass-loss rates of around 30\%. In other words, such changes in distance would not change the overarching trends and conclusions found in this work.

 \subsection{Fitting Procedure}\label{fittingprocedure}

 \begin{sidewaystable*}
\begin{center}
\caption{Sample of analysed OB stars in this work and the initial set of stellar and wind parameters from the literature (`ref' column) used as input values in our simulations. log g indicates gravities without centrifugal correction. For HD\,66811 ($\zeta$\,Pup) we provide two entries, based on different distances (\citetalias{Puls2006}).
}
\begin{tabular}{l l c c c c c c c c c c l}
\hline\hline
Source & Spec.Type &	\teff & log g  & \Rstar &\xhe & \vinf 		& $\beta$	& \Mdot  						  & $d$		&	ref	& Alt. Name\\
		&			   &(kK)  &			& (\Rsun)& (number)		    &(km s$^{-1}$) & 			& (10$^{-6}$ \Msun yr$^{-1}$) &	(kpc)	& 		&		\\\hline
HD66811    & O4 I (n)f  	&39.0   	& 3.60 	& 18.6		& 0.2 	& 2250 	& 0.9	& 4.2	 	&  0.46		& 8		& $\zeta$\,Pup\\
			&				&			&			& 29.7		&		&			&		& 8.5		&	0.73		& 8		&\\
CyOB2\#11 & O5 If+     	& 36.5  	& 3.62 	& 23.6 	& 0.1 	& 2300 	& 1.1	& 5.0 	& 1.71  	& 8 	&\\   
HD210839  &O6 I (n)f    	& 36.0  	&3.55 		& 23.3 		& 0.1 	& 2250 	& 1.0	& 3.0  	& 1.077 	& 8 	& $\lambda$\,Cep\\  
HD152408  &O8 Iafpe    	& 33.0		& 3.20 	& 30.0		& 0.2 	& 955 		& 2.1	& 8.4  	& 1.68 		& 7a 	&\\ 
HD151804   &O8 Iab      	& 32.8		& 3.25		& 35.0 	& 0.1   & 1450  	& 2.0	& 6.4  	& 1.8  		& 7a 	& V973\,Sco\\
HD149404  &O9 Ia        	& 34.0  	& 3.55 	& 37.0 		& 0.1 	& 2450  	& 1.3  	& 1.95 	& 1.3		& 9  	&\\
HD30614    &O9.5 Ia     	& 29.0  	& 3.0		& 20.7 	& 0.1	& 1550 	& 1.15	& 1.5   &0.8	 	& 8 	& $\alpha$\,Cam\\\hline 
HD37128    &B0 Ia 	   	& 27.0		& 2.9		& 24.0 	& 0.2	& 910 		& 1.5	& 2.5	& 0.363	& 2	 	& $\epsilon$\,Ori\\
HD38771    &B0.5 Ia     	& 26.5 	& 2.9 		& 22.2		& 0.2	& 1525		& 1.5	& 0.9  	& 0.4 		& 2 	&$\kappa$\,Ori\\ 
HD154090  &B0.7 Ia     	& 22.5 	& 2.65		& 36.0  	& 0.2  	& 915 		& 1.5	& 0.95 	& 1.096 	&2 		&$\kappa$\,Sco\\ 
HD193237  &B1 Ia        	& 16.7  	& 2.0  		& 97.14 	& 0.3  	& 195  		& 2.5	& 5.4  	& 1.7 		&7b	&P\,Cyg\\
HD24398   &B1 I         	& 20.26	& 2.95 	& 23.0  	& 0.2  	& 1295 	& 1.5	& 0.23 & 0.28 	& 11 	&$\zeta$\,Per\\
HD169454  &B1.5 Ia+ 	   	& 20.4 	& 2.35 	& 74.0  	& 0.1   & 850  	& 2.1	& 6.4  	& 1.54 		& 7a	&\\ 
HD152236  & B1.5 Ia+   	& 18.0 		& 2.0  		& 112.4  	& 0.2 	& 390 		& 2.0	& 6.0  	& 1.995 	& 2 	&$\zeta^{1}$\,Sco\\
HD41117     &B2 Ia       	& 19.0  	& 2.35		& 61.9  	& 0.2  	& 510  		& 2.0	& 0.9  	& 1.5 		& 2 	&$\chi^{2}$\,Ori\\
HD194279  &B2 Ia        	& 19.0  	& 2.3  		& 44.7  	& 0.2  	& 550  	& 2.5	& 1.05 	& 1.202 	& 2 	&\\ 
HD198478  &B2.5 Ia+ 	   	& 17.5  	& 2.12 		& 49.0  	& 0.1-0.2	& 200....470 & 1.3	& 0.117.....0.407  &0.83	& 4& 55\,Cyg\\
HD80077   &B2/B3 Iae  	& 15.1 		& 1.85 		& 243.0 	& 0.35	& 250 		& 3.0	& 1.18  & 3.6  		&  1, 3, 6b  &\\
HD53138    &B3 Ia       	&15.5 		& 2.05 	& 65.0  	&  0.2 	& 865  	& 2.0 	& 0.36 & 1.096   & 2 &$o^{2}$\,CMa\\ 
CyOB2\#12  &B3-4Ia    	&13.7 		& 1.7   	& 246.0 	& 0.2  	& 400  	& 1.5	& 15.0 	& 1.75		&1 		&\\\hline
HD24912 &O7.5 III (n)((f))& 35.0 		& 3.5   	& 24.2  	& 0.15 	& 2450  	& 0.9 	& 2.3  & 0.854   	& 8 	& $\xi$\,Per\\
HD36861    &O8 III((f)) 	& 33.6 	& 3.56  	& 14.4  	&0.1  	& 2400  	& 0.9	&	0.4 & 0.5 		& 8 	& $\lambda$\,Ori\,A\\
HD37043    &O9III       	& 31.4 		& 3.5    	& 17.9  	&0.12  	& 2300 	& 0.9	&0.8./0.25 & 0.5 & 8		&$\iota$\,Ori\\\hline
HD149757  &O9.5 Vnn		& 32.0 	& 3.65  	& 8.9   	&0.17 	& 1550 	& 0.8	& $\lesssim$ 0.18 & 0.154 & 10a,10b &$\zeta$\,Oph\\
HD149438  &B0.2 V		&31.9 		& 3.99  	& 5.3 		&0.12 	& 2400 	& 0.8	& $\lesssim$ 0.0614 & 0.167  & 6, 11 & $\tau$\,Sco \\	 
 \hline\hline
\end{tabular}
\tablebib{[1] \citet{Clark2012}; [2] \citet{Crowther2006}; [3] \citet{Marco2008}; [4] \citet{Markova2008}; [5] \citet{Morel2004}; 
[6] \citet{Mokiem2005}; [7a] \citet{Najarro1995}; [7b] Najarro (priv. comm.); [8] \citet{Puls2006}; [9] \citet{Raucq2016}; [10a] \citet{Repolust2004}; [10b] \citet{Repolust2005}; [11] \citet{SimonDiaz2006}.
}
\label{tablestellarparams}
\end{center}
\end{sidewaystable*}

We compare continuum emission models with the observed SED for the 25 OB stars in the sample. In Table\,\ref{tablestellarparams} we summarise the stellar and wind parameters of the sample from the literature (references therein), used as input values in our simulations. The best-fit model for each source is obtained from a simple maximum likelihood method ($\chi^2$) as follows:  
  
First, we de-reddened the observed flux using the extinction law provided by \cite{Cardelli1989}. By comparing the observed VJHK fluxes with theoretical flux emission predictions we derived values for the colour excesses E(B-V) and their corresponding R$_{\rm V}$.

Secondly, we computed the stellar radius. For a given distance, the initial value of the stellar radius is adapted to match the VJHK de-reddened observed fluxes. Since flux is diluted by (\Rstar/$d$)$^2$ (see Eq.\!\,\ref{eqn.9}), the ratio between radius and distance represents a scaled flux factor, which has to be conserved for any distance. In Table\,\ref{tabledistances} we present the final used distances, and the corresponding derived stellar radiii and de-reddening parameters.
 
Thirdly, we estimated mass-loss rates required to reproduce the continuum distribution across the electromagnetic spectrum.
Since the optical depth and the diluted flux scale with \Mdot/\Rstar$^{3/2}$, this ratio has to be conserved for a given (\Rstar/$d$). 

Initially, all clumping factors in our simulations are set to the minimum value, \clin= \clmid= \clout= \clfar= \fclmin=1 (unclumped wind) and the mass-loss rate is adapted to reproduce the observed fluxes throughout the wind (1.05 \Rstar\ $<$ r $<$ 100 \Rstar). Note that the emitted flux decreases on average with increasing wavelength and, therefore, radio thermal emission represents the lowest absolute flux values for a certain source\footnote{Assuming thermal emission from single, non-magnetic stars. }. As such, the derived \Mdot\ represents the maximum possible mass-loss rate (\Mdotmax) consistent with radio observations assuming a minimum clumping condition (\clfar=\,\fclmin=\,1). In the cases where radio fluxes are well-defined but somehow larger than at shorter wavelengths, the estimated \Mdotmax\ relies instead on the consistent observed fluxes at those wavelengths (usually the FIR domain). The same is true  
for those sources whose radio observations are not well constrained, although in this cases the adopted \Mdotmax\ is only an upper limit. 

Finally, the radial stratification of the clumping is derived by adapting the values of the clumping factors to match the continuum emission. The best-fit solution for each source corresponds to the emission model (\Mdotmax, \Rstar, d and \clin, \clmid, \clout, \clfar) providing the minimum value of the statistics fitting function $\chi^2$. Unlike \citetalias{Puls2006}, we follow two different approaches: \textit{i)} \textit{Fixed-regions approach}. Here, we fix the defined radial boundaries (see Table\,\ref{tablewindregion}) for all sources. This allows us to investigate consistently the global behaviour of the clumping stratification for our sample, comparing the derived average clumping factors across the wind regions. \textit{ii)} \textit{Adapted-regions approach}. In these simulations, the limits of the wind regions can be adapted, in addition to the clumping factors ---when necessary--- to secure the best possible fit compared to fixed-region approach. This allows us to  probe also the boundaries defined above for the different wind regions.

\subsection{Additional considerations}\label{additionalconsiderations}

In order to be consistent, besides setting \rout\,=\,15\,\Rstar\ in the fixed-regions approach in our simulations, we also probe \rout\,= 10\,\Rstar, since this limit was used by \citetalias{Puls2006} when analyzing low or intermediate density winds (where, in their case, the profile shape of \Ha\ served as a proxy for wind-strength.)

Additionally, due to the relative scarcity of observations in the sub-mm regime, we computed the maximum clumping values for Region 4 still compatible with flux emission at shorter and longer wavelengths (\cloutmax\ in Tables\,\ref{tableclfactors} and \ref{tableclbestfit}). For the sources previously analysed by \citetalias{Puls2006}, and with new distance estimates from \textit{GAIA} (asterisked values in Table \ref{tabledistances}), we first tested the derived \Mdotmax\ and clumping stratification by these authors against the obtained FIR flux values and the additional measured fluxes at MIR and radio ranges found in the literature since \citetalias{Puls2006}. In a second step, we scaled the stellar radius and the mass-loss rate to the new distance according to Eq.\ref{eqn.8} and derived the best-fit solution to all data as described above.

\subsection{Prototypical examples}\label{prototypical_examples}

\begin{figure}[!ht]
\begin{center}
\includegraphics[width=9.5cm]{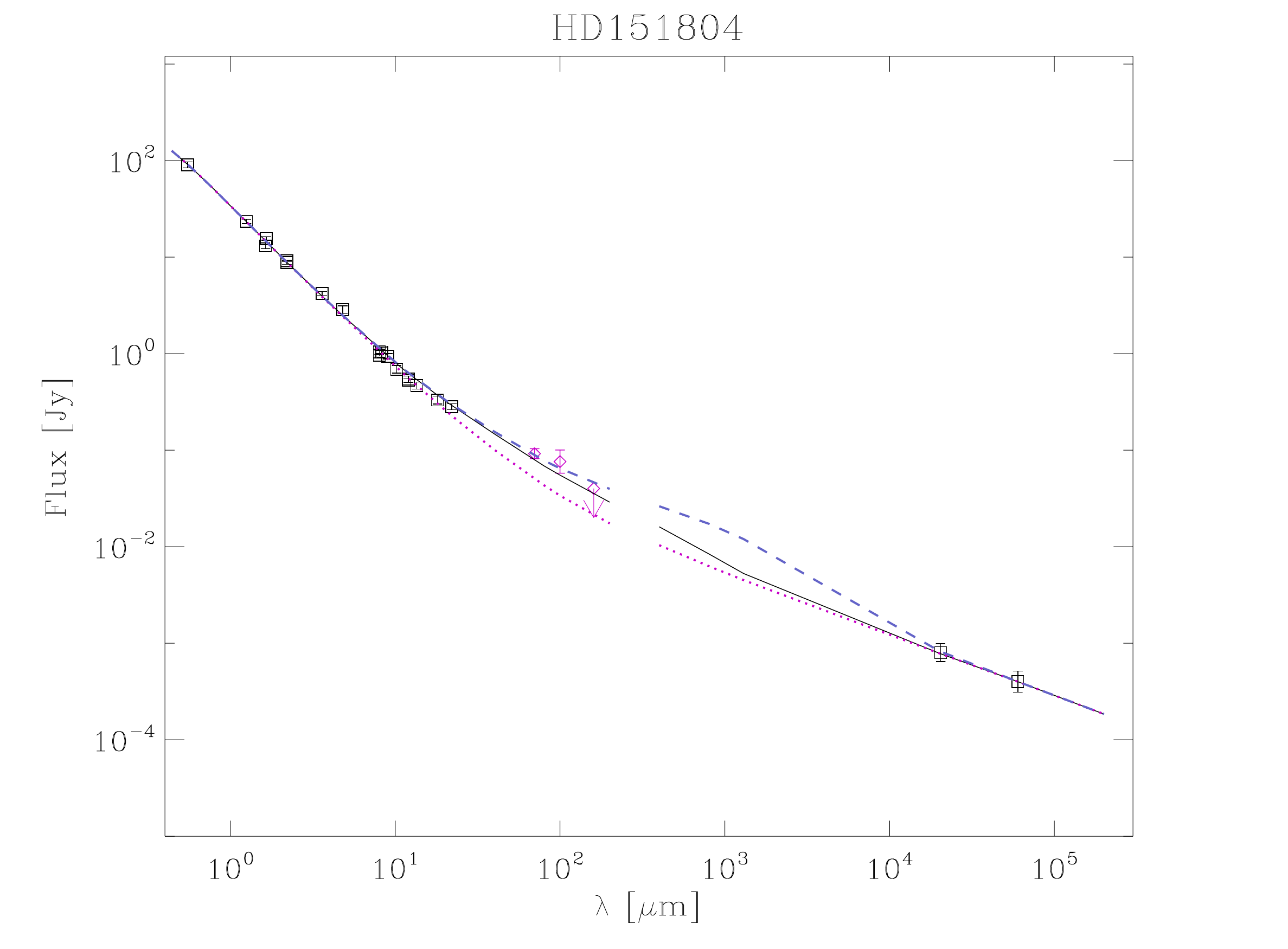}
\includegraphics[width=9.5cm]{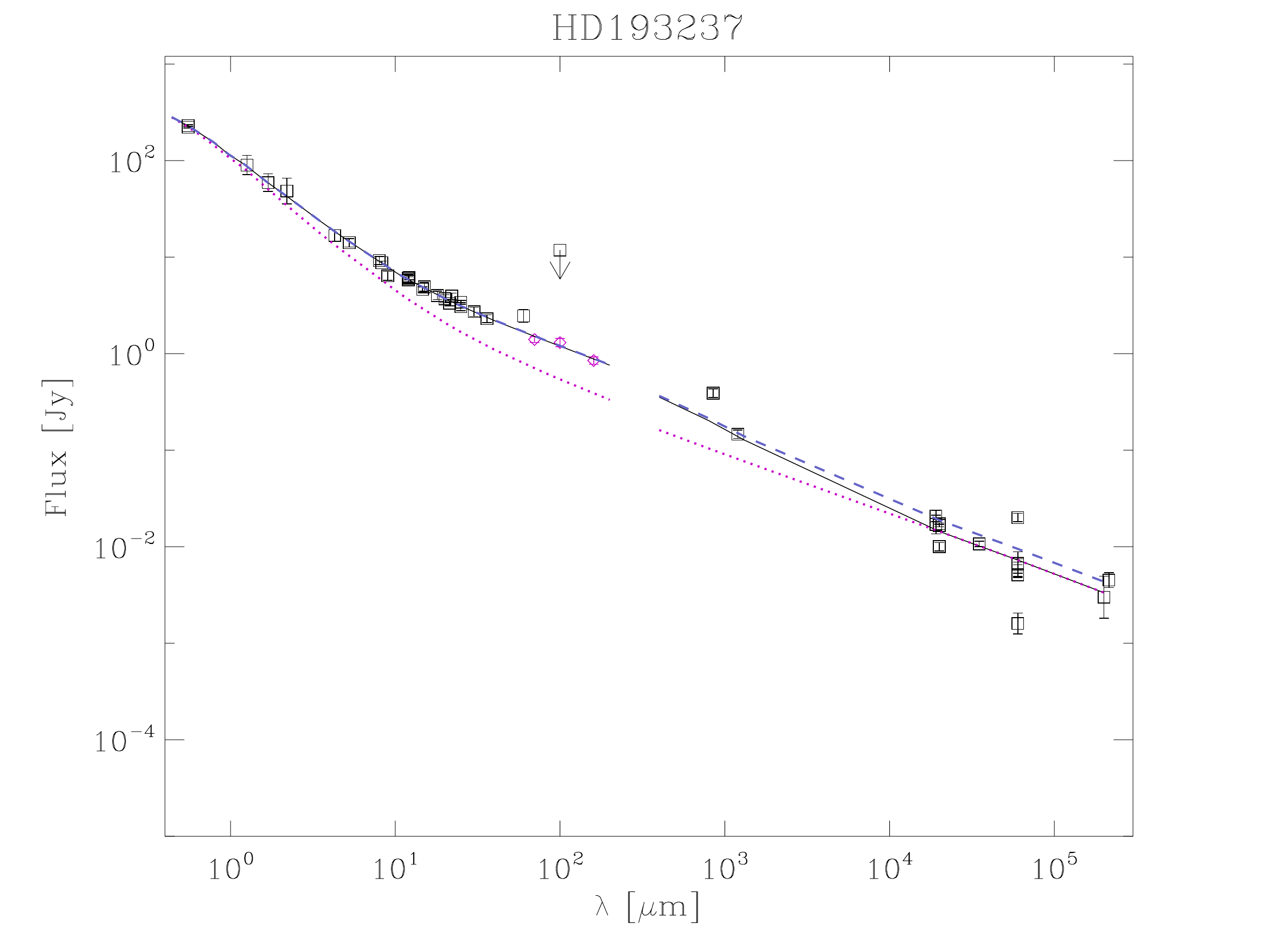}
\includegraphics[width=9.5cm]{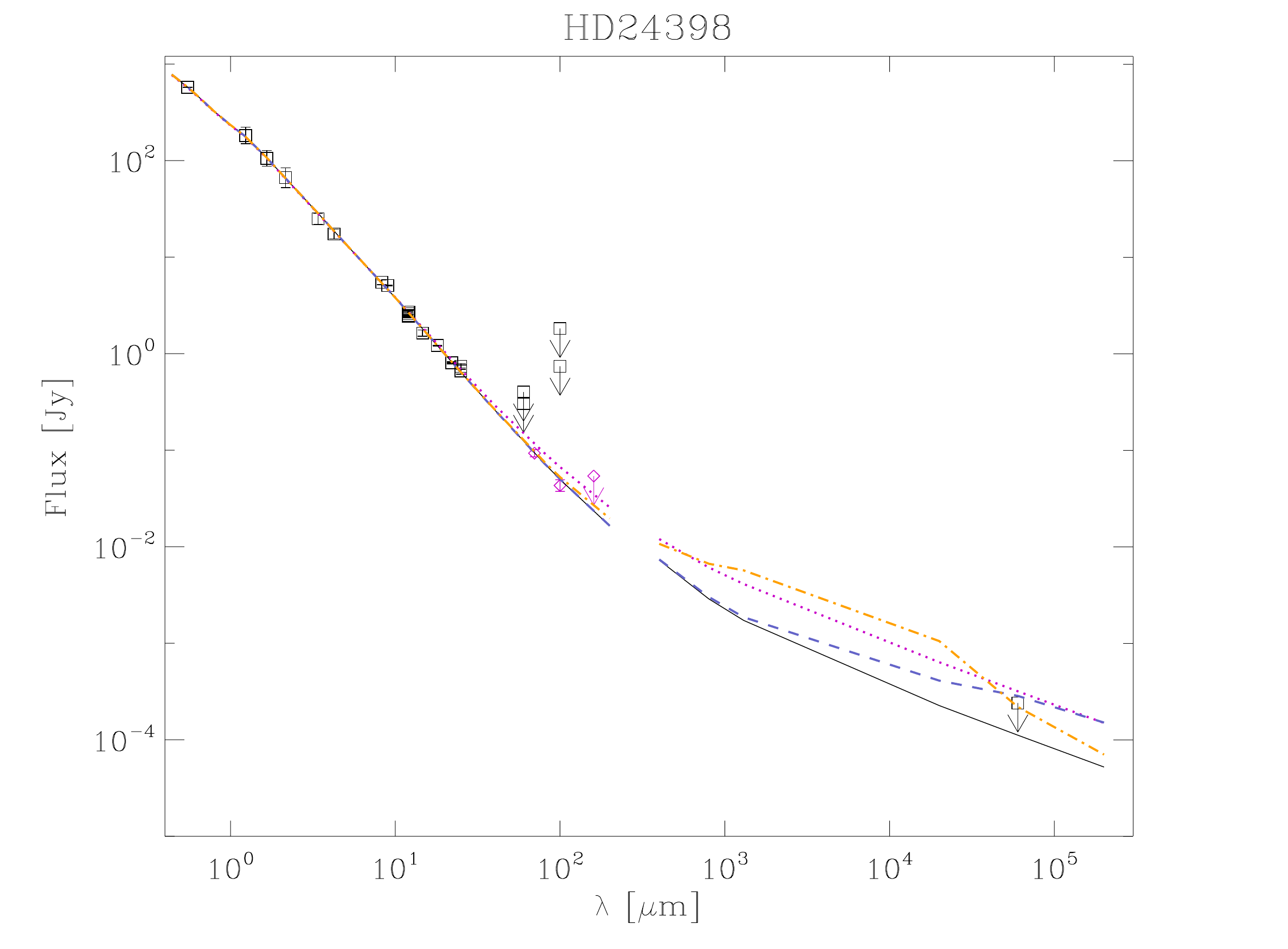}
\caption{From Top to Bottom, observed and best-fit fluxes vs. wavelength in the fixed-regions approach for HD\,151804, HD\,193237 (P\,Cyg), and HD\,24398  ($\zeta$ Per). Solid lines correspond to the best-fits; magenta-dotted, blue-dashed and orange-dashed-dotted lines correspond to different models (see text).}
\label{fig_examples}
\end{center}
\end{figure}   

In this section we briefly describe three prototypical examples of the \Mdotmax\ and clumping factor estimates through the fitting procedures in the fixed-region approach. They cover different possibilities with respect to the flux values, and how arising peculiarities are reflected in the table entries displaying the results for the whole sample (Tables \ref{tableclfactors}, \ref{tableclbestfit}, \ref{tablebetachange}). From top to bottom Figure\,\ref{fig_examples} displays observed and synthetised fluxes for HD\,151804, HD\,193237 (P\,Cyg), and HD\,24398  ($\zeta$ Per). For a detailed discussion of every star in the sample, see Appendix\,\ref{appendixA}.\\

\noindent{\it HD\,151804}

This is one of the most straightforward cases, where the radio flux is a unique, determined value providing a well constrained \Mdotmax. For \Mdotmax\,=\,6.4 \massloss, the synthesised flux perfectly matches the observed values across the wind without the need for clumping (\clin\,=\,\clout\,=\,\clfar\,=\,1), except in the intermediate wind region (Region 3). The synthesized fluxes with all clumping factors set to unity are displayed as a magenta-dotted line. To reproduce the observed fluxes in the FIR, we increase to \clmid\,=\,3.2 (best-fit, solid line). Although the best-fit for this star has \clout\,=\,1, given the observational gap in the mm-regime a maximum value of the clumping degree in Region\,4 \cloutmax\,=\,8 (blue dashed-line) is still consistent with the data at shorter wavelengths. This \cloutmax\ is typed between parentheses in the corresponding entry for this star in Table\,\ref{tableclfactors}. \\

\noindent{\it HD\,193237 (P\,Cyg)} 

This known LBV is well sampled throughout the wind. It is well established that the observed variability on timescales on the order of months at mm and radio wavelengths (\citealt{Abbott1981}, \citealt{Vandenoord1985}, \citealt{Bieging1989}, \citealt{Scuderi1998}, \citealt{Ofek2011}, \citealt{Perrot2015}) is not related to P\,Cyg being a non-thermal source, but rather to changes in the ionisation stage of the outer wind (recombined outermost wind model, \citealt{Najarro1997b}; see also \citealt{Pauldrach1990}). In addition, the observed larger IRAS and SCUBA flux measurements at 60 and 850\,\micron\ may be due to spatial resolution effects (such as contamination by the characteristic LBV nebula surrounding P\,Cyg), whereas the lower value at 6\,cm observed by \citet{Bieging1989} has not been confirmed by newer measurements. In view of this, we estimated \Mdotmax\ for P\,Cyg for being consistent with all measured fluxes from 2 to 21\,cm, regardless of the upper IRAS\,60\,\micron\ and SCUBA\,850\,\micron\ values and the lower flux at 6\,cm, obtaining   
\Mdotmax\,=\,12.8\,\massloss.

With this mass-loss rate the synthesised flux underestimates the data at shorter wavelengths (magenta-dotted line). Thus, the clumping degree needed to be increased to reproduce the observed SED, yielding \clin\,=\,3, \clmid\,=\,3.5, \clout\,=\,5 and \clfar\,=\,1 (best-fit, solid line). To match the larger well-defined fluxes observed in the radio-regime the clumping degree in Region\,5 had to be increased to \clfar\,=\,1.5 (blue-dashed line). The corresponding clumping factors for these two models for \clfar\ are separated by a forward slash (/) in Table\,\ref{tableclfactors}, with the number on the left referring to the best-fit solution. The well-sampled mm-regime of this star further allows us to constrain the clumping degree in Region\,4 to \cloutmax\,=\,\clout\,=\,5. \\

\noindent{\it HD\,24398  ($\zeta$ Per)}

Only an upper limit of the flux at radio wavelengths is available for this star. An estimate of \Mdotmax\ by matching this limit is not consistent with FIR flux values, since any model with such \Mdotmax\ would overestimate the FIR flux (magenta-dotted line). Instead, an upper limit $\Mdotmax \lesssim 0.18\,\Msunyr$ derived from the FIR flux values is consistent with the radio upper flux limit. In this case the observed SED is perfectly matched by the minimum clumping degree throughout the whole wind, \clin\,=\,\clmid\,=\,\clout\,=\,\clfar\,=\,1 (best-fit, solid black line). It is possible, however, to still match the upper radio flux limit if increasing up to \clfarmax\,=\,5 (parenthesed value in Table\,\ref{tableclfactors}; blue-dashed line). Although the best-fit for this star is for \clout\,=\,1, given the observational gap at $\lambda >$\,100\,\micron, \cloutmax\,=\,45 (orange-dashed-dotted line) is still consistent with the data (also between parentheses in Table\,\ref{tableclfactors}).

\section{Analysis and results}\label{analysis}

In this section we present the derived maximum mass-loss rates, the corresponding minimum radial clumping stratification, and their associated uncertainties. 

The results obtained in the fixed- and the adapted-regions approaches (see previous section) are summarised in Tables\,\ref{tableclfactors} and \ref{tableclbestfit}. A detailed description of the fits for the individual objects can be found in Appendix \ref{appendixA}. 

\subsection{General findings}

Overall, we find that the minimum $\chi^2$ of the fits in the fixed-regions approach is comparable to that obtained with the adapted-regions approach. As such, unless specified otherwise, the following general findings refer to the best-fit solutions in the fixed-regions approach (Table\,\ref{tableclfactors}).  A definite \Mdotmax\ was derived for 17 of the 25 sources. For the remaining 8 objects, we only provide upper limits of \Mdotmax\ consistent with the observations, due to lack of detections in the radio regime. 

\begin{figure}[!ht]
\begin{center}
\includegraphics[width=9cm]{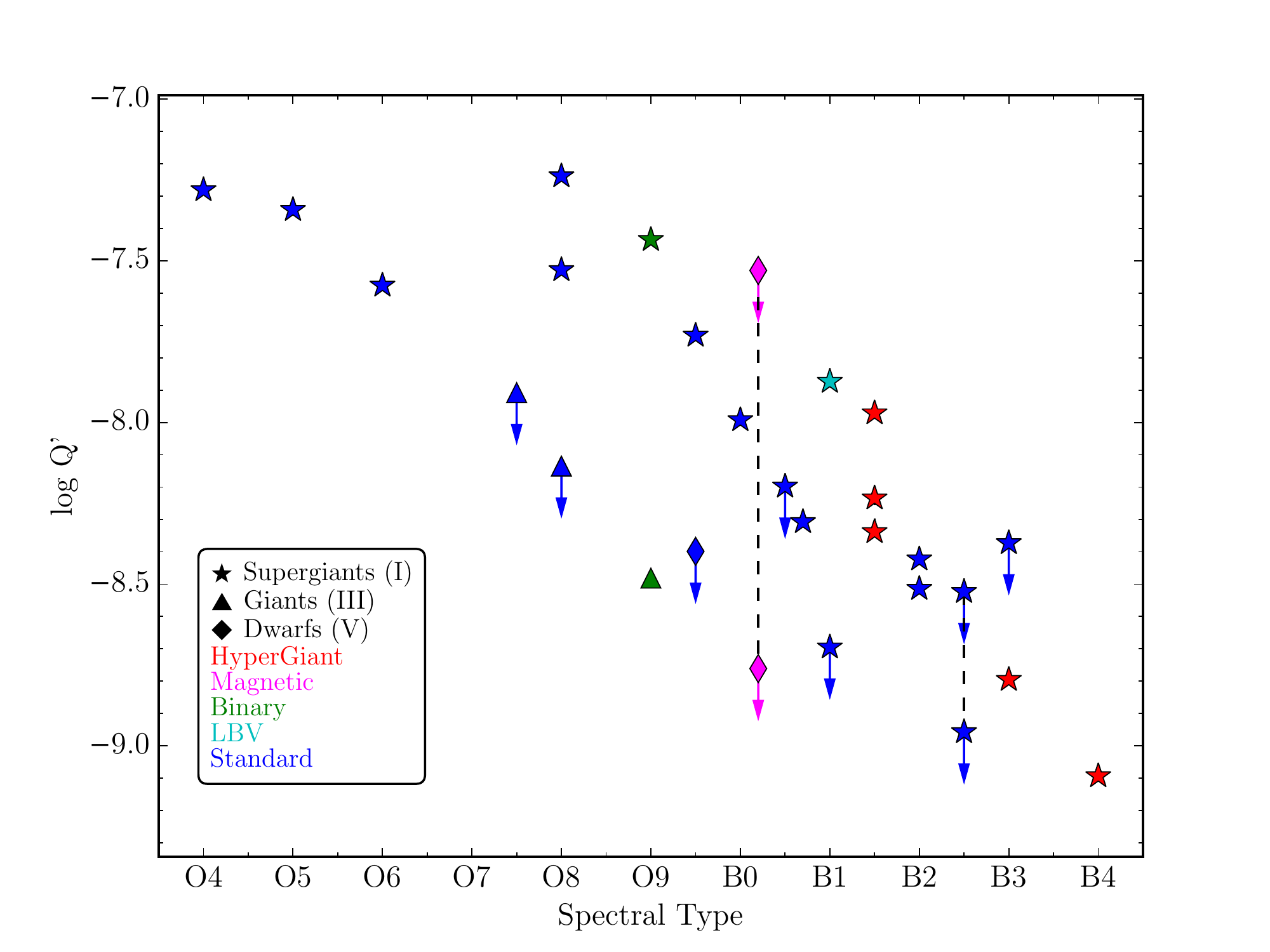}
\caption{Derived invariant log $Q'$ = log (\Mdotmax/\Rstar$^{1.5}$) for our sample as a function of spectral type, in units of \Msun\ yr$^{-1}$ \Rsun$^{-1.5}$. Different symbols and colours represent the luminosity class coverage in the sample and the nature of the sources, respectively. Arrows indicate objects with upper limits for \Mdotmax, and symbols joined by dashed-lines correspond to sources with two possible solutions for \Mdotmax.}
\label{figmdot_spec}
\end{center}
\end{figure}   

Figure\,\ref{figmdot_spec} displays log $Q'$ = log (\Mdotmax/\Rstar$^{1.5}$) as a function of spectral type and luminosity class. We can see that log $Q'$ decreases with luminosity class and spectral type, and that the values departing from this trend correspond to the non-standard sources in our sample, i.e. confirmed binaries, magnetic stars, LBVs and eBHGs (see coloured symbols in Figure\,\ref{figmdot_spec}). In particular, this figure shows that non-standard OB stars seem to be displaced towards higher log $Q'$ values and later spectral types when compared to standard OB\,Supergiants\footnote{Note that in the case of eBHGs, which are late type by definition, this might be due to a selection bias.}.

Due to the low number of OB\,Giants and Dwarfs in our sample, it is not possible to clearly observe a similar behaviour as that seen for the OB\,Supergiants. Overall, the general trend and the estimated values of \Mdotmax\ and log $Q'$ for the complete sample appear consistent. Indeed, the large values obtained for some of the non-standard objects agree with some previous estimates (\citealt{Clark2012}, \citealt{Crowther2006}). For instance, for the eBHG star HD\,152236, we derived \Mdotmax\,=\,6.2\,\massloss, a value close to the (unclumped) mass-loss rates obtained by these authors, 6.33\,\massloss\ and 6.0\,\massloss, respectively.

\begin{figure*}[t]
\begin{center}
\includegraphics[width=8.5cm]{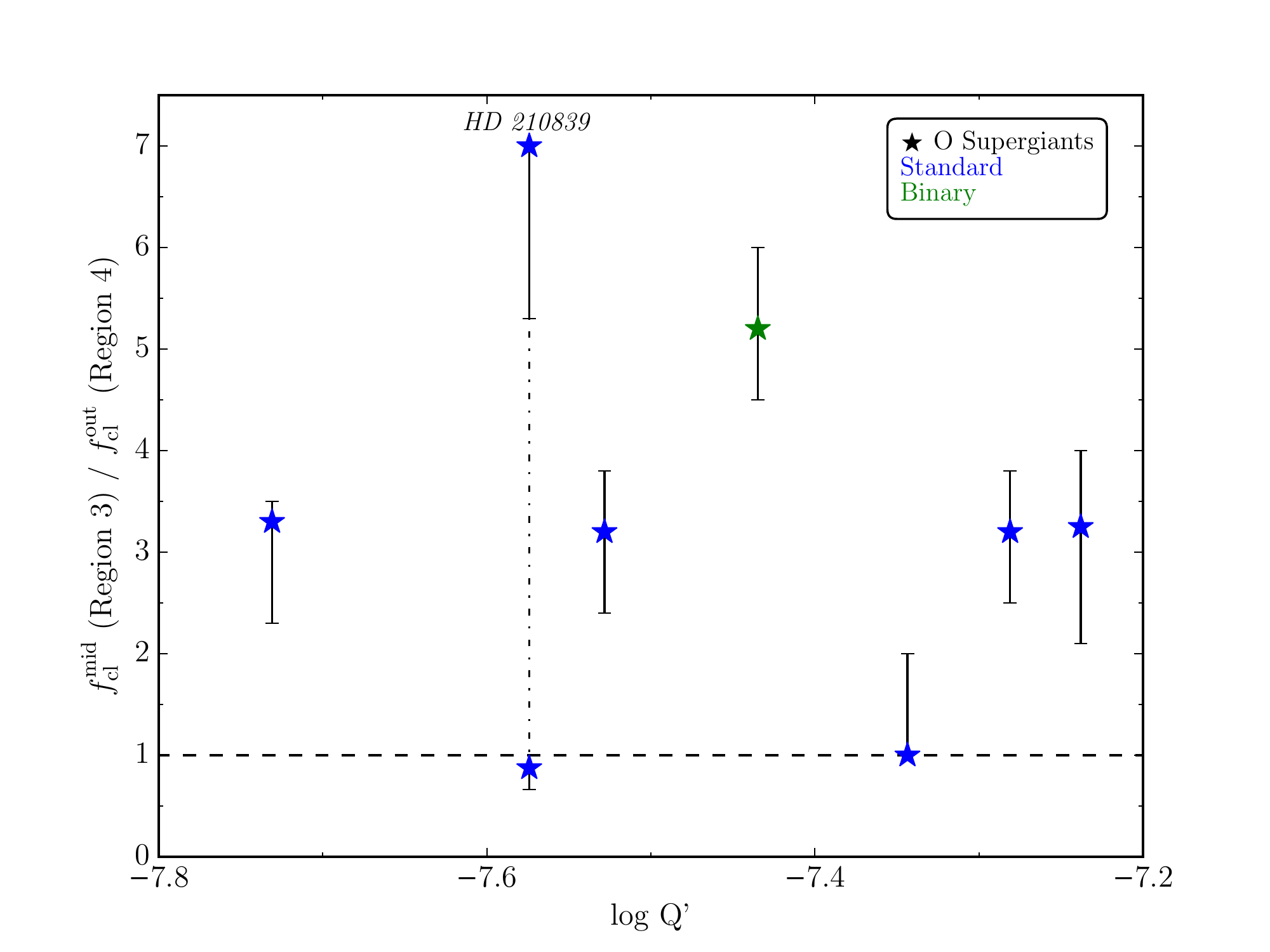}
\includegraphics[width=8.5cm]{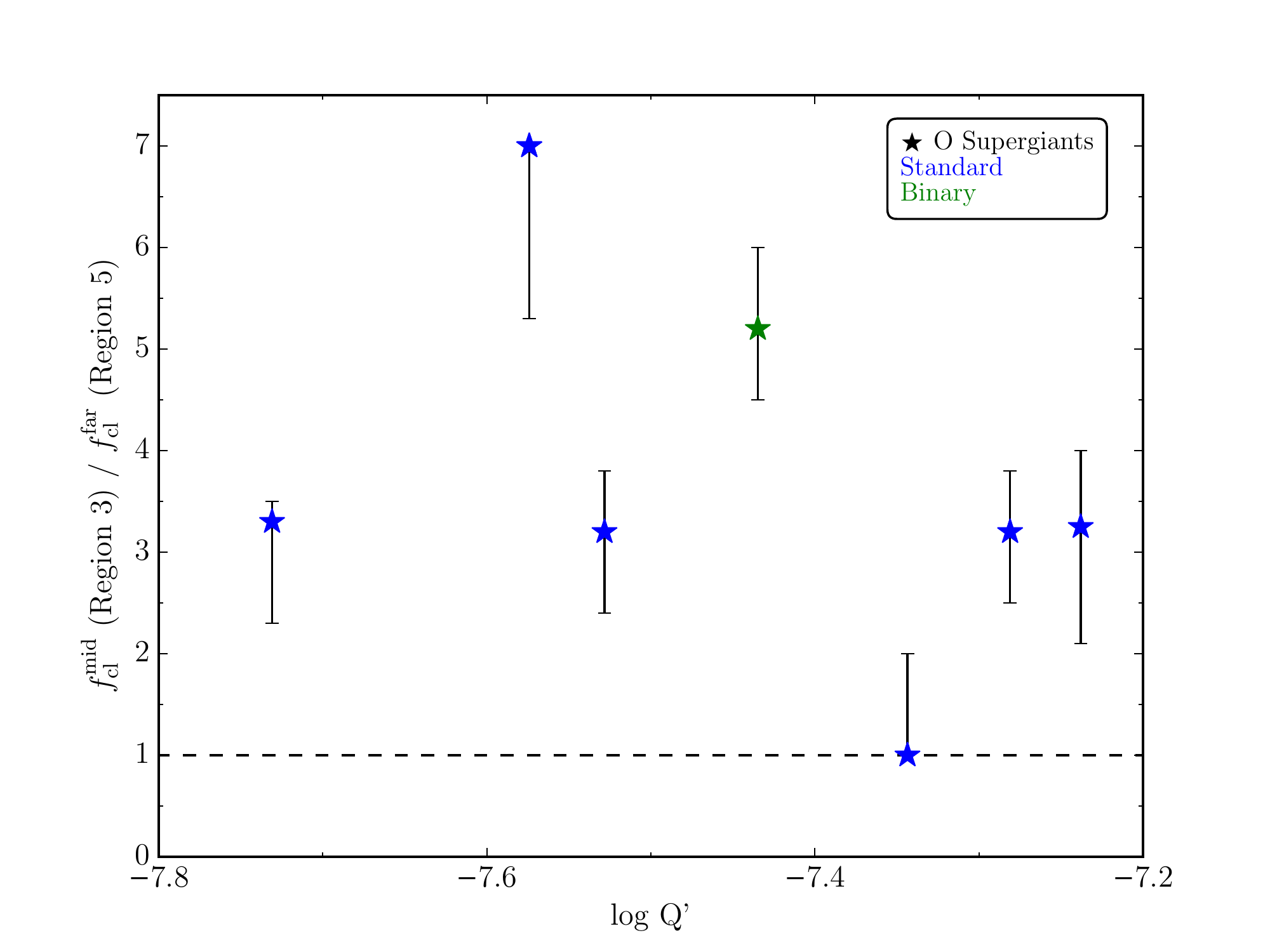}
\includegraphics[width=8.5cm]{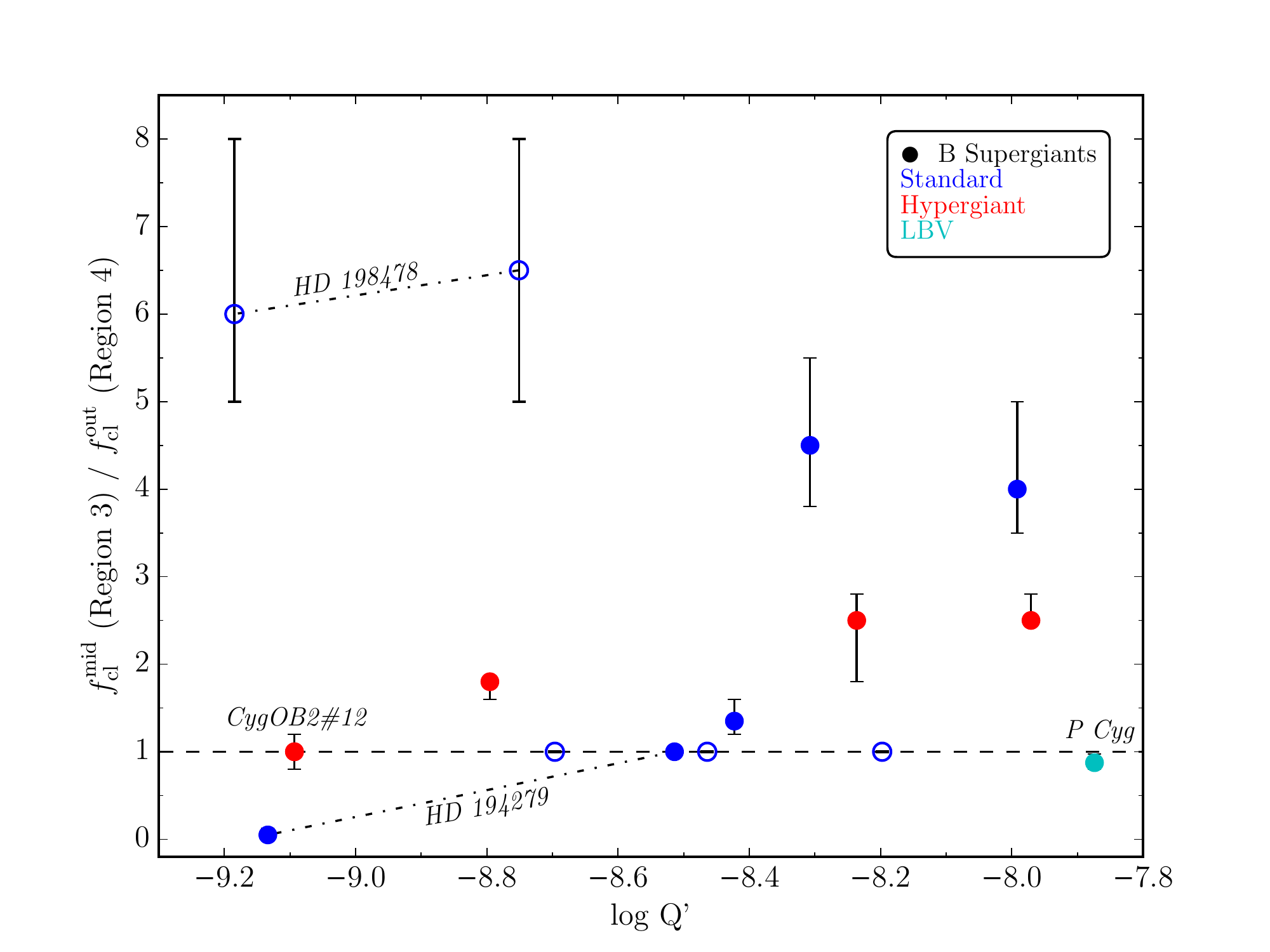}
\includegraphics[width=8.5cm]{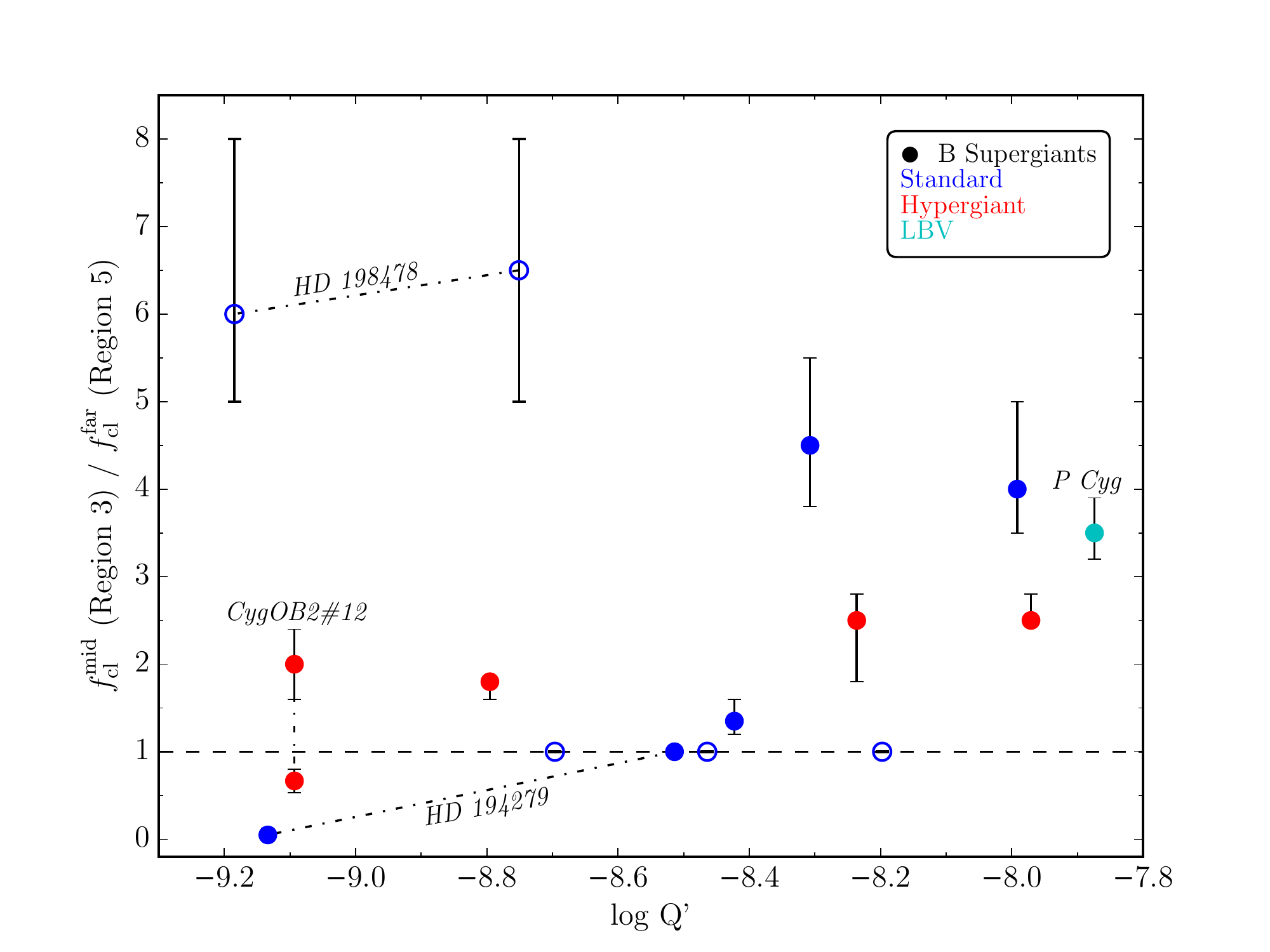}
\includegraphics[width=8.5cm]{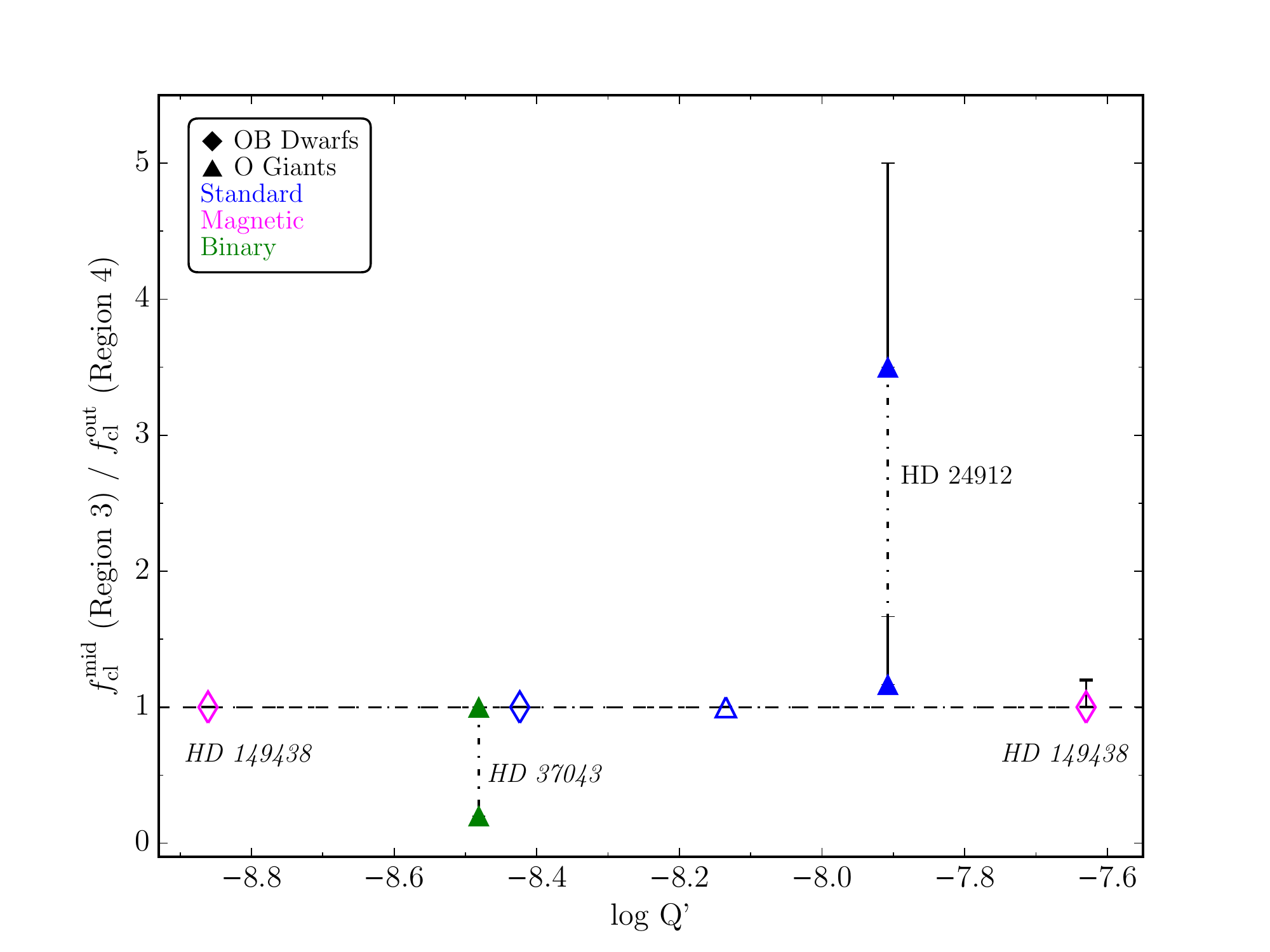}
\includegraphics[width=8.5cm]{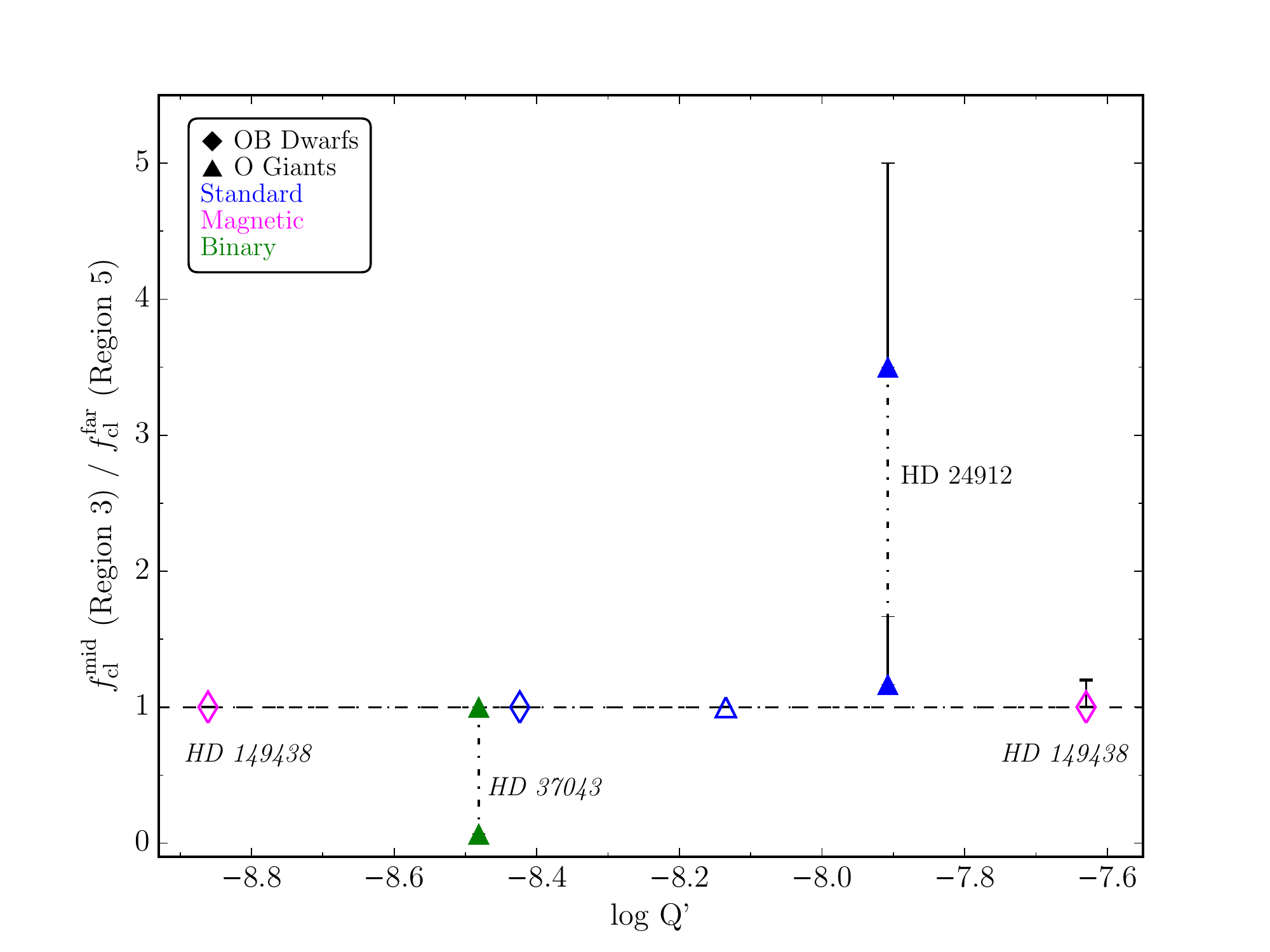}
\caption{From Top to Bottom, clumping factor ratio \clmid/\clout\ (\textit{Left}) and \clmid/\clfar\  (\textit{Right}) for the OB\,Supergiants, Dwarfs and Giants derived in this work, as a function of the invariant log $Q'$ = log (\Mdotmax/\Rstar$^{1.5}$), in units of \Msun\ yr$^{-1}$ \Rsun$^{-1}$. Symbols and colours as in Figure\,\ref{figmdot_spec}, except for empty symbols, which indicate sources with upper limits for \Mdotmax. Dotted-dashed lines correspond to sources with two solutions for either \Mdotmax\ or clumping factors at the outermost wind regions.  
}
\label{figlogQclfactors}
\end{center}
\end{figure*}

The derived minimum clumping structures derived for r\,$\gtrsim$\,2\,\Rstar\ are displayed in Table\,\ref{tableclfactors}. Only one source, CygOB2\#11, remains unconstrained in the intermediate wind region (\clmid\,$\lesssim$\,3) since only upper flux limits were detected at 70, 100 and 160\,\micron. In Figure\,\ref{figlogQclfactors} we present the ratio of the clumping factors in the intermediate and outer part of the wind (\clmid/\clout; left panels) and in the intermediate and outermost wind regions (\clmid/\clfar; right panels) as a function of the wind density (log $Q'$), for the OB\,Supergiants (top and middle panels, respectively) and for the OB\,Giants and Dwarfs (bottom panels). From top to bottom, Figure\,\ref{figlogQclfactors} shows that the clumping degree decreases or remains nearly constant with increasing radius throughout the wind regardless of luminosity class or wind density. Overall, the minimum clumping degree in Region\,4 (\clout) is similar to that in the outermost Region\,5 (\clfar). 
OB\,Giants and Dwarfs are the sources with generally the lowest, most homogeneous clumping across all wind regions. OB\,Supergiants, on the other hand, display a more radially structured clumped wind. 

There are four exceptions to these general trends, where the clumping structure seems to increase with increasing distance to the photosphere. These are P\,Cyg (HD\,193237), CygOB2\#12, $\iota$\,Ori (HD\,37043) and $\tau$\,Sco (HD\,149438). Note that all these are non-standard objects: respectively, a LBV, an eBHG with a recently measured variable radio flux at 21\,cm (\citealt{Morford2016}), a confirmed binary, and the well-known magnetic B\,Dwarf $\tau$\,Sco. 

\begin{table*}
\caption{Maximum mass-loss rates and minimum clumping factors for the sample as derived in the fixed-regions approach (see Sec.\,\ref{modeling}). Sources are sorted as in Table\,\ref{tabledata}. \Rstar\ corresponds to the estimated stellar radius for the assumed distance (see Table\,\ref{tabledistances}), whereas $\beta$ and \vinf\ are taken from the literature (references in Table\,\ref{tablestellarparams}). Clumping factors denoted by \clin, \clmid, \clout, \clfar, and \rinc, \rmidc, \routc\ and \rfarc\ are corresponding wind-region (cwr) boundaries. The maximum possible clumping factor in Region\,4 still consistent with the data denoted by \cloutmax. Boldfaced names refer to stars with the best constrained clumping factors for all wind regions. Values separated by a forward slash (/) indicate alternate well-defined solutions, with the best-fit solution being always the number on the left. Values in parentheses refer to upper limits of the maximum clumping factor in the cwr (see Sec.\,\ref{prototypical_examples}). } 
\tiny{
\begin{center}
\begin{tabular}{l|c|c|c|c|c|c|cc|c|l}\hline\hline
\multirow{2}{*}{Source}  			 		&\Mdotmax		&  \Rstar	  &	\multirow{2}{*}{$\beta$}	&	\vinf	& Reg. 2 & Reg. 3  &\multicolumn{2}{c|}{Reg. 4} & Reg. 5	& \multirow{2}{*}{Comments}\\	
  				 							& (10$^{-6}$\Msunyr)  				&  (\Rsun)                       & 							&	(kms$^{-1}$)				&\clin       & \clmid      & \clout & \cloutmax & \clfar &\\\hline 	
\textbf{HD66811} ($\zeta$\,Pup)        & 4.2                 & 18.6 	  & 0.7		&	2250	& 5.0        & 3.2         & 1.0   	  &  1.2  		& 1.0      &\\
											& 8.5		          & 29.7 	  & 		&			& 5.0        & 3.2         & 1.0   	  &  1.2  		& 1.0      &\\                           
CygOB2\#11  					  			& 5.05			     & 23.15   	  & 1.1		&	2300	& 1.0         & 1.0 (3.0)   & 1.0      &  (15.0) 		& 1.0      & \\   
HD210839 ($\lambda$ Cep)    			& 1.3                & 13.35	      & 1.0		&  2250	& 1.0         & 7.0         & 1.0/8.0 & 5.0/12.0   & 1.0      & \\   
\textbf{HD152408}    			 			& 9.5                & 30.0 	  & 2.1		&	 955	& 1.0         & 3.25       &  1.0     & 3.0   		& 1.0      &\\                   
HD151804 (V973 Sco)   					& 6.4                & 36.0	      & 2.0		&	1450	& 1.0         & 3.2         & 1.0      & (8.0)  		& 1.0      & \\                   
HD149404    							    & 8.27               & 37.0   	  & 1.3		&  2450    & 1.0         & 5.2         & 1.0      & (6.0) 		& 1.0      & Binary  \\                          
HD30614 ($\alpha$ Cam)      				& 1.75               & 20.70	  & 1.15    &  1550	& 6.0        & 3.3          & 1.0      & (2.0) 		& 1.0      & \\\hline                                       
\textbf{HD37128} ($\epsilon$ Ori)		& 1.25			     & 24.68	  &	1.5		&	1910	& 5.0		 & 4.0		    & 1.0	   & 1.0 		& 1.0	   & \\
HD38771 ($\kappa$ Ori)  & $\lesssim$0.7\tablefootmark{a}   & 23.0       &	1.5		&	1525	& 8.0        & 1.0          & 1.0  	   & (8.0)  		& 1.0      &  \\                            
HD154090 ($\kappa$ Sco)     				& 1.2                 & 39.0		  & 1.5		&	915		& 2.0        & 4.5         & 1.0       & (8.0) 		& 1.0      &   \\                                       
\textbf{HD193237} (P Cyg)   			& 12.8		   		 & 97.14	  & 2.5		&	195		& 3.0        &	3.5         & 5.0      & 5.0 		& 1.0/1.5  & LBV    \\                      
HD24398 ($\zeta$ Per) 					& $\lesssim$0.18\tablefootmark{b}  & 21.12	       &	 1.5		&	1295	& 1.0     	 & 1.0          & 1.0       & (45.0) 		& 1.0 (5.0)&   \\                               
\textbf{HD169454}    			 			& 10.4                 & 98.2  	  & 2.1		&  850		& 1.0        & 2.5          & 1.0      & 2.5  		& 1.0       & eBHG   \\                                   
\textbf{HD152236}($\zeta^{1}$ Sco)   	& 6.2                 & 104.0	  &	2.0		&  390		& 1.0      	 & 2.5          & 1.0      & 1.0   		& 1.0       & eBHG \\                          
HD41117 ($\chi^2$ Ori)     				& 1.8   		      & 61.0	  & 2.0		&	510		& 7.3    	 & 1.35         & 1.0      & (6.0)   	     & 1.0      &  \\                    
HD194279	  								& 2.12     		      &	78.4	  & 2.5		&	550	& 1.0	   	 & 1.0	         & 1.0      & (30.0)   	& 1.0	    &  \\
			  								& 0.5$^d$	 	      &			  &			&			& 1.0        & 1.0  		    & 20.0     & (100.0)		& 20.0	    &   \\	
HD198478\tablefootmark{1} (55 Cyg)	&$\lesssim$ 0.14-0.38\tablefootmark{a} 	 & 35.8 	  & 1.3	   &200-470 & 40.0-45.0 &  6.0-6.5 & 1.0      & (10.0)       & 1.0       &\\    
HD80077   								& 3.45              &  167.0 	  & 3.0		&	250	& 2.5   	 & 1.8          & 1.0      & (6.0)        & 1.0       &   eBHG \\                                                
HD53138 ($o^{2}$ CMa) 					& $\lesssim$1.8\tablefootmark{c}    &  65.0  	  & 2.0	    &	865	& 2.0   	 & 1.0           & 1.0      & (18.0)       & 1.0 (50.0) &    \\                                         
\textbf{CygOB2\#12}  					&  3.0\tablefootmark{d}          & 	240.0	  & 3.0		&	400	& 1.0     	 & 10.0         & 10.0     & 12.0       & 5.0/15.0   & eBHG \\
						 					&  1.02\tablefootmark{d}          & 	116.7	  & 3.0		&	400	& 1.0     	 & 10.0         & 10.0     & 12.0       & 5.0/15.0   & \\\hline                                                                           
HD24912 ($\xi$ Per)    					& 1.4 			      &	23.4   	  & 0.9		&	2450	& 3.5     	 & 3.5          & 1.0/3.0  &  (5.0)	     & 1.0/3.0    &   \\                              
HD36816 ($\lambda$ Ori A)  				& $\lesssim$0.16\tablefootmark{a}    & 7.8  	  & 0.9		&	2400	& 1.0     	 & 1.0 		    & 1.0       & (2.0)        &1.0         &    \\                        
HD37043 ($\iota$ Ori)    					& 0.25$^d$   	      & 17.9  	  & 0.9		&  2300	& 1.0         & 1.0     		& 1.0/5.0  &  10.0      & 1.0/15.0  & Binary \\\hline
HD149757 ($\zeta$ Oph)    				& $\lesssim$0.07\tablefootmark{b}    & 8.9 		  &	0.8		&	1550	& 1.0         & 1.0       	& 1.0       &  (2.0)		 & 1.0 (8.0) &\\ 
HD149438 ($\tau$ Sco)    				& $\lesssim$0.315\tablefootmark{a} & 5.65 	  &	0.8     &   2400  	& 1.0         & 1.0          & 1.0       & (7.0) 		 & 1.0       & Magnetic \\                                      
 			  								& $\lesssim$0.0185\tablefootmark{a} &		  &			&			& 300.0	  & 1.0/300.0	& 1.0		& (300.0)	 &1.0 (300.0)&  \\\hline\hline
\end{tabular}
\label{tableclfactors}
\end{center}
}
\mbox{\tablefoottext{1}{Value interval for \Mdotmax\ and clumping factors derived from the stellar parameters in the literature (see Table \ref{tablestellarparams} and Appendix A.2 for further discussion.)}}\\
\tablefoottext{a}{Upper limit of \Mdotmax, derived from upper radio-flux limits.}\\
\tablefoottext{b}{Upper limit of \Mdotmax, derived from well-defined FIR fluxes, since upper radio-flux limits are inconsistent with all the data. }\\
\tablefoottext{c}{Upper limit of \Mdotmax\ derived from well-defined FIR fluxes. No radio fluxes measurements available.}\\
\tablefoottext{d}{Well-defined \Mdotmax, derived from FIR fluxes, since well-defined radio fluxes are not consistent with all the data.}
\end{table*} 

\begin{figure*}[t]
\includegraphics[width=9cm]{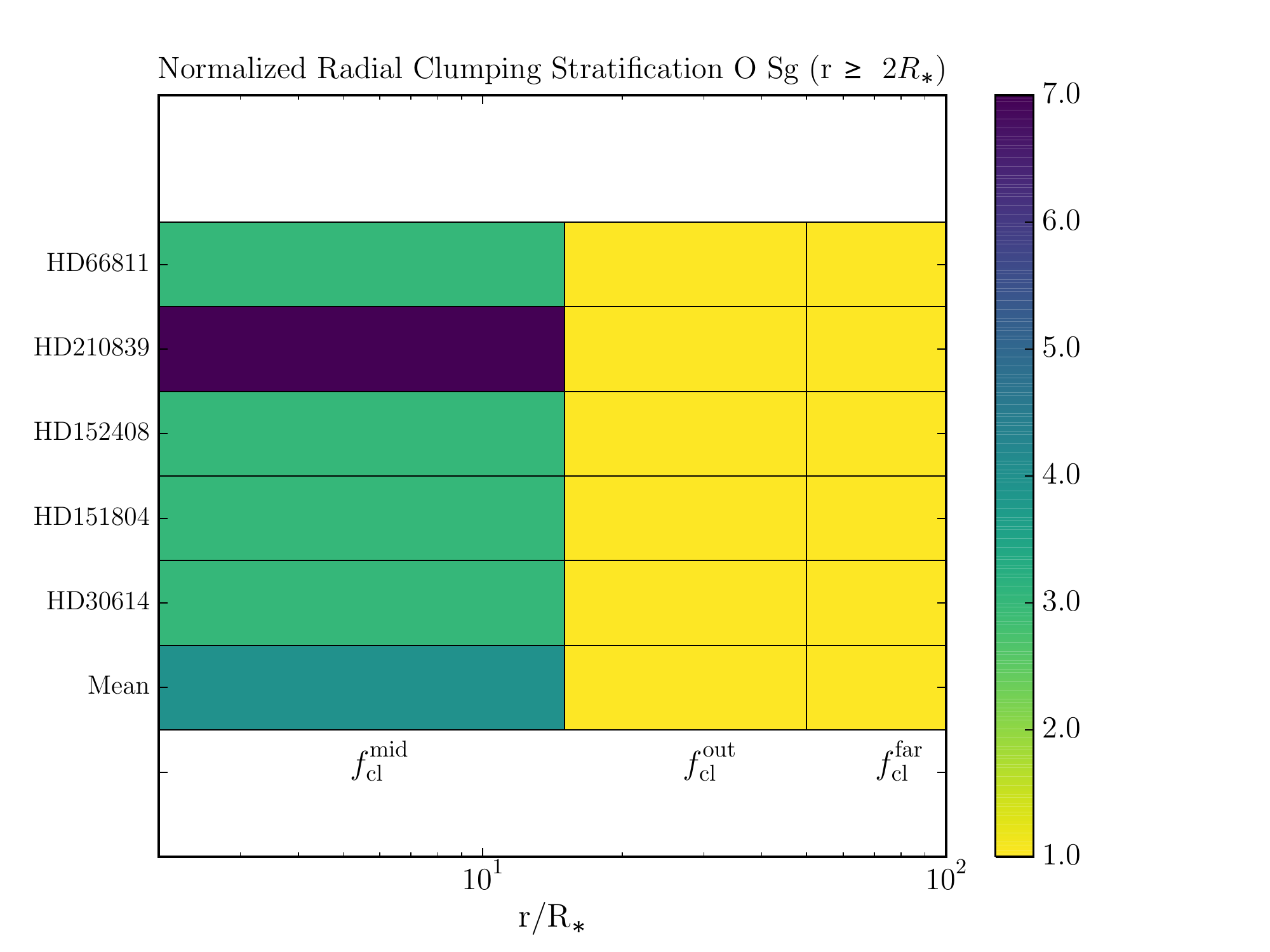}
\includegraphics[width=9cm]{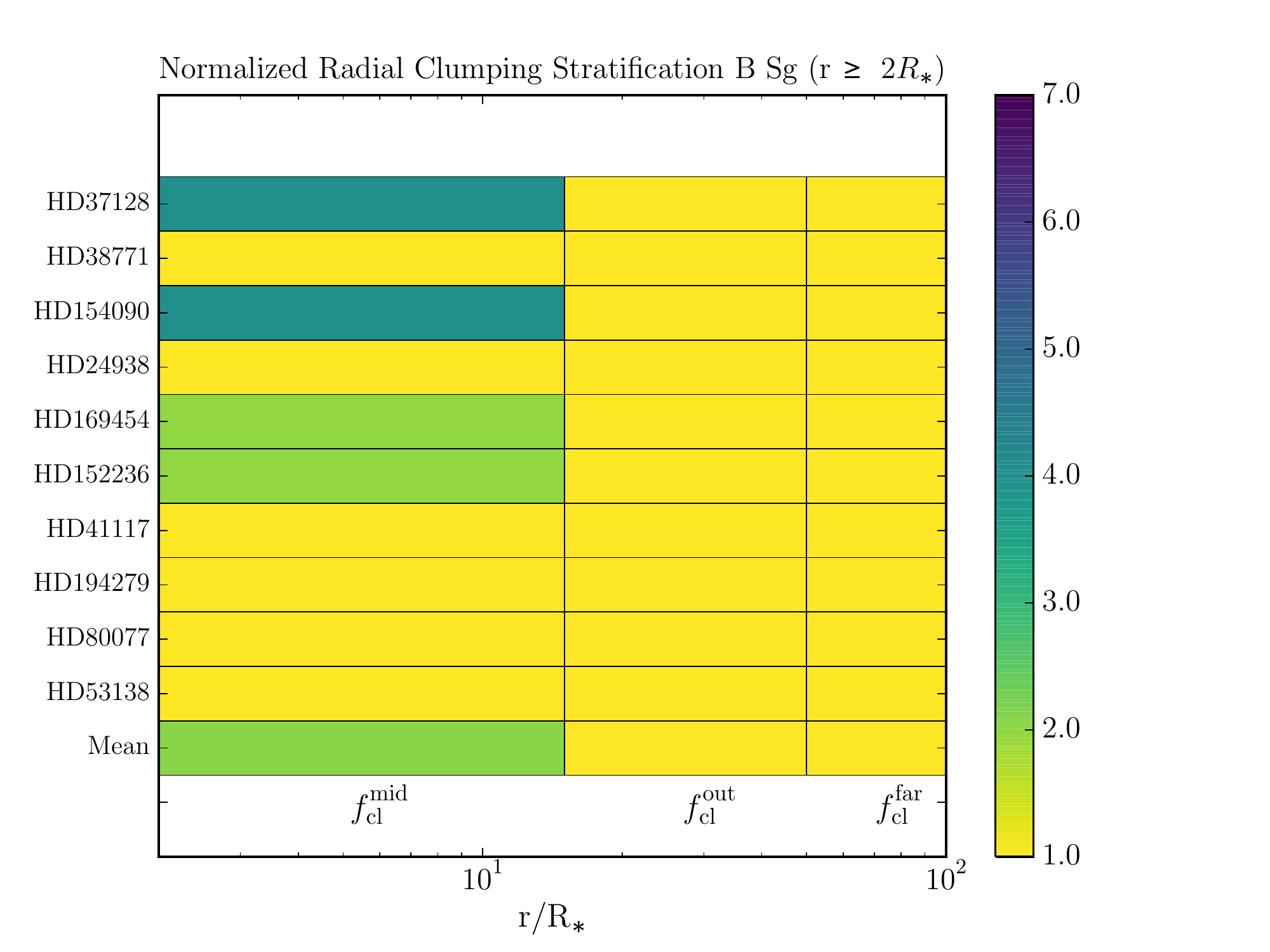}
\includegraphics[width=9cm]{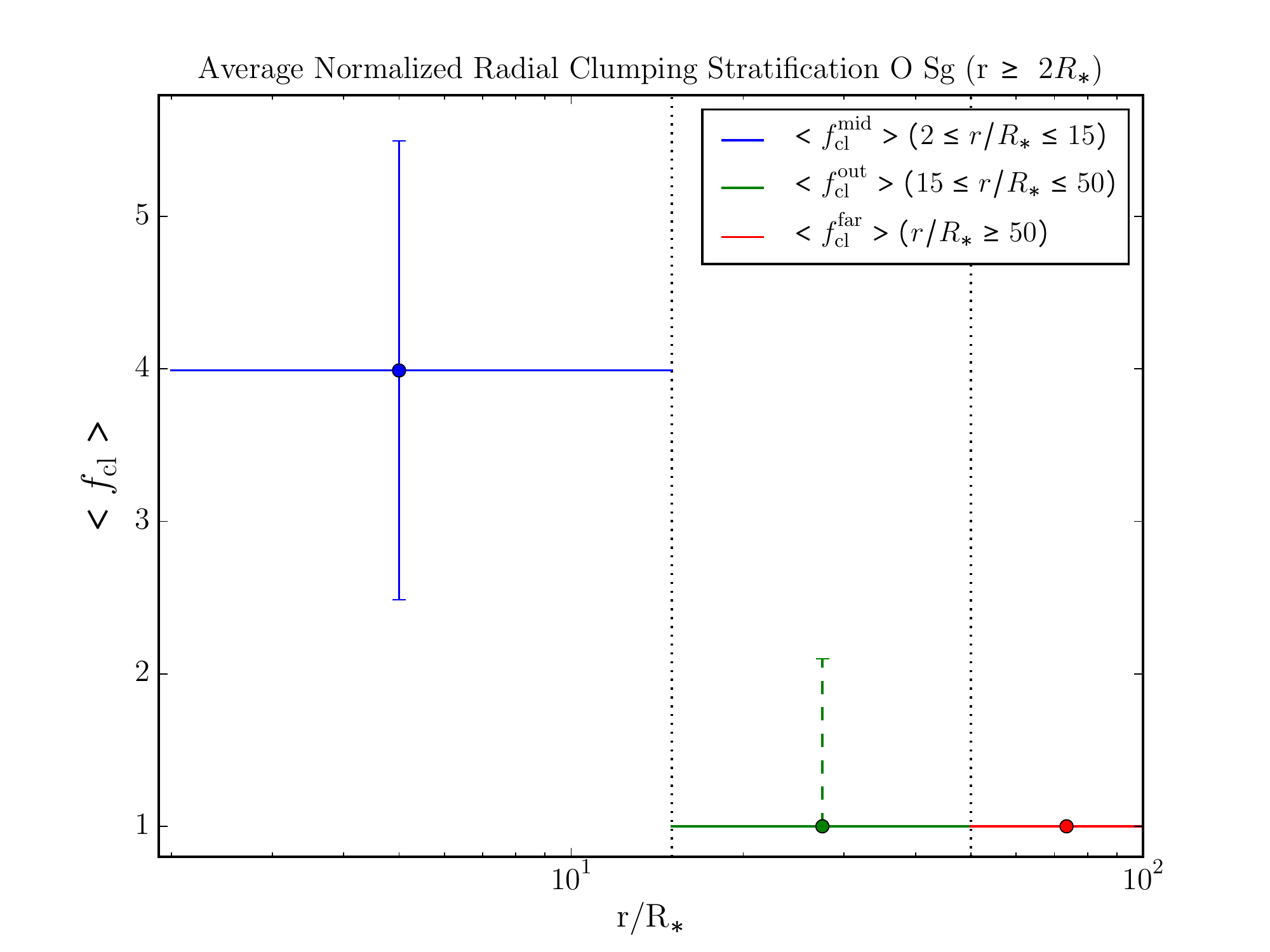}
\includegraphics[width=9cm]{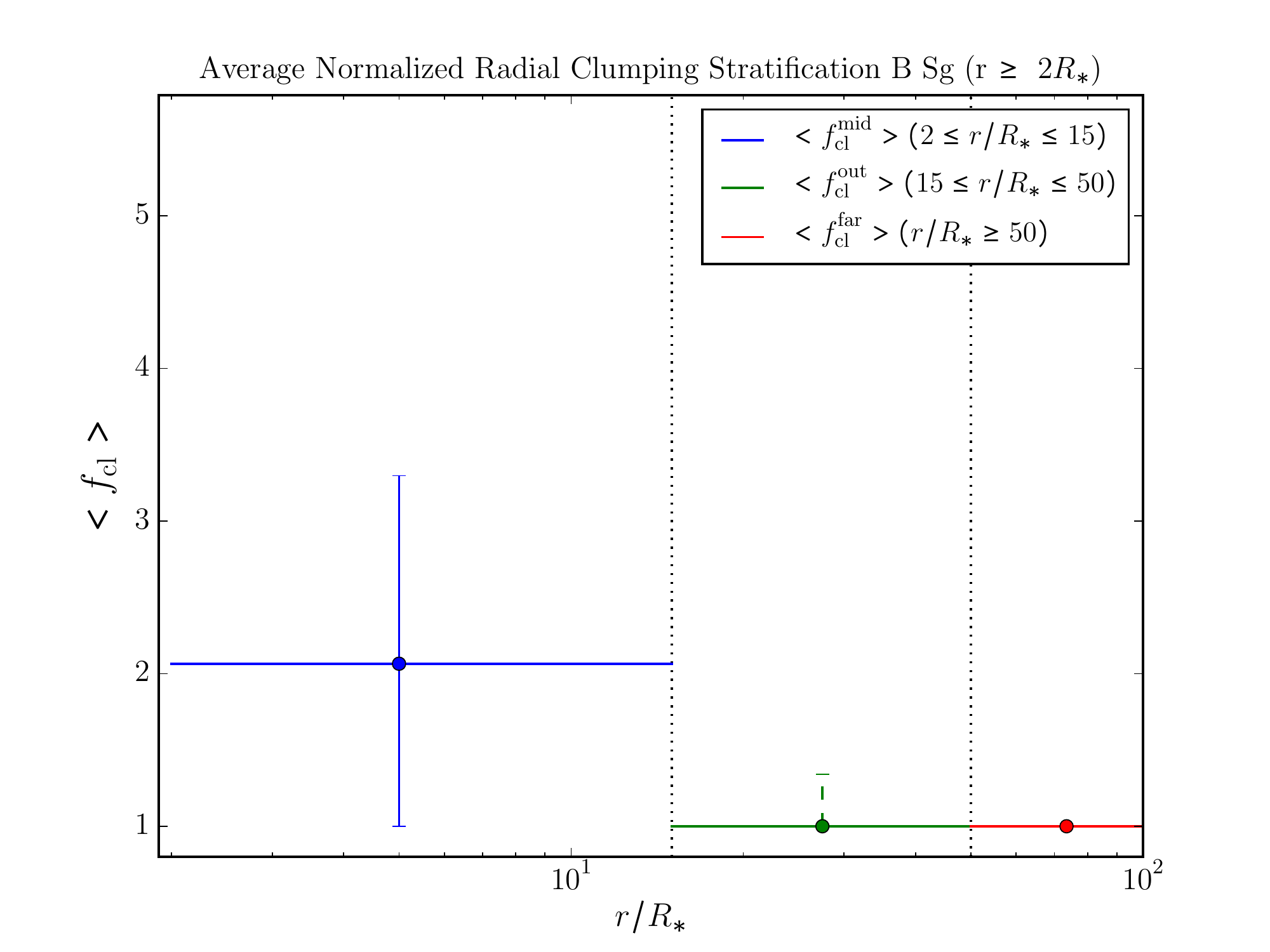}
\caption{\textit{Top:} Individual minimum and average values of the clumping factors for r\,$\geq$\,2\,\Rstar\ derived in the fixed-regions approach for the subsample of O (\textit{Left}) and B (\textit{Right}) Supergiants (see Sec.\,\ref{clOBSg}). \textit{Bottom:} Average minimum clumping stratification of the subsample as in the top panel, with error bars. Solid error bars correspond to the standard deviation of 
$\langle\clmid\rangle$, whereas dashed error bars correspond to $\langle\clout\rangle$, and represent the mean of the best constrained \cloutmax\ listed in Table\,\ref{tableclfactors} (bold names). Vertical dotted lines represent the boundaries of the defined wind regions.}
\label{figclaverage}
\end{figure*}

\begin{table*}
\caption{Maximum mass-loss rates, minimum clumping factors and alternative boundaries of the defined wind regions (adapted-regions approach, see Section 3) providing the best-fit models for some objects in the sample. Column descriptions and values in parentheses or separated by forward slashes as in Table \ref{tableclfactors}. $r'_{\rm in}$ indicates values different from the nominal, adopted \rinc, and \rfar\ is always set to 50\,\Rstar. }
\tiny{
\begin{center}
\begin{tabular}{l|c|c|c|c|c c|c c|c c|c|c}\hline\hline
\multirow{2}{*}{Source}  &  \Mdotmax		& \Rstar & \multirow{2}{*}{$\beta$} 	& \vinf & \multicolumn{2}{c|}{Reg. 2}&\multicolumn{2}{c|}{Region 3}	 &\multicolumn{2}{c|}{Reg. 4	}& Reg. 5	& \multirow{2}{*}{Comments} \\	
				&   (10$^{-6}$ \Msun\,yr$^{-1}$)&(\Rsun) &	   		& (km s$^{-1}$)	& \clin\ ($r'_{\rm in}$) & \rmid &   \clmid &	\rout & \clout  & \cloutmax & \clfar \\\hline                      
\textbf{HD66811} ($\zeta$\,Pup)      	& 4.2                  			  &18.6	  	& 0.7		& 2250   & 5.0 (1.12)      &2.0   &      3.1    &   12.0    &  1.0   &  1.2	 &	1.0   & \\                                     
											& 8.5                  			  &29.7	  	& 			&		   & 5.0 (1.12)      &2.0   &      3.1    &   12.0    &  1.0   &  1.2	 &	1.0   & \\        
\textbf{HD152408}    						& 9.5                 			  &	30.0 	&	2.1		&955	   & 1.0    			& 3.5  &       4.0   &   15.0    &  1.0   &  3.0	 & 1.0    & \\                   
HD151804    								& 6.4                  			  &36.0	&	2.0		& 1450	   & 1.0    			& 5.0  &       8.5   &   12.0    &  1.0   &  1.0	 & 1.0    &\\                   
HD149404    								& 8.27                 			  & 37.0  	&	1.3		& 2450	   & 1.0    			& 2.5  &       7.0   &   15.0     & 1.0   &  (6.0)	 & 1.0    & Binary\\                         
HD30614 ($\alpha$ Cam)    				& 1.75           				  & 20.70	& 1.15		& 1550	   & 6.0 (1.1)       & 4.0  &       1.0    &   15.0    &  1.0   &  (2.0)	 & 1.0    &\\ \hline                   
\textbf{HD37128} ($\epsilon$ Ori)     	& 1.25                  			  & 24.7	& 1.5		& 1910	   & 5.0			& 2.0  &       8.0   &   6.0     &  1.0   &  1.0    & 1.0	   &\\                         
HD38771 ($\kappa$ Ori)   & $\lesssim$0.7\tablefootmark{a}      		 & 23.0	&	1.5		& 1525    & 20.0   	    & 1.55  &       1.0   &   15.0    &  1.0   &  (8.0)	 & 1.0     & \\                            
HD154090    								& 1.2                  			  & 39.0 	&	1.5		& 915	   & 4.0 (1.4)     	& 2.0  &       3.5   &   15.0     & 1.0   &  (8.0)    & 1.0     & \\                                                    
\textbf{HD169454}    						& 10.4                  			  & 98.2 	&	2.1		& 850	   & 1.0     	    & 4.0   &       4.5   &   15.0    &  1.0   &  2.5	  & 1.0    & eBHG \\                                  
\textbf{HD152236} ($\zeta^1$ Sco)       & 6.2                			  	  & 104.0 	&	2.0		& 390	   & 1.0  			& 2.0  &       3.0   &   10.0     &  1.0   &  1.0    &	1.0	   & eBHG\\ 
HD53138 ($o^{2}$ CMa)      			    &$\lesssim$ 1.8\tablefootmark{c}                		  & 65.0	& 2.0		& 865	   & 3.0  			& 1.3   &       1.0   &   15.0     &  1.0   &  (18.0)  & 1.0 (50.0)	   &  \\                                      
\textbf{CygOB2\#12} 						& 3.0$^d$                  			  & 240.0	& 3.0		& 400     & 1.0       		& 2.5  &      12.0   &   8.0      &  10.0  & 	12.0  & 5.0/15.0 & eBHG\\
											& 1.02$^d$                   		   & 116.7	& 3.0		& 400     & 1.0       		& 2.5  &      12.0   &   8.0      &  10.0  & 	12.0  & 5.0/15.0 & \\\hline                         
HD24912 ($\xi$ Per)     					& 1.4                  			  & 23.4	& 0.9		& 2450	   & 5.0 (1.1) 	     & 4.0  &       1.0   &   15.0     &  1.0/3.0 & (5.0)  & 1.0/3.0  &\\\hline\hline                               
\end{tabular}
\label{tableclbestfit}
\end{center}
}
\mbox{\tablefoottext{a}{Upper limit of \Mdotmax, derived from upper radio-flux limits.}}\\
\tablefoottext{c}{Upper limit of \Mdotmax\ derived from well-defined FIR fluxes. No radio fluxes measurements available.}\\
\tablefoottext{d}{Well-defined \Mdotmax, derived from FIR fluxes, since well-defined radio fluxes are not consistent with all the data.}
\end{table*}

\subsection{Clumping in the intermediate wind of OB\,Supergiants}\label{clOBSg} 

The relatively large number of OB Supergiants in the sample allows for a statistically significant analysis of the average radial clumping structure of these stars at r\,$\gtrsim$\,2\,\Rstar. To do this, we use the fixed-regions approach and remove from the analysis those OB\,Supergiants with a peculiar nature and/or whose derived clumping properties significantly depart from the general trend (HD\,193237, Cyg0B2\#12, HD\,149404, HD\,198478; see Table\,\ref{tableclfactors}). Due to the large uncertainty in PACS flux estimates for HD\,194279 at 70 and 100\,\micron, the scarcity of data at mm wavelengths and the lack of evidence of it being a non-thermal emitter, we consider the first derived solution to be more likely for this target (first entry in Table\,\ref{tableclfactors}; for discussion see Appendix \ref{apxBSg}).

Figure\,\ref{figclaverage} displays the individual (top panels) and the average (bottom panels) minimum radial clumping stratification for the OSG and BSG subsamples (left and right, respectively). The OB\,Supergiants show a similar or larger clumping degree in the intermediate wind region than in the outermost ones (Regions 4 and 5). We estimate that for OSGs the average minimum clumping factor for the intermediate wind region is $\langle\,\clmid\,\rangle$\,$\sim$\,4.0\,$\pm$\,1.5, whereas for BSGs it is $\langle\,\clmid\,\rangle$\,$\sim$\,2.1\,$\pm$\,1.1 (errors represent standard deviations).

Concerning the outermost wind regions, due to the scarcity of sub-mm and mm observations it is possible to derive precise clumping properties in Region\,4 for only 7 sources (boldfaced names in Tables \ref{tableclfactors} and \ref{tableclbestfit}). For the rest of the objects, \clout\,=\,1 generally provides the best-fit solution, but larger values may also be possible (see \cloutmax\ col. in Table\,\ref{tableclfactors}). We compute an average minimum clumping factor for OSGs and BSGs of $\langle\clout\rangle$\,=\,1\,(+2) and 1\,(+1.3), respectively. Unlike for $\langle\clmid\rangle$, the number in parentheses here do not represent formal errors, but are instead rough indications based on the average ratio $\langle\cloutmax\rangle$ of those 7 OB\,Supergiants whose clumping properties are fully constrained.

\subsection{Clumping factor uncertainties}\label{errorfcl}

The mass-loss rate and clumping factor derivations have been made using literature-values of $\beta$ and \vinf. 

As discussed above, the synthesized and observed fluxes depend mostly on the local quantities \Mdot$\sqrt \fcl$ (for a given \Rstar). Assuming an (almost) perfect fit, the derived clumping factors scale inversely with the mass-loss rate as $\delta \fcl/\fcl \approx -2 \,\delta\dot M/\dot M$. Taking into account that uncertainties in \Mdotmax\ depend directly on the errors in radio fluxes (temporal variability and/or intrinsic errors) with $2 \,\delta\dot M/\dot M \approx 1.5 \, \delta F_{\nu}/F_{\nu}$, it follows that the uncertainty in clumping factors also scales inversely with flux as $\delta \fcl/\fcl \approx -1.5 \, \delta F_{\nu}/F_{\nu}$. 
We computed an average uncertainty $\sim$\,15\% for \Mdotmax\ due to flux uncertainties, which leads to an average uncertainty in derived clumping factors of $\approx$\,30\%. For those sources with uncertainties in \Mdotmax\ above average, the errors in the clumping factors can reach 50\%. 

Other factors affecting the derived \Mdotmax, and thus $f_{\rm cl}$, are the assumed helium fraction and the ionisation stage. We performed several simulations varying both parameters. We found that a factor 2 difference in the helium fraction leads to a variation in \Mdotmax\ of $\sim$\,16\%, quite similar to the uncertainty found ($\sim$\,15\%) when the ionisation state in the region probed in the mm and radio domain is different than assumed. From these simulations, we estimate that the average uncertainty of the clumping factors derived in this work, due to uncertainties in \Mdotmax, is in between 30\% to 40\%. 

\subsection{Clumping dependence on $\beta$}\label{betasection}
\begin{figure}[t]
\begin{center}
\includegraphics[width=9cm]{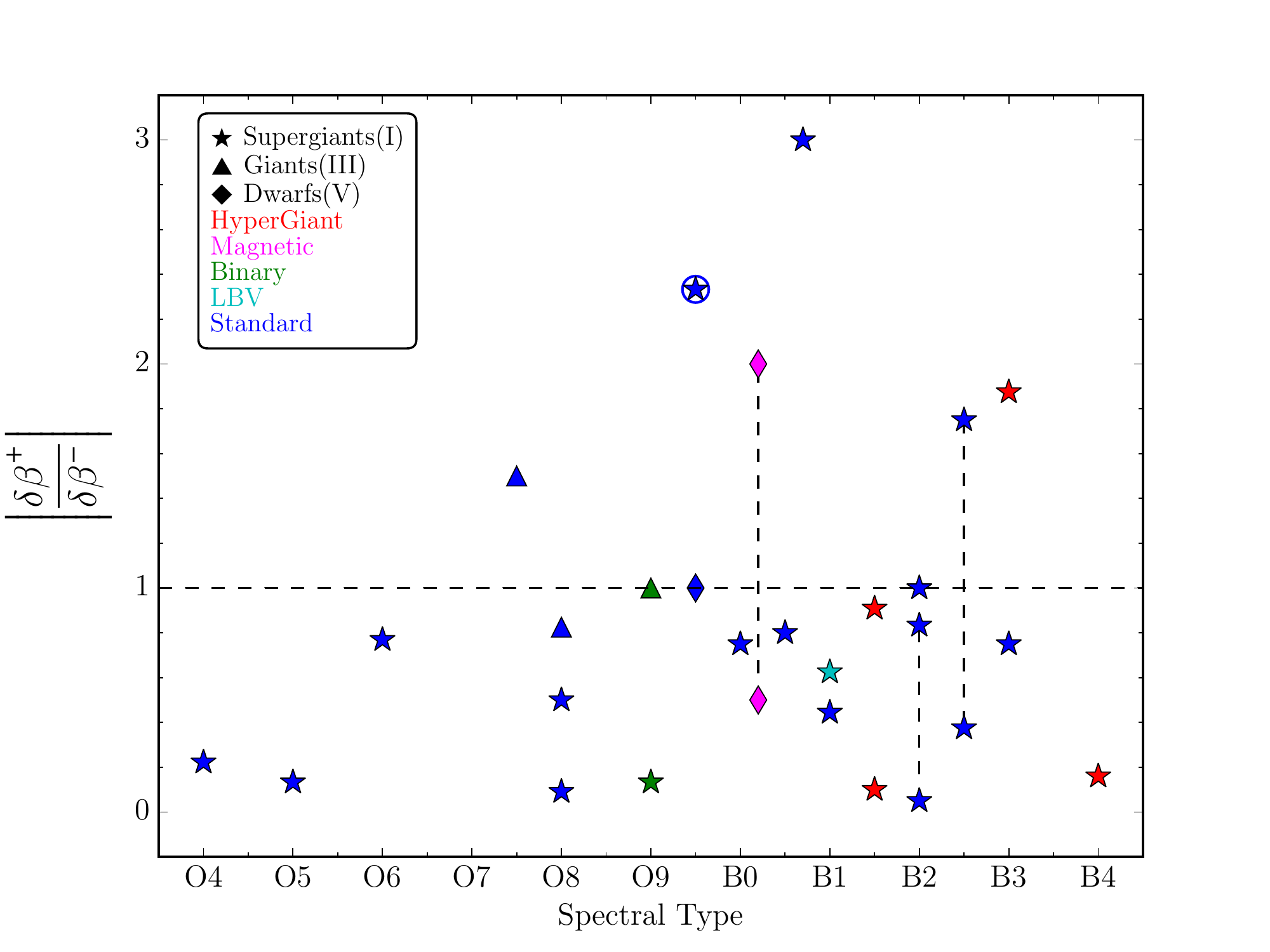}
\caption{Absolute ratio between the semi-widths ($\delta \beta^+$, $\delta \beta^-$) of the validity interval of $\beta$ as a function of spectral type for the best-fit solutions derived in the fixed-regions approach in this work (see Table \ref{tableclfactors}). Symbols and colours as in Fig. \ref{figmdot_spec}. The circled symbol indicates that the ratio was divided by 3 for display purposes. }
\label{ratiodiffbeta}
\end{center}
\end{figure}

 Whereas the clumping degree in the outermost wind regions (r\,$\gtrsim$\,15\,\Rstar) does not depend on the exponent of the velocity field, $\beta$, the opposite is true for the inner wind (r\,$\lesssim$\,2\,\Rstar). Since the $\beta$ and \vinf\ used in our analysis come from a range of different studies (see references in Table\,\ref{tablestellarparams}), they are not really coherently derived. Thus, some analysis regarding the impact of $\beta$ on the derived clumping factors is warranted, and carried out below using two experiments.
In the first one, we computed the boundaries of the validity interval of $\beta$ for which a given best-fit model still reproduced the data with a goodness of fit within 10\% of the corresponding minimum $\chi^2$. The resulting $\beta$-intervals are presented in Table\,\ref{tableinterval} as 
\begin{equation}
\beta^+=\beta_0+\delta\beta^+\  {\mathrm{and}\ } \,\beta^-=\beta_0+\delta \beta^- \label{eqn.10},  
\end{equation}

\noindent with $\beta_0$ being the used $\beta$ in our simulations. We found that the limits of the intervals are not symmetric with respect to the used $\beta$ (except in four objects). Moreover, our best-fit solutions are less sensitive to a decrease than to an increase of $\beta$, regardless of luminosity class, spectral type or ``nature'' of the source (Figure\,\ref{ratiodiffbeta}). Thus, our best-fit solutions are valid for average changes of $\beta$ of $\langle\,|\delta \beta^- /\beta_0|\,\rangle \approx 24\% $ and $\langle\,|\delta \beta^+ /\beta_0|\,\rangle \approx 16\% $. 

In the second experiment we computed new fitting models. We varied the clumping factors (and \Mdotmax, if necessary) for different values of $\beta$, until a best-fit solution was found. This test estimates the errors of the derived clumping factors as: 

\begin{equation}
 \fcl\,(\beta')=\fcl\,(\beta_0)+\delta \fcl = \fcl\,(\beta_0) + (\delta \fcl/\delta\beta')\,\delta\beta' \label{eqn.11}, 
 \end{equation}
 
\noindent with $\delta \beta' = \beta'-\beta_0. $ In Table\,\ref{tablebetachange} we present the results of these simulations for  $\fcl\,(\beta')$, with $\beta'= (\beta_1, \beta_2)$, where $\beta_1\lessapprox\beta^-$, $\beta_2\gtrapprox\beta^+$, and $\beta_0$ is the used $\beta$ in our simulations. 
\begin{table}[!h]
\begin{center}
\caption{Validity interval around the used $\beta$ in our simulations ($\beta_0$; see Sec.\,\ref{errorfcl}). The two entries for HD\,194279, HD\,198478 and HD\,149438 correspond to each of the two derived solutions presented in Table\,\ref{tablebetachange}.}
\begin{tabular}{|l|c|c|}
\hline\hline
Source &		[$\beta^{-} : \beta^{+}$] & \ \ $\beta_0$ \ \ \\\hline
HD66811  	 	&   [0.61\ :\ 0.72]   &	0.7	 \\%
CygOB2\#11 	&   [0.8\ :\ 1.14]   &	1.1	\\%
HD210839    &  [0.87\ :\ 1.1] 	& 1.0	\\%
HD152408    &   [1.5\ :\ 2.1]  	& 2.05 \\%
HD151804     &   [1.8\ :\ 2.1] 	& 2.0	\\%
HD149404    &   [1.15\ :\ 1.32] 	&   1.3 \\%
HD30614      &   [1.1\ :\ 1.5] 	&  1.15 \\\hline
HD37128     &   [1.3\ :\ 1.65]		&	1.5	\\%
HD38771      &  [1.25\ :\ 1.7] &1.5 \\%
HD154090    &  [1.45\ :\ 1.65] &1.5  \\%
HD193237    &  [1.7\ :\ 3.0] &2.5\\%
HD24398     & [1.05\ :\ 1.7]&1.5\\%
HD169454    & [1.9\ :\ 2.12] & 2.1 \\%
HD152236     & [1.45\ :\ 2.5]&2.0\\%
HD41117       & [1.5\ :\ 2.5] 	&2.0\\%
HD194279	  & (.....\ :\ 2.6]&2.5\\%
			  & [1.3\ :\ ....)	&2.5\\%
HD198478    & [1.1\ :\ 1.65]&1.3\\%
			  &[0.9\ :\ 1.45]&1.3\\%
HD80077     & [2.6\ :\ 3.75]&3.0	\\%
HD53138     &  [1.8\ :\ 2.15]&	2.0	\\%
CygOB2\#12 \ \ \ & [1.75\ :\ 3.2]	&3.0	\\\hline
HD24912      & [0.8\ :\ 1.05]&0.9	\\%
HD36816      & (....\ :\ 1.23] & 0.9 \\%
HD37043     & (....\ :\ 1.3]   & 0.9 \\\hline
HD149457		& (....\ :\ 1.1] & 0.8\\%
HD149438    &  [0.7\ :\ 0.85] &  0.8 \\%
			  &  [0.75\ :\ 0.9]&  0.8 \\\hline
\end{tabular}
\label{tableinterval}
\end{center}
\end{table}

Note that for some sources the values of the alternative $\beta_1$ and $\beta_2$ are larger (or lower) than the typical limits derived from spectral line fitting (though always consistent with the theoretical limit $\beta$\,$>$\,0.5). Nonetheless, our approach allows us to check the validity of the analysis as a function of $\beta$. From this experiment we found that, as expected, \textit{i)} large clumping factors, $f^{\rm in,mid}_{\mathrm{cl}}$, correspond to low values of $\beta'$\ and vice versa, i.e. $\delta \fcl \propto -\delta \beta'$; \textit{ii)} changes in $\beta$ do not uniformly (or symmetrically) affect all the defined clumping factors, and also depend on luminosity class. 
Thus, for OB\,Supergiants average values $\langle\,\left|\delta \beta'\right|\,\rangle \sim$15\% -- 30\% lead to average uncertainties of about 15\% -- 60\% in \clin\ and $\sim$\,15\% in \clmid. On the other hand, for the sources with weaker winds in our sample, i.e OB\,Giants and Dwarfs, average values of $\langle\,\left|\delta \beta'\right|\,\rangle \sim$15\% -- 40\% translate into average uncertainties of about 10\% -- 15\% for \clin, and up to 8\% for \clmid; and \textit{iii)} for a few sources, an increase of $\beta$ cannot be compensated by a decrease of clumping factors, since the clumping factors for the used $\beta$ reach unity, so that instead a decrease in \Mdotmax\ is required (last column in Table\,\ref{tablebetachange}).

Note that the objects with weaker winds and upper limits in \Mdotmax\ also require a clumping increase in the outermost wind region, \clfar\ (\fcl$^{\mathrm{max}}$) $>$ 1, in order to be able to consistently reproduce upper radio-flux limits. 
Our results agree with the average uncertainties estimated for \clin\ and \clmid\ by \citetalias{Puls2006} ($\sim$\,30\% and  $\sim$\,20\%, respectively) by means of \Ha\ profiles and IR/radio fluxes analyses in parallel. 

Considering these experiments, the average error due to $\beta$ uncertainties for the estimated values of \clmid\ is $\sim$\,30\%, even in the extreme case that the actual values of the wind acceleration parameter $\beta$ would be significantly different from those adopted in this work.

\begin{table*}[!ht]
\caption{Clumping factors and maximum mass-loss rates (in units of 10$^{-6}$\Msunyr) derived for alternative values ($\beta_1$ and $\beta_2$) of the used $\beta$ in our simulations in the fixed-regions approach (Table\,\ref{tableclfactors}; labeled here as `\textbf{used}'). Entries in the column labeled `\Mdotmax\ ($\beta_2$)' indicate sources for which only a value lower than the estimated maximum mass-loss rate for the \textbf{used} $\beta$ is consistent with all observations (see text); in these cases, the outermost wind must be clumped as indicated by the column labeled `\clfar ($\beta_2$)'. Usually \clout\ did not require further modification except for the asterisk-marked sources (see table foot text). Superscript-letters and values in parentheses or separated by forward slashes as in Table\,\ref{tableclfactors}.
}
\tiny{
\begin{center}
\begin{tabular}{l|c|ccc|ccc|ccc|c|c}\hline\hline
\multirow{2}{*}{Source}&	\Mdotmax &       & $\beta$&		& 			&\clin\		&	 & &\clmid\	&	&\multirow{2}{*}{\clfar\  ($\beta_2$)}	& \multirow{2}{*}{\Mdotmax ($\beta_2$)}\\	
  						 &		\textbf{used} 		       &$\beta_{1}$  &  \textbf{used} 		 & $\beta_{2}$	&  \clin ($\beta_{1}$)		&	\textbf{used}	& \clin ($\beta_{2}$)& \clmid ($\beta_{1}$)&\textbf{used}&\clmid\ ($\beta_{2}$) & & \\\hline   
						 	
HD66811  	   & 4.2	  & 0.5 &	0.7	 &   0.9  & 7.0  & 5.0	 & 3.7 	&	3.8 &	3.2	& 2.5& &\\
			   & 8.5	  & 0.5 &	0.7	 &   0.9  & 7.0  & 5.0	 & 3.7 	&	3.8 &	3.2	& 2.5& &\\	
CygOB2\#11  & 5.05	  & 0.7  &	1.1	 &   1.15 & 1.0   & 1.0	 & 1.0	&  2.0 &  1.0	& 1.0& &4.5\\        
HD210839     & 1.3   	& 0.8  & 1.0	 &	1.15	  & 2.0   & 1.0	 &	1.0	&  7.0 &  7.0  &  5.3& &\\    
HD152408    & 9.5    	& 1.4  & 2.05  &  2.15  & 1.0   & 1.0	 & 1.0	&  4.0 & 3.25 & 2.1 & &\\       
HD151804    & 6.4    	& 1.7  & 2.0	 &  2.2  &	 1.0	   & 1.0 & 1.0  &  3.8 &  3.2	& 2.4& &\\          					 
HD149404    & 8.27  	& 1.1   &   1.3  &  1.4   &	 1.5	   & 1.0 & 1.0  &  6.0 &	5.2	& 4.5& &\\          					 
HD30614     & 1.75   	& 1.0   &  1.15  &  1.6   &	 8.0   & 6.0 & 4.0 &  3.5 &	3.3	& 2.3& &\\\hline  	 			
HD37128     & 1.25   	& 1.25	&	1.5	 &	1.7	  & 6.0   & 5.0  & 4.0 &	5.0	&	4.0	& 3.5 & &\\	      
HD38771     & $\lesssim$0.7$^a$  &1.4	&1.5    &1.85  & 10.5	& 8.0  & 6.0  	&1.0 &1.0     &1.0  & &\\                            
HD154090    & 1.2 			 & 1.1	&1.5    & 1.8   & 4.0 	& 2.0	& 1.3   &5.5 & 4.5	   & 3.8   & & \\                                       
HD193237    & 12.8              &1.7	&2.5	&3.2    & 3.5  & 3.0	& 2.5	&3.9 &3.5    & 3.2 &  &  \\                               
HD24398     & $\lesssim$0.18$^b$ &0.9	&1.5	&1.8 	& 2.5	& 1.0	&  1.0  & 1.0 &1.0    & 1.0   & 1.0 (10.0) &$\lesssim$ 0.151\\                               
HD169454    & 10.4			 	 &1.8   & 2.1   & 2.15 & 1.0	& 1.0	&  1.0 	&2.8 & 2.5	   &2.5 & & 10.1\\                                   
HD152236    & 6.2             & 1.3   &2.0    &2.6 & 2.5 	& 1.0	& 1.0   &2.8 & 2.5  & 1.8 & & \\                         
HD41117      & 1.8              	& 1.4	&2.0	& 2.6 &  12.0	& 7.3   & 4.8  	&1.6  & 1.35 & 1.2	& &  \\                    
HD194279	  & 2.12			& 2.0	&2.5	&2.7  &  1.0	& 1.0	& 1.0  	&1.0 &1.0	   &1.0  & & 1.95\\
			  &	0.5$^d$				& 1.2	&2.5	&2.7  &  2.5	& 1.0	& 1.0  	&1.0 &1.0	   &1.0  & & \\
HD198478   &	$\lesssim$0.38$^a$ 	&0.8	&1.3	&2.0  & 90.0	&40.0	&20.0	& 8.0	&6.5	&5.0 & &	 \\
			  &$\lesssim$0.14$^a$  	&0.8	&1.3	&2.0  & 90.0	&45.0	&28.0	& 8.0	&6.0	&5.0 & &	\\
HD80077     & 3.45                 & 2.55		& 3.0 	& 3.8		&3.0  &  2.5  	&  2.2 & 1.8 & 1.8	& 1.6  & &\\                                                
HD53138     & $\lesssim$1.8$^c$	   & 1.7	& 2.0	&2.2	&3.3   & 2.0  & 1.8   & 1.0   &1.0	& 1.0  & &   \\                                         
CygOB2\#12\tablefootmark{*}& 3.0$^d$		& 1.6	& 3.0	&3.2    & 1.0	& 1.0  & 1.0   & 12.0	&10.0	& 8.0 & &\\
								& 1.02$^d$		& 1.6	& 3.0	&3.2    & 1.0	& 1.0  & 1.0   & 12.0	&10.0	& 8.0 & &\\\hline                                                                                             
HD24912    & 	1.4		       &0.7	& 0.9	&1.1    & 5.0  & 3.5  & 2.0   & 5.0	& 3.5	& 3.5 & &\\                              
HD36816    & $\lesssim$0.16$^a$      & 0.7    & 0.9  &1.25   & 1.0  &  1.0  & 1.0	&	1.0	&	1.0	&	1.0 &1.0 (3.5) &$\lesssim$0.1\\                        
HD37043\tablefootmark{**} & 0.25$^d$               & 0.7   & 0.9   &1.4    & 1.0   & 1.0	&1.0	&1.0	&1.0	&	1.0 &  2.05/23.0 & 0.195\\\hline  
HD149757    & $\lesssim$0.07$^b$     &  0.7	& 0.8	 &1.15   & 1.0   & 1.0   & 1.0	& 1.0	&1.0	&1.0 &1.0 (15.0) &$\lesssim$ 0.036\\ 
HD149438\tablefootmark{***} & $^1\lesssim$0.315$^a$  &  0.6  &  0.8   &1.0    & 1.0   &1.0    &1.0	&1.6	&1.0	&15.0 & 1.0 (10.0) & $\lesssim$0.115\\
			  & $^2\lesssim$0.0185$^{a}$ &  0.7  &  0.8   &1.0    &400.0 &300.0&250.0& 1.0/300.0 & 1.0/300.0 &1.0/350.0 & 1.0 (350.0) &$\lesssim$0.016\\\hline                                      
\end{tabular}
\label{tablebetachange}	
\end{center}  
} 
\mbox{\tablefoottext{*}{\clout  ($\beta_{1}$) = 9}}\\
\tablefoottext{**}{\clout ($\beta_{2}$) = 1.0/10.0}\\
\tablefoottext{***}{$^1$ \clout ($\beta_{2}$) = 1.0 (10.0); $^2$\clout ($\beta_{2}$) = 1.0 (350.0)}
\end{table*} 

\section{Discussion}\label{discussion}

The previous section showed that the clumping structure can be well described relative to the outermost wind in Region 5 (radio domain). Assuming for now minimum values \clfar= \fclmin= 1, this allows us to discuss in detail the relative radial clumping-behaviour as well as the derived upper limits \Mdotmax. 

\subsection{Clumping properties of OB stars}
\subsubsection{Radial stratification at r\,$\gtrsim$ 2\,\Rstar}

A key finding of our analysis is that clumping at r\,$\gtrsim$ 2\,\Rstar\ presents similar radial stratification regardless of the strength of the wind, always fullfilling the condition $\clmid \gtrsim \clout \gtrapprox \clfar = \fclmin$ (see Tables \ref{tableclfactors} and \ref{tableclbestfit}). 

The specific behaviour from the intermediate to outermost wind seems to depend on luminosity class and spectral type, though, on average, the decrease in \fcl\ from the intermediate to outermost regions is steeper for OSGs (a factor 4) than for BSGs (a factor 2). Moreover, the clumping properties of the few OB\,Dwarfs and Giants in our sample show a smoother, more homogeneous behaviour than OB\,Supergiants, with similar clumping degrees at the intermediate and outer wind regions for most sources. 

This varying clumping degree for different luminosity classes and spectral types is supported not only by larger values of relative clumping factors, but also by the fact that the clumping stratification seems to be confined to a narrower region in OSGs and some BSGs, as revealed by the use of our adapted-regions approach (see Sec.\,\ref{modeling}). For the majority of the sample the boundaries used in the fixed-regions approach seem to describe the radial clumping stratification rather well. However, for some OSGs in our sample the best-fit solution was found after adapting mainly \rmid, and in a few cases also \rin\ and/or \rout. This is also the case for half of the BSGs (one B0\,I and all eBHGs) and for one of the O\,Giants (see Table\,\ref{tableclbestfit}). Changes in the extension of the wind regions are balanced by changes in the clumping factors, thus the overall clumping stratification is rather similar but more precisely described. Moreover, in all simulations we observed that the quality of the fits barely changed when varying \rout\,=\,15\,\Rstar\ to \rout\,= 10\,\Rstar, with the exception of OSGs, for which \rout\,=\,12\,--\,15\,\Rstar\ always provided the best-fit. Keeping in mind that the average minimum clumping degree at this radius is similar to the clumping  degree in the outermost wind region, r\,$>$\,50\Rstar, this implies that the clumping-degree of these sources drops quickly after reaching its maximum value at r\,$\approx$ \,1.1\,--\,6\,\Rstar. 

There are a few exceptions to this general behaviour, involving non-standard sources such as a binary system (HD\,37043) and variable thermal emitters (P\,Cyg, CygOB2\#12). The other binary source in our sample, HD\,149404, seems to follow the general trend, however the characteristic radio flux variability associated to binarity could not be tested for, since only one flux measurement at 3.6\,cm is available. Therefore, this object could eventually behave as the other non-standard objects in the sample. This discrepancy in the clumping behaviour between standard and non-standard objects suggests that those sources showing  an increasing clumping degree from the intermediate to outermost regions do not reflect intrinsic wind-clumping differences. Instead, this might rather be explained by different physical conditions in the outer wind when compared to standard sources (colliding winds, binarity, magnetic fields, ionisation changes, etc), which can significantly modify the flux emission in the mm and the radio regimes. If such a correlation was confirmed by further studies, an analysis of radial clumping stratifications in the intermediate to outermost wind regions would help to discriminate between objects of different natures (\eg\ standard or peculiar).

\subsubsection{Comparison with empirical and theoretical studies}\label{comparison}

Our derived clumping structures at r\,$\gtrsim$\,2\,\Rstar\ agree well with previous studies. For instance, \citetalias{Puls2006} found a similar dependency of the clumping degree with  luminosity class when comparing the intermediate and outermost wind regions. In addition, for those stars in common with their sample, our flux estimates from PACS observations at 70, 100, and 160\,\micron\ allowed us to here better constrain clumping in the intermediate wind region, lowering \citetalias{Puls2006}'s previous 
upper-limit estimates for this region.

Moreover, we find qualitatively similar results for the clumping structure to those reported by \cite{Najarro2011} and \cite{Clark2012}, using a different clumping parametrisation (\citealt{Najarro2008}). Overall, these authors find a sharp decrease in the clumping degree for OB\,Supergiants beyond r\,$\gtrsim$\,1.5\,--\,2\,\Rstar, while OB\,Giants and Dwarf stars present a roughly constant ---possibly unclumped--- value throughout the wind. If our clumping factors in the outermost wind region (radio regime) are normalised to a similar value as those in \cite{Najarro2011} and \cite{Clark2012} the wind structure of the common targets can be compared quantitatively. For HD\,66811 ($\zeta$\,Pup) and HD\,36861, the agreement is excellent, despite the different luminosity class and clumping properties derived for these two stars. For HD\,152236 ($\zeta^1$\,Sco), and CygOB2\#12, our absolute values of clumping factors are very similar in the inner and outermost wind regions, and a factor 3 lower at r\,=\,2\,\Rstar. Finally, for the other two OB\,Supergiants in common, HD\,30614 and HD\,37128 ($\alpha$\,Cam and $\epsilon$\,Ori, respectively), the derived clumping values differ considerably by more than a factor 6 at r\,=\,2\,\Rstar, but they converge quickly for larger radii. Nevertheless, this apparent discrepancy might be due to the lack of reliable FIR flux observations in those studies, which resulted in poorly constrained degrees of clumping in this wind region. Interestingly, in all sources in common with \cite{Najarro2011} and \cite{Clark2012}, the best-fit model is found in the adapted-regions approach. This would imply that the physical conditions in the wind of these stars change considerably over small distance increments, suggesting a more confined or structured wind.
On the other hand, \cite{Sundqvist2011} analysed the clumping structure in $\lambda$\,Cep by means of radiation-hydrodynamic (RH) simulations and empirical, stochastic models (including effects of optically thick clumping). Here the agreement with our work is again very good once we scale our clumping factors accordingly, although we find here that $f_{\rm cl}$ reaches its maximum at a slightly larger radius (r\,$\gtrsim$\,2\,\Rstar) when compared to their simulations (r\,$\approx$\,1.2\,--\,1.5\,\Rstar). 

The above results further allow us to compare to theoretical predictions from LDI simulations. Overall, our results agree quite well with 1D simulations for OSGs by \citet{Sundqvist2013} and \citet{Driessen2019} that account for limb-darkening and/or photospheric perturbations (see also discussion below). The clumping factor in these models typical peaks at $\sim$\,1.5\,--\,2\,\Rstar\ and decreases beyond, in general agreement with our empirical findings here. Moreover, our empirical study also tentatively agrees with recent simulations by \citet{Driessen2019}, which show that the winds of BSGs should be overall less clumped than OSGs. However, all these LDI simulations only reach r\,=\,10\,\Rstar\ and would need to be extended to higher radii in order to confirm this. 

\subsubsection{The innermost wind region}

There are also a few interesting aspects of the clumping structure in the inner-to-intermediate wind region transition. We find that for several OB\,Supergiants in our sample (13/20), the clumping degree seems to increase from the inner to the intermediate wind region (\clin\,$\lesssim$\,\clmid). This agrees with \citetalias{Puls2006}, who found the same trend for all their sample but one star, HD\,66811 ($\zeta$\,Pup).   
Such a trend implies that the maximum \fcl\ would occur at r $\sim$\,2\,--\,6\,\Rstar, depending on the source. However, this is not really a general trend, since for a non-negligible fraction of our sample (7/20) we find the opposite behaviour (\clin\,$>$\,\clmid). In other words, the exceptional behaviour of $\zeta$\,Pup found by \citetalias{Puls2006} is shared among several stars in our sample (and also those studied by \citealt{Najarro2008, Najarro2011}, \citealt{Clark2012} and \citealt{Sundqvist2011}). This would imply that in these stars the clumping onset occurs very close to the base of the wind (r\,$\lesssim$\,1.2\,\Rstar), and that the maximum clumping is achieved slightly below the intermediate wind region (r\,$\sim$\,1.5\,--\,2\,\Rstar). 

This discrepancy could, however, be influenced by the $\beta$-\clin\ degeneracy problem (see Sec.\,\ref{betasection}). Lowering/increasing the values of $\beta$ requires a corresponding increase/decrease in clumping factors, which is larger in the inner region than in the intermediate one (see Table\,\ref{tablebetachange}). Therefore, for stars with \clin\,$<$\,\clmid\ to become reversed (i.e., \clin\,$>$\,\clmid), a very large decrease of $\beta$ would be needed, and vice versa. It is reasonable to assume some uncertainties in the used $\beta$ values, however it seems unlikely that the estimated values of $\beta$ are off by such a large amount in so many stars. That is, the existence of two different trends in our sample may point to intrinsic differences in the efficiency of the different mechanisms governing the onset of clumping. 

As also discussed above, these observed trends may be compared to the different theoretical LDI simulations by \cite{Sundqvist2013} and \cite{Driessen2019}. Namely, depending on the different initial conditions of the simulations, and whether the modeled star is an OSG or BSG, the onset, the peak, and the overall predictions for the radial stratification of clumping may differ. For example, for high-luminosity objects it is possible that the radiative acceleration will exceed gravity already in deep sub-surface layers (e.g., at the so-called `iron opacity-bump'), which might then trigger a turbulent atmosphere that may also affect quantitative predictions for clumping factors \citep{Cantiello2009, Jiang2015}. Effects on clumping factors from a perturbed photosphere were investigated by \citet{Sundqvist2013} (see also \citealt{Feldmeier1997}), who indeed found that this can affect predictions both for the inner-wind clumping and for the clumping stratification as a function of radius. As mentioned above, the empirical results for the outer wind regions found here seem overall consistent with these simulations. We note, however, that since these models had an outermost wind radius $\sim 10\,R_\ast$ it remains to be investigated how far out in the wind effects from a potentially turbulent photosphere could persist.  

As such, the observational constraints obtained here will provide a very sound base for future theoretical models targeting more detailed investigations of this radial behaviour. Observationally, to understand and to properly describe the physical mechanisms that determine the clumping degree in the innermost wind region, further investigations using multi-wavelength analyses, continuum and line fits, including both $\rho$-dependant (resonance lines) and $\rho^2$-dependant (\Ha\ and NIR lines + V to radio continuum emission) clumping diagnostics are required.

\subsection{Upper-limit mass-loss rates and comparison to theory across the bi-stability jump}\label{massOBSg}

In the following, we compare our empirical estimates of maximum mass-loss rates (\Mdotmax)
to the theoretical predictions by \citetalias{Vink2000} and \citetalias{Vink2001} (\Mdotth), since they are the mass-loss rate recipes most commonly used in key applications such as stellar evolution\footnote{Current evolutionary models implement Vink's equations in several ways. For a brief summary see e.g. \citealt{Martins2013} and \citealt{Keszthelyi2017}.} These authors provide simple recipes to estimate mass-loss rates for various ranges of effective temperatures, depending on the so-called first and second `bi-stability jumps'. These jumps occur in the models when iron recombines first from Fe\,{\sc iv} to Fe\,{\sc iii} (first jump) and then to Fe\,{\sc ii} (second jump); since the lower iron ionisation states have more efficient driving lines, crossing these jumps from the hot sides result in significant increases in predicted mass-loss rates. 

The recipe suggested in \citetalias{Vink2000} (and \citetalias{Vink2001}), switches from `Fe\,{\sc iv}' to `Fe\,{\sc iii}' in the range $T_{\rm eff}^{\rm jump1}$\,=\,27.5\,--\,22.5\,kK and then further to the `Fe\,{\sc ii}' branch around  $T_{\rm eff}^{\rm jump2}\sim$\,12.5\,--\,18.5\,kK, depending on the mean wind density (see eqn.\,6 in \citetalias{Vink2000}, and discussion in their section\,5.3). This means that the central jump temperature that defined the first and the second bi-stability jump regions are $T_{\rm eff}^{\rm jump1}\sim$\,25\,kK and $T_{\rm eff}^{\rm jump2}\sim$\,15\,kK, respectively. However, studies by \cite{Lamers1995}, \cite{Crowther2006}, \cite{Markova2008}, \cite{Petrov2014} and \cite{Vink2018} indicate that the first jump, if any, is around 22\,--\,20\,kK, whereas the second jump, according to \cite{Petrov2016}, might be below 9\,kK. Since our OB\,Supergiant sample covers effective temperatures from 39\,000\,K to 13\,700\,K (O4\,--\,B3/B4) we can empirically investigate the mass-loss behaviour across the predicted bi-stability jumps using our data.

Let us note already here that, very recently, \citet{Bjorklund2020} presented new theoretical \Mdot\ predictions for O-stars. These are based on steady-state hydrodynamics and NLTE line radiative transfer in the co-moving frame (instead of a prescribed $\beta$ velocity law and NLTE Sobolev-Monte-Carlo line-transfer computations, as in \citetalias{Vink2000} and \citetalias{Vink2001}), and consistently find lower \Mdot\ values than \citetalias{Vink2000} and \citetalias{Vink2001}. However, since these models have not yet been extended to BSGs, we opt here to compare our findings only to those by \citetalias{Vink2000} and \citetalias{Vink2001} (\Mdotth), rather than using some proxy such as the wind-momentum luminosity relation. 

We compare our empirical results \Mdotmax\ with theoretical predictions \Mdotth\ as they are used in the often-cited grids of evolutionary models in codes such as Geneva (e.g. \citealt{Ekstrom2012}, \citealt{Yusof2013}), Bonn (e.g. \citealt{Brott2011}, \citealt{Kohler2015}) and MESA (Modules for Experiments in Stellar Astrophysics; \citealt{Paxton2011}). 

 First, we specifically use the definitions of \tjumphot\ and \tjumpcold\ from equations 15 and 6 from \citetalias{Vink2001} and \citetalias{Vink2000}, respectively, and the mass loss recipe corresponding to the location of the source in the temperature/bi-stability jump space is applied (see \citetalias{Vink2000} and \citetalias{Vink2001}). This implementation corresponds to the bi-stability jump regions and the mass-loss rates recipe as outlined above, which is indeed used in the Geneva code (hereafter Geneva approach). For a second test, we adopt a similar approach to that used in the MESA and Bonn codes (hereafter MESA approach): \tjumphot\ is obtained as a function of the density of the wind via metallicity (eqn.\,14 and 15 by \citetalias{Vink2001}), \tjumpcold\ is set to 10\,kK, and the mass loss prescriptions for the `hot' and `cold' side of the first jump are used (eqns.\,24 and 25 by \citetalias{Vink2001}). For \teff\ between 27.5 and 22.5\,kK, we use the same interpolation method than in MESA, and developed by \cite{Brott2011}. Finally, to investigate \Mdotth\ as a function of the temperatures of the first and second jumps reported by several authors (see above), we carry out a third test, following the MESA approach but fixing also \tjumphot\ at 22\,kK (hereafter Fixed-jumps approach). It is worth to mention that, whereas the Bonn and MESA codes define \tjumpcold\ below 12.5\,--\,10\,kK --switching to different mass loss predictions\footnote{The switch to a different wind density scenario and mass loss prescription in these codes following different criteria as \teff\ reaching the defined \tjumpcold.}-- and \tjumphot\ depending on the wind density of the source via metallicity, the Geneva code defines them via $\Gamma_{e}$ (eqn.\,5 by \citetalias{Vink2000}). Since our sample are galactic OB stars, this means that \Mdotth\ estimates will be barely affected by which definition of \tjumphot\ we use in this analysis.\footnote{For our sample, \tjumphot$\approx$\,25\,kK (26.1\,kK--24.1\,kK) when eqn.\,5 and eqn.\,4 (\citetalias{Vink2000}) are used, and \tjumphot$\approx$\,25.9\,kK when eqn.\,14 and eqn.\,15 (\citetalias{Vink2001}) are applied instead.}

 \begin{figure}[ht]
\centering
\includegraphics[width=7.5cm]{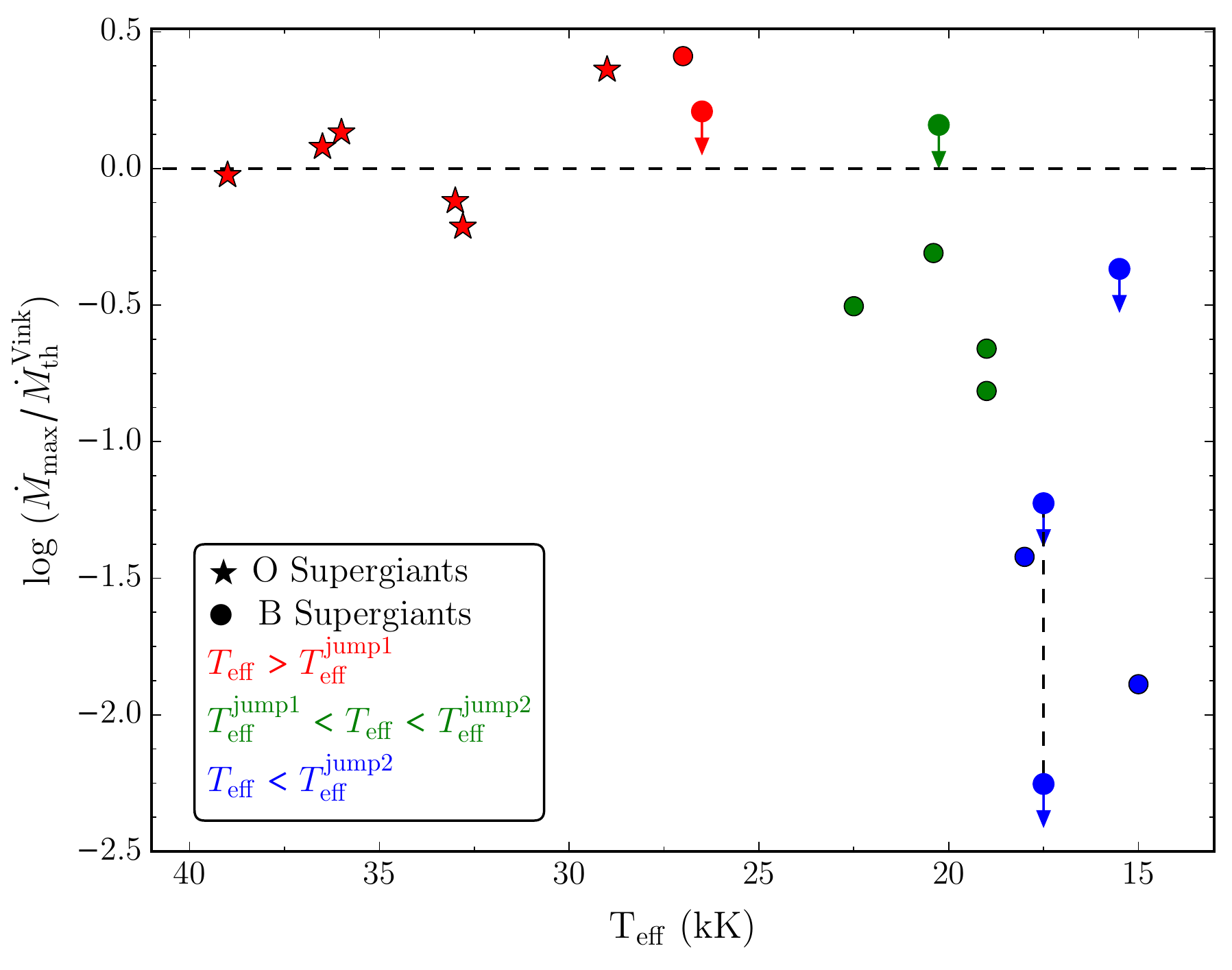}
\includegraphics[width=7.5cm]{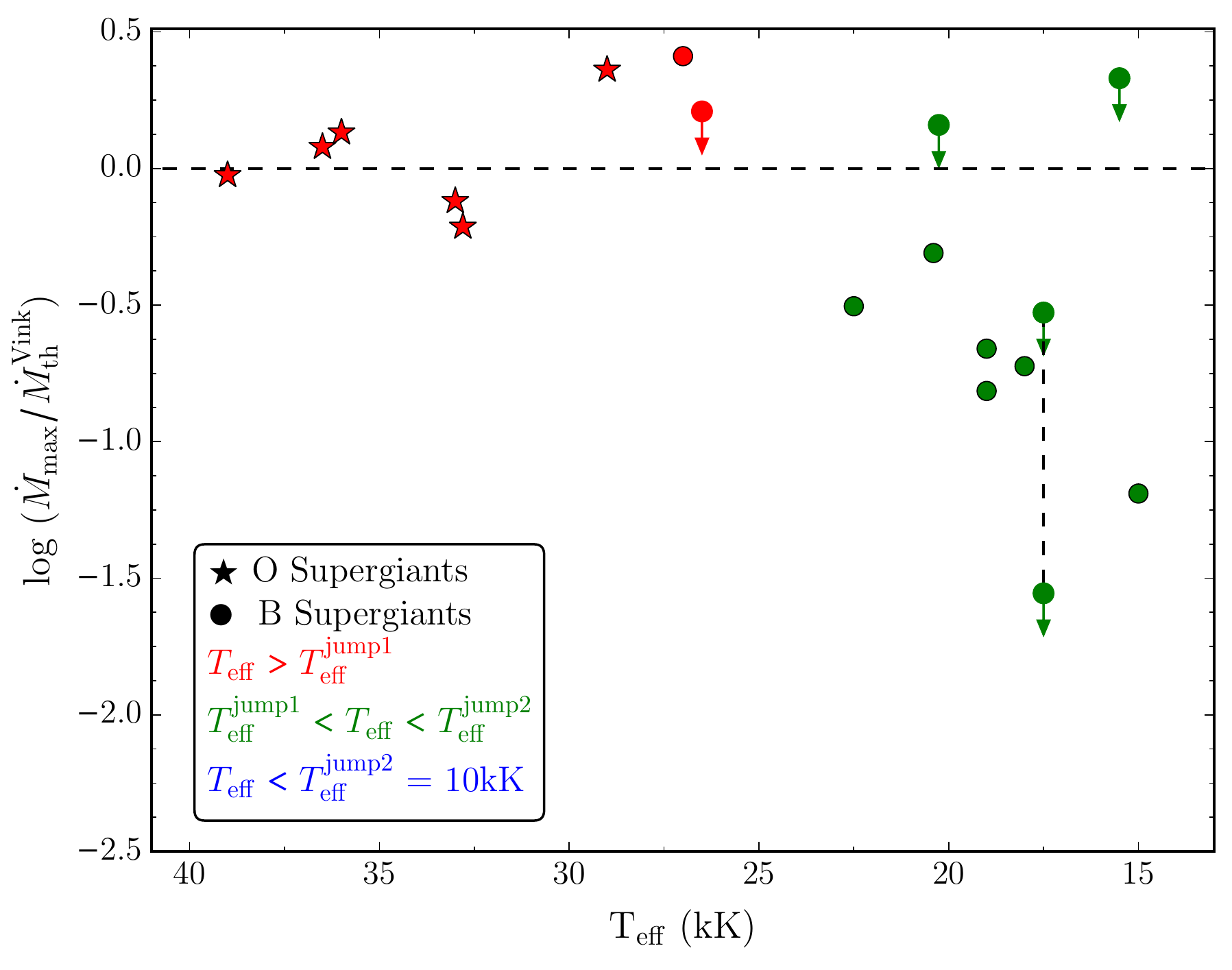}
\includegraphics[width=7.5cm]{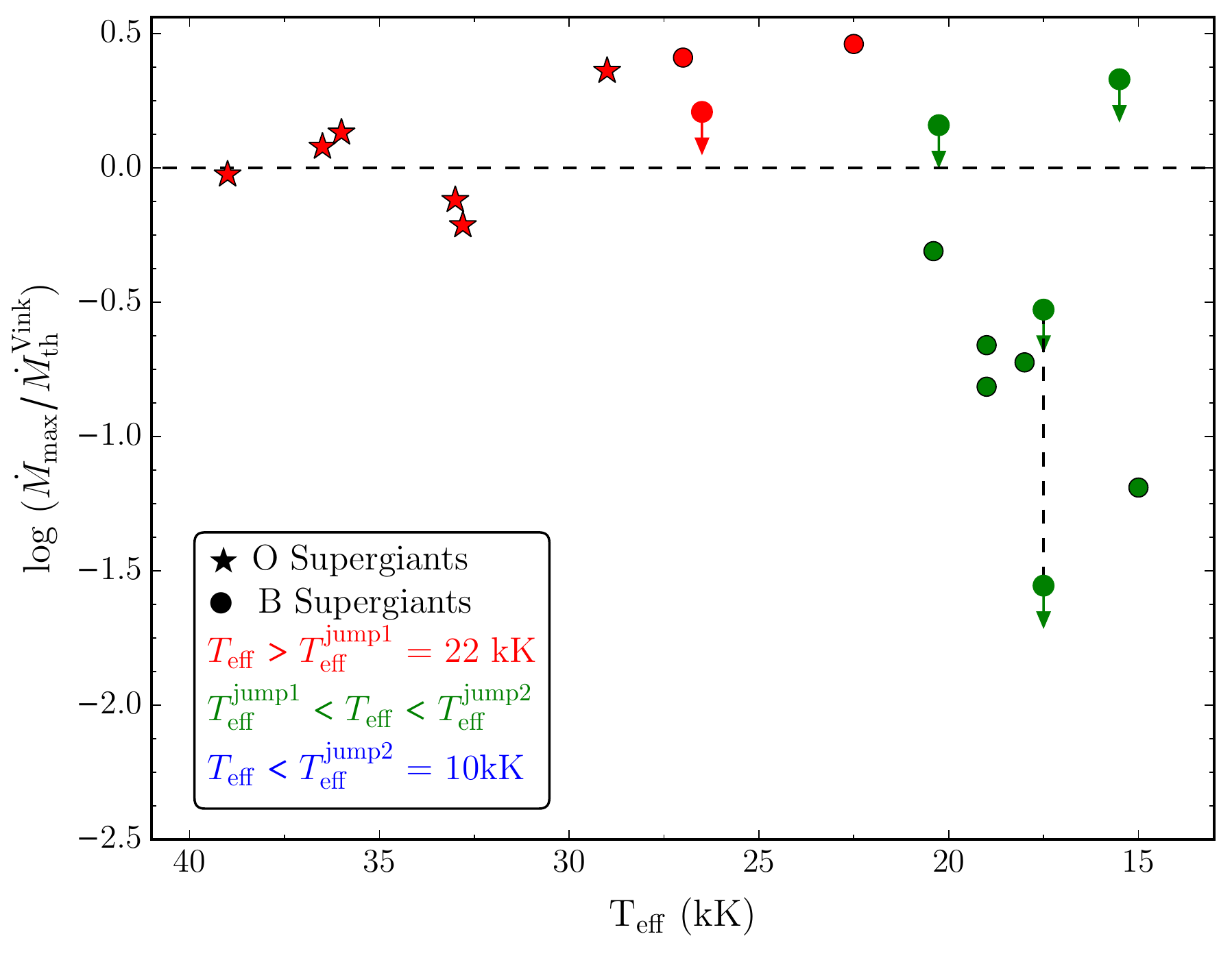}
\caption{From Top to Bottom, empirical to theoretical mass-loss rates ratio, in logarithmic scale, as a function of effective temperature for the OB\,Supergiants subsample. Empirical mass-loss rates correspond to the \Mdotmax\ derived in this work. Theoretical mass-loss rates, \Mdotth, correspond to the mass-loss rates computed via recipes from \citetalias{Vink2000} \& \citetalias{Vink2001} for different definitions of the temperatures of the jumps (see Sec.\,\ref{massOBSg}); \textit{top:} Geneva approach, \textit{middle:} MESA approach, and \textit{bottom:} Fixed-jumps approach (see Sec.\,\ref{massOBSg}). Different colours indicate at which side of the bi-stability jumps the sources are located. Arrows and symbols as in Figure\,\ref{figmdot_spec}.}
\label{figmdotmax}
\end{figure}

Figure\,\ref{figmdotmax} shows the empirical to theoretical mass-loss ratio (\Mdotmax/\Mdotth) as a function of \teff\ for the OB\,Supergiants subsample,  for the three different implementations of  \citetalias{Vink2000} \& \citetalias{Vink2001} described above. The non-thermal and the variable thermal emitters (HD\,149404, P\,Cyg, CygOB2\#12) were excluded of this analysis.

There is a clear trend in the sample, showing higher levels of discrepancy between \Mdotmax\ and \Mdotth\ for lower effective temperatures, regardless the used definition for the temperature of the bi-stability jumps. For the hotter OSGs, \Mdotmax\ agree within errors (Section\,\ref{errorfcl}) with the \Mdotth\ recipes. But as $T_{\rm eff}$ is lowered reaching the BSG regime, the difference between empirical and theoretical mass-loss rates increases significantly. Indeed, for the coolest BSGs \Mdotmax\ is up to almost 2 orders of magnitude lower than the prediction when using the Geneva approach, and close to 1.5 orders of magnitude when using the MESA approach.
 
At first glance, these very big discrepancies for luminous BSGs may seem surprising. However, the \Mdotmax\ values derived in this work are (on average) actually not that different than many others obtained by various studies present in the literature. For example, \citet{Haucke2018} derived \Mdotmax\ for a sample of BSGs by means of \Ha\ fitting using (unclumped) {\sc fastwind} models. For the 4 stars that are present in both samples, having $T_{\rm eff}$\,=\,17\,500\,--\,25\,000\,K, we obtain an average $\dot{M}^{\rm Haucke}_{\rm max}/\dot{M}^{\rm this work}_{\rm max} \approx 0.8$. Moreover, inspection of their table A.1, where they also list rates derived by a few other studies, seem to not reveal any major systematic discrepancies. Similarly, \citet{Crowther2006} also used \Ha\ to derive \Mdotmax\ (using unclumped {\sc cmfgen} models); here, for the 9 overlapping stars, all within $T_{\rm eff}$\,=\,15\,500\,--\,29\,000\,K, we find an average $\dot{M}^{\rm Crowther}_{\rm max}/\dot{M}^{\rm this work}_{\rm max} \approx 1.0$. As such, although the scatter is large it seems unlikely that the big discrepancies between observations and theory found for BSG mass-loss rates should be caused by any systematic effect specific for our study here (even though considerable uncertainties exist for various individual objects, see Table\,\ref{tableclfactors} and discussions in Appendix\,\ref{appendixA}).  

\begin{figure*}[ht]
\centering
\includegraphics[width=15cm]{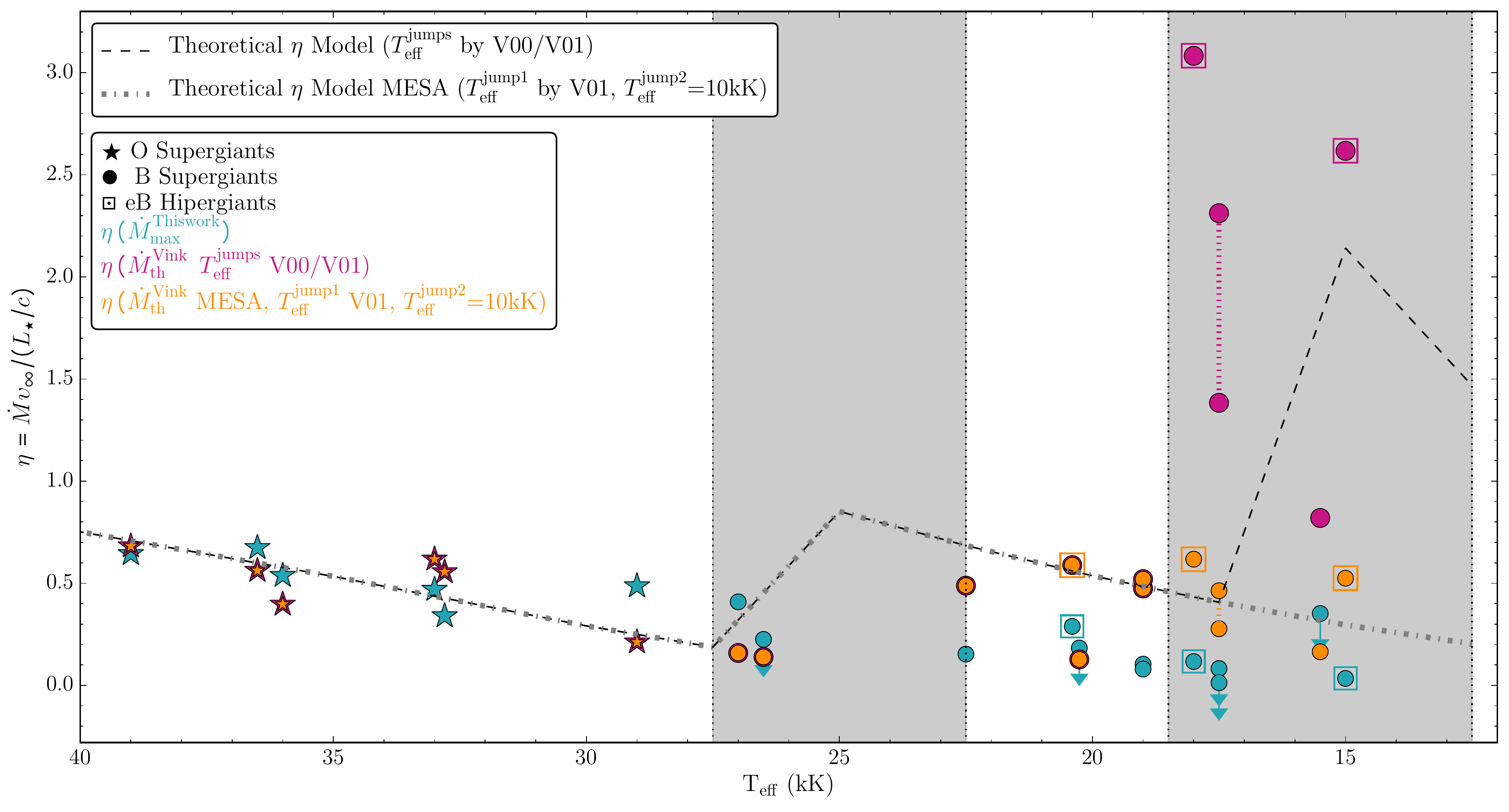}
\includegraphics[width=15cm]{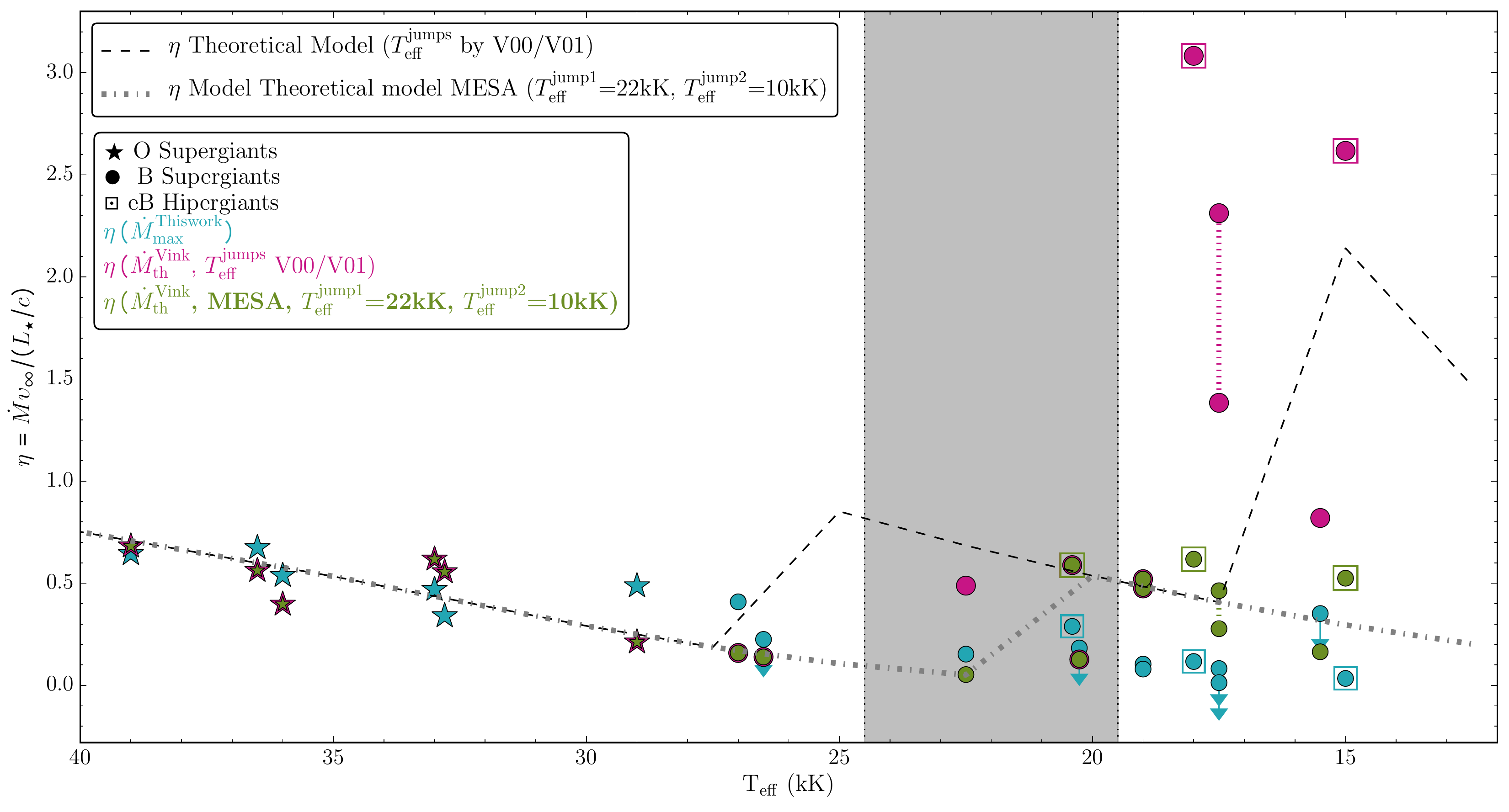}
\caption{\textit{Top:} Empirical-maximum (\Mdotmax) and theoretical (\Mdotth) wind performance numbers, $\eta = \dot M$ \vinf/$(L_{\ast}/c)$, as a function of \teff, for our OB\,Supergiants subsample, in the Geneva (magenta) and MESA (orange) approaches.  Dashed and dotted-dashed lines correspond to theoretical predictions for a source with log L/\Lsun = 5.75 and \Mstar = 45 \Msun\ for solar metallicity respectively based on Geneva and MESA implementations of \citetalias{Vink2000} and \citetalias{Vink2001}. The shadowed regions represent the first (27.5\,--\,22.5\,kK) and second (18.5\,--\,12.5\,kK) bi-stability jump zones as defined by \citetalias{Vink2000}. 
\textit{Bottom:} Same as top panel, but showing theoretical wind performance numbers in the Geneva (magenta) and Fixed-jumps (light green) approaches, and a dotted-dashed line marking the theoretical model based on the Fixed-jumps implementation of \citetalias{Vink2000} and \citetalias{Vink2001}. The shadowed region represents the first bi-stability jump zone (24.5\,--\,19.5\,kK) for a central temperature of the jump of 22\,kK. In both panels the early B-type Hypergiants in our sample are indicated with squares, whereas arrows indicate upper limits of \Mdotmax. Finally, arrows, dotted lines, and symbols as in Figure\,\ref{figmdotmax}.
}
\label{figwindeff}
\end{figure*}

As just one specific example, for HD\,198478 with $T_{\rm eff}$\,=\,17\,500\,K we derive an interval (see Table\,\ref{tableclbestfit}, and discussion in Appendix\,\ref{appendixA}) \Mdotmax\,=\,0.14\,--\,0.38\,$\times\,10^{-6} \Msunyr$. This agrees well with the interval \Mdotmax\,=\,0.12\,--\,0.41\,$\times\,10^{-6} \Msunyr$ obtained by \citet{Markova2008} from their unclumped \Ha\ analysis. Direct application of Geneva approach discussed above to HD\,198478 places this star at the cool side of the second bi-stability jump (see dotted-lined source in Figures\,\ref{figmdotmax} and \ref{figwindeff}); as such, even when using the empirical upper-limit  \vinf\,=\,470\,\kms\ from Table\,\ref{tableclfactors}, the recipe predicts a very large \Mdotth\,=\,6.38 \,$\times\,10^{-6} \Msunyr$, causing the large discrepancy visible in Figure\,\ref{figmdotmax} (see top panel).
Even if the star is placed in the `Fe\,{\sc iii}' region, i.e. when MESA and Fixed-jumps approaches are used, the recipe would still return a rate, $\Mdotth\,= 5.025\,\times\,10^{-6}\,\Msunyr$, that is more than an order of magnitude higher than the empirical upper limit (middle and bottom panel in Figure\,\ref{figmdotmax}). This shows that the reason we find large discrepancies between \Mdotmax\ and \Mdotth\ for BSGs, but not for OSGs, is because of the large predicted \Mdotth\ increase when crossing the bi-stability jump regions. 

In fact, when inspecting the so-called wind-performance number $\eta = \dot M$ \vinf/$(L_{\ast}/c)$ for our sample (Figure\,\ref{figwindeff}) we find a gradually decreasing trend --albeit again with significant scatter-- with decreasing $T_{\rm eff}$, and do not see any evidence for sudden increases, or a secondary local maximum (e.g. \citealt{Benaglia2007}), at any of the predicted bi-stability limits. This is quite similar to the findings by \citet{Crowther2006}, \citet{Markova2008} and \citet{Haucke2018}, who also did not find any empirical evidence for a sudden mass-loss increase with decreasing $T_{\rm eff}$. Note that, although the difference between theoretical predictions and empirical values is reduced when setting the central temperatures for the first and second bi-stability jumps at $\sim$22~kK and $\sim$10~kK respectively, the observed discrepancies for B~Supergiants are still considerable.   

These large discrepancies may thus point toward a problem with current theoretical mass-loss rate predictions around the first bi-stability jump, regardless of its exact location, rather than toward issues related to empirical mass-loss determinations for BSGs derived by different authors and methods. This suggests that the basic mechanism responsible for increasing the mass-loss rate at the bi-stability jump (essentially recombination of iron) needs to be revisited in future work, preferably by locally consistent hydrodynamic calculations that do not assume a pre-described $\beta$ velocity law. This is important also in view of the quite big role this bi-stability \Mdot\ increase plays in general stellar evolution modelling (see \eg\ \citealt{Vink2010, Keszthelyi2017}). 

\subsubsection{Impact of clumping in scaling the derived upper-limit mass-loss rates}

As mentioned above, the empirical values of the maximum OSG mass-loss rates derived from radio fluxes agree (within uncertainties) with the theoretical predictions by V00 and V01 (see also \citetalias{Puls2006}). However, 'real' agreement would only be true if the outermost wind of OSGs were unclumped. We note again that our \Mdotmax\ estimates were obtained normalising the clumping stratification  to the outermost wind region by setting \clfar\,=\,\fclmin\,=\,1.
Therefore, if instead this region is affected by clumping, downward corrections become necessary. Hydrodynamical wind models extending all the way to the radio-emitting regions suggest typical factors of about \clfar $\approx$ 4\,--\,9 (\citealt{Runacres2002, Runacres2005}). These models, however, were calculated assuming an unperturbed lower boundary at the stellar surface, and so it remains to be seen whether effects from, e.g., a turbulent photosphere could persist all the way out to the radio region (see also discussion in Sect. 5.1.3). Assuming for now the outer-wind values from \citet{Runacres2002, Runacres2005}, the \Mdotmax\ estimates of this paper would need to be scaled down by a factor $\approx$ 2\,--\,3. We re-emphasise that these typical \clfar $\approx$ 4\,--\,9 only refers to the outermost wind. Clumping factors predicted for the inner wind regions are typically significantly higher; for example, in the O-star simulations by \citet{Driessen2019} $f_{\rm cl}$ peaks around $r/R_\ast \sim 1.5-2$ and the average value for the inner wind ($r/R_\ast = 1-2$) is $f_{\rm cl} \approx 15-20$. Thus, O-star mass-loss rates derived from unclumped models of inner-wind diagnostics (e.g. using optical recombination lines like H$\alpha$) should be scaled down more than corresponding radio rates. This is consistent with the empirical finding here that clumping factors on average are higher in the inner wind regions than in the outer ones. Typical reductions of $\approx 2-3$ would agree well with the new theoretical mass-loss rate predictions for O-stars by \citet{Bjorklund2020} (see also \citealt{Sander2017a} and \citealt{Sundqvist2019}), as well as with various multi-wavelength empirical studies such as \citet{Najarro2011} (see figs. 12 and 13 in \citealt{Bjorklund2020}).  As demonstrated by Figure \ref{figmdotmax}, however, the situation is very different for BSGs, where \Mdotth\ can be overestimated by as much as 2 orders of magnitude. Of course, if the outermost wind of BSGs should be significantly clumped, this discrepancy would become even larger. For example, assuming the same clumping degree in the outermost wind as for OSGs (but see \citealt{Driessen2019}), current theoretical \Mdot\ predictions by V00 \& V01 (the standard rates used in most evolutionary models), would be overestimated by factors 6\,--\,200 in case of the Geneva implementation.

\section{Summary and conclusions}\label{summary}

We have constrained clumping properties of the intermediate stellar winds for a sample of 25 OB\,Supergiants, Giants and Dwarfs stars, by using {\sc Herschel-PACS} flux measurements at 70, 100 and 160\,\micron. For the analysis, we followed the approach developed, and further tested, by \citetalias{Puls2006}, assuming optically thin clumping. Additional available continuum observations in the literature from optical to radio wavelengths further allowed us to perform a consistent analysis and to derive robust estimates for the (minimum) wind-clumping structures at r\,$\gtrsim$\,2\,\Rstar, as well as maximum mass-loss rate estimates (\clfar\,=\,\fclmin\,=\,1). 

Accurate radio emission measurements allowed us to derive precise \Mdotmax\ values and to infer robust clumping structures for 17 of the 25 stars in our sample. In addition, mm flux observations of 7 of these stars led to better constraints of the outer wind region (15\,$\lesssim$\,r/\Rstar\,$\lesssim$\,50), providing a precise determination of the radial stratification of the clumping at r\,$\gtrsim$\,2\,\Rstar. Our major findings can be summarised as follows:

\begin{itemize}

\item The stellar wind at r\,$\gtrsim$\,2\,\Rstar\ of most of the stars in our sample fulfills the clumping stratification condition \clmid\,$\gtrsim$\,\clout $\gtrapprox$\,\clfar\,=\,\fclmin\ regardless of the strength of the wind. The exceptions correspond to non-thermal or variable thermal sources, such as HD\,37043 (binary), HD\,193237 (LBV) and CygOB2\#12 (eBHG).\\ 

\item The clumping-degree drop from the intermediate (\clmid) to the outer wind region (\clout) depends on spectral type and luminosity class: on average, \clmid\ is $\approx$\,4 times larger than \clout\ for OSGs, $\approx$\,2 times larger for BSGs, and similar to  \clout\ for OB\,Dwarfs and Giants. \\

\item Our findings agree well with the empirical clumping properties at r\,$\gtrsim$\,2\,\Rstar\ derived by \cite{Najarro2011} and \cite{Clark2012} following a different parametrisation. In addition, our results overall support the hydrodynamical OSG models by \cite{Sundqvist2013}, where clumping starts to decrease at r\,$\approx$\,2\,--\,6\,\Rstar, and, tentatively, the recent 1D LDI-simulations of OSG and BSG winds by \cite{Driessen2019}, which predict lower amounts of clumping in BSGs. \\

\item We found that for 8 OB\,Supergiants in our sample \clin\,$>$\,\clmid. This significantly extends the findings of \citetalias{Puls2006} in just one star of their sample ($\zeta$\,Pup), and is in agreement with the empirical clumping properties by \cite{Najarro2011} and the theoretical predictions for OSGs by \cite{Sundqvist2011} and \cite{Sundqvist2013}. This suggests that such a behaviour, rather than being an exception, could imply the existence of two trends characterised by different physical conditions at the base of the wind.\\

\item For OSGs the derived upper-limit mass-loss rates, \Mdotmax, agree with the theoretical predictions by \citetalias{Vink2000} and \citetalias{Vink2001} for unclumped winds, whereas BSGs show a discrepancy which severely increases with decreasing effective temperature: the estimated \Mdotmax\ start to differ from theoretical recipes in the predicted first bi-stability transition zone, and up to 1.5-2 orders of magnitude lower for the coolest B~Supergiants below the first bi-stability jump. Since the empirical scaling invariant is $\sim \dot{M} \sqrt{f_{\rm cl}}$ and our derived mass-loss rates are upper limits assuming an unclumped radio-emitting wind ($f_{\rm cl}^{\rm far}=1$), any clumping in this outermost region would only increase this discrepancy.\\ 

\end{itemize}

A key conclusion of our analysis regards the upper-limit mass-loss rates of OSGs and BSGs derived from radio emission. Although the actual empirical \Mdot\ will depend on the level of clumping in the outermost wind, these upper limits should be quite robust since radio emission is a relatively 'clean' diagnostic. This thus allows us to perform important empirical testing of theoretical mass-loss predictions across the so-called bi-stability jumps (see previous sections). 

If the absolute value of clumping in the outermost wind region of OB\,Supergiants was \clfar\,=\,4\,--\,9, as suggested by the hydrodynamic O-star models by \citet{Runacres2002, Runacres2005}, the theoretical mass-loss rate recipes by \citetalias{Vink2000} and \citetalias{Vink2001} would be overestimated by a factor 2\,--\,3 for OSGs; as discussed above, this would then agree well with the recent theoretical O-star mass-loss predictions by \citet{Bjorklund2020}. On the other hand, the consequences for BSGs across the bi-stability regions are dramatically \textit{independent of their clumping properties, and temperatures of the jumps}, since these objects require downward \Mdotth\ corrections of up to 1.5-2 orders of magnitude, even in the case BSGs were not as clumped as OSGs (\citealt{Driessen2019}). Thus, this finding calls for an urgent re-investigation of the role recombination of iron-like elements plays in determining the mass-loss rates of objects that cross the bi-stability region, as well as a careful analysis of corresponding effects for stellar evolution models \citep{Keszthelyi2017}.

Regarding future diagnostic studies, the next step is to further investigate the clumping degree also in the inner wind, and to compute actual, absolute values of $f_{\rm cl} (r) $. To achieve this our team is collecting archival and missing observations throughout the spectral range, including X-Ray emission. Multi-wavelength analysis, optical to radio continuum fitting as applied in this work, and UV to NIR spectroscopy in combination with state-of-the-art model atmosphere codes including an adequate treatment of wind clumping, will then allow for precise determinations of clumping factors and the actual mass-loss rates of OB stars.

\begin{acknowledgements}
M.M.R-D gratefully acknowledges J.A. Fern\'andez-Ontiveros for his invaluable insights and help at earlier stages of the paper. This research was partially supported by from a FPI-INTA fellowship, and the Spanish MICINN through grants AYA2008-06166-C03-02, AYA2010-21697-c05-01 and FIS2012-39162-C06-01. F.N and M.M.R-D acknowledge financial support through Spanish grants
ESP2017-86582-C4-1-R and PID2019-105552RB-C41 (MINECO/MCIU/AEI/FEDER) and from the Spanish State Research Agency (AEI)
through the Unidad de Excelencia ``Mar\'ia de Maeztu''-Centro de Astrobiolog\'ia (CSIC-INTA) project No. MDM-2017-0737.
JOS acknowledges support from the Odysseus program of the Belgian Research Foundation Flanders (FWO) under
grant G0H9218N, support from the KU Leuven C1 grant MAESTRO C16/17/007, and previous support from the European Union Horizon 2020 research and innovation program under the Marie-Sklodowska-Curie grant agreement No 656725.

\end{acknowledgements}

\bibliographystyle{aa}
\bibliography{mmrdbiblio.bib}

\appendix

\section{Fitting individual objects}\label{appendixA}

In the following, we provide detailed comments on our fits to the individual objects. For those sources previously analysed by \citetalias{Puls2006} through the same procedure, we also provide the best solutions derived by these authors. The fit-diagrams of OSGs are displayed in Figure\,\ref{figOSupergiants}, BSGs in Figures\,\ref{figBSupergiants1} and \ref{figBSupergiants2}, O\,Giants in Figure\,\ref{figOGiants}, OB\,Dwarfs in Figure\,\ref{figOBDwarfs}, and the two confirmed binaries in Figure\,\ref{figOBbinaries}. The results of the fixed-regions approach are summarised in Table\,\ref{tableclfactors}, and those of the adapted-regions approach in Table\,\ref{tableclbestfit}. Distances, colour excesses and extinction parameters are listed in Table\,\ref{tabledistances}.

\subsection{O\,Supergiants}\label{apxOSg}

\textit{HD\,66811 ($\zeta$\,Pup)}

This source was previously analysed by \citetalias{Puls2006}. Therefore, we follow a procedure like that also described in Section\,\ref{fittingprocedure} of the main text: first we test the clumping stratification derived by these authors (see their Table 7) against new PACS flux observations at 70, 100 and 160\,\micron\ (magenta diamonds in diagrams) together with additional observations at MIR (AKARI \& WISE catalogs) and at 2 and 6\,cm (\citealt{Morton1978, Morton1979}, \citealt{Bieging1989}) available in the literature (green circles in diagrams). Since all radio measurements are well determined, \Mdotmax\ is also definite. 

In Figure\,\ref{figOSupergiants}, we plot the two derived solutions in the fixed- (solid line) and adapted-regions (magenta-dotted line) approaches, and \citetalias{Puls2006}'s best-fit solution (blue-dashed line) for the shorter distance: \Mdotmax\,=\,4.2\massloss, \clin\,=\,5, \clmid\,=\,1.5, \clout\,=\,1.4 and \clfar\,=\,1 with \rin\,=\,1.12 \Rstar, \rmidc, \routc\ and \rfarc. We note that the clumping factor at the intermediate region \clmid\ derived by \citetalias{Puls2006} leads to an underestimation of the FIR fluxes.

Our best solution is obtained in the adapted-regions approach for the same values of \Mdotmax\ and \clin\ as derived by \citetalias{Puls2006}, and a larger intermediate clumping factor. Provided the clumped wind is restricted to 1.12\,$\lesssim$\,r/\Rstar\,$\lesssim$\,12, then derived values are those given in the first entry in Table\,\ref{tableclbestfit}. For the fixed-regions approach, the best possible solution is also achieved for the same values of \Mdotmax\ and clumping factors, but with a slightly worse $\chi^2$, 
see the first entry  in Table\,\ref{tableclfactors}. 
Identical results are obtained for the alternative solution provided by \citetalias{Puls2006} (see first entry in their tables 1 and 7) corresponding to the larger distance of $\zeta$\,Pup, $d$\,=\,0.73\,kpc: \Mdotmax\,=\,8.5\massloss, \clin\,=\,5, \clout\,=\,\clfar\,=\,1  and \clmid\,=\,3.1 and 3.2 for the adapted- and fixed-regions approaches. Despite the scaled values of \Mdotmax\ and \Rstar\ for this alternative distance, the radial clumping structure and optical depth invariant, $Q'$, are conserved.

As pointed out above, the IRAS measurements used by \citetalias{Puls2006} were not colour-corrected, whereas we performed the colour correction for IRAS observations at 12, 22 and 60\,\micron, leading to slightly lower flux values than those used by \citetalias{Puls2006}. However, this difference does not lead to significant changes in the derived clumping structure. 
Note that, although this source is well constrained, only additional mm and radio observations can help disentangling whether the ``odd'' measurements at 160\,\micron\ and 2\,cm are corrected (or not), or if they perhaps respond to variable thermal emission from the source.\\

\noindent\textit{CygOB2\#11}

This source was previously analysed by \citetalias{Puls2006}. Since the measured fluxes at 3.5 and 6\,cm  are well determined, \Mdotmax\ is definite.  However, only upper limits for the fluxes at 70, 100 and 160\,\micron\ can be estimated from the PACS observations, thus the clumping properties in the intermediate wind region remain unconstrained. Despite this, we tested the best solutions derived by \citetalias{Puls2006} against additional measured fluxes at MIR wavelengths from the literature (Spitzer, WISE and MSCX6 catalogs). There is a recent estimated distance for this source by \textit{GAIA}, $d$\,=\,1.72\,kpc, which is roughly the same value as used by \citetalias{Puls2006}, $d$\,=\,1.71\,kpc; as such, only a very small scaling-factor for \Mdotmax\ and \Rstar\ is needed to account for the updated distance.

The corresponding panel in Figure\,\ref{figOSupergiants} displays the best-fit solution by \citetalias{Puls2006} (magenta-dotted line) and our derived best-fit solution (solid line), corresponding to the fixed-regions approach.  The maximum mass-loss rate estimated by \citetalias{Puls2006} is consistent with all radio fluxes and the upper flux limits at FIR wavelengths (flux emission model well below the upper limits). However, the clumping structure derived by the authors
starts to depart from observations beyond 10\,\micron, overestimating the measured fluxes up to 25\,\micron\ and marginally matching the measured flux at 30\,\micron. Our best-fit solution is achieved in the fixed-regions approach for a constant and minimum clumping degree throughout the entire wind, i.e. \clin\,=\,\clmid\,=\,\clout\,=\,\clfar=\,1. The maximum mass-loss rate is only marginally different from \citetalias{Puls2006}'s estimate. 

Although our FIR flux estimates are only upper limits, the maximum clumping degree consistent with observations at the intermediate wind region is \clmidmax\,=\,3, a factor $\approx$\,1.7 lower than the value derived by \citetalias{Puls2006}. Note that only additional observations at FIR up to $\sim$\,1.3\,mm can help properly constrain the clumping properties at the intermediate wind region, although our simulations point to a lower clumping degree than previous estimates in that region.   
Finally, the large measured flux at 10.9\,\micron\ (\citealt{Leitherer1982}) is still not reproduced by any model, and thus either is affected by some problem (i.e. artifacts, contamination, etc.), or it could be related to warm dust emission. The former is more likely since the nearby flux estimates at 9 and 11.02\,\micron\ are consistent with a lower flux emission model for the source, and with the rest of flux observations at MIR. \\

\noindent\textit{HD 210839 ($\lambda$ Cep)}

\begin{figure}[!htp]
\begin{center}
\includegraphics[width=9cm]{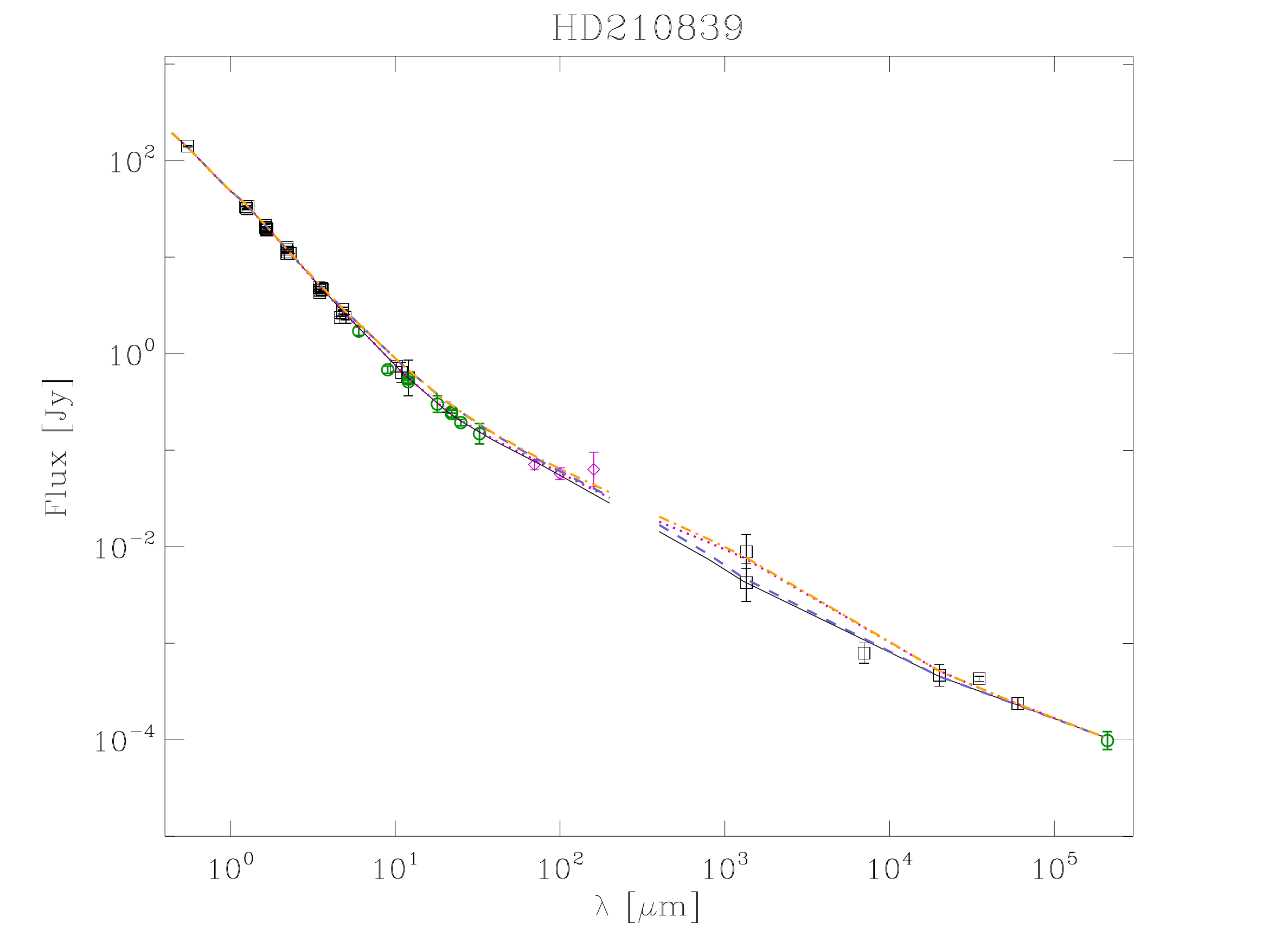}
\caption{Comparison between the best-fit solutions for HD\,210839 by \citetalias{Puls2006} (blue-dashed and orange-dotted-dashed lines) and our best-fit solutions (solid and magenta-dotted lines) for the same value of the distance (see text). Magenta diamonds are our measured FIR fluxes at 70, 100 and 160\,\micron. Black squares and green circles indicate flux values from the literature. For those sources in common with \citetalias{Puls2006}, green circles indicate new available data added to the analysis. Arrows indicate upper limits.}
\label{hd210839p06}
\end{center}
\end{figure}

This source was also analysed by \citetalias{Puls2006}, and a new distance estimate from \textit{GAIA} is available. 
Figure\,\ref{hd210839p06} displays the best-fit solutions from this work and by \citetalias{Puls2006} for the distance used by these authors. In Figure\,\ref{figOSupergiants} we present our best-fit models for the \textit{GAIA} distance used in this work.

From the well determined radio measurements, \citetalias{Puls2006} estimated a well defined maximum mass-loss rate for this source. Due to the two different SCUBA flux measurements at 1.3\,mm, the authors provided two alternative solutions for the clumping structure in wind region 4. Thus, for \Mdotmax\,=\,3\massloss\ ($d$\,=\,1.08\,kpc) their best-fit solution assumes \clin\,=\,6.5, \clmid\,=\,10 and \clfar\,=\,1 with \rinc, \rmid\,=\,4\,\Rstar, \routc\ and \rfarc. With \clout\,=\,1 (blue-dashed line in Figure\,\ref{hd210839p06}), the lower SCUBA value, the 0.7\,cm and the radio flux values are reproduced, whereas with \clout= 8, the upper SCUBA and all radio fluxes are well fitted, but the 0.7\,cm flux is overestimated (orange dashed-dotted line in Figure\,\ref{hd210839p06}). Using our new flux values, Figure\,\ref{hd210839p06} shows that the \citetalias{Puls2006} mass-loss rate is consistent with the added flux at 21\,cm, whereas their clumping stratification marginally matches the flux measurements at MIR and FIR wavelengths, but perfectly fits radio observations. In addition, the new measurement at 160\,\micron\ is consistent with the upper SCUBA value, which does not help to properly constrain the clumping degree for Region\,4. However, both 160\,\micron\ and SCUBA fluxes have large errors\footnote{The spatial resolution of the PACS photometer at 160\,\micron\ is lower than at 70 and 100\,\micron, and the PSF-FWHM is large enough to enclose  extended emission from its surrounding. Indeed, a close inspection of PACS images reveals a strong background emission at 160\,\micron.}, preventing us from drawing any further conclusions.

In the following, the described clumping structure refers to the fixed-regions approach. The best-fit solution is achieved for the minimum possible clumping degree for the inner and outermost wind regions, and a larger clumping for the intermediate one, \clin\,=\,\clout\,=\,\clfar\,=\,1 and \clmid\,=\,7 (solid line in Figures\,\ref{hd210839p06} and \ref{figOSupergiants}). In Region\,4, although a minimum clumping factor provides the best-fit solution, the clumping degree could range from 1 to 8, with a maximum value also ranging from 5 to 12. Thus, with \clout\,=\,1 or \clout\,=\,8, the lower (solid line in Figures\,\ref{hd210839p06} and \ref{figOSupergiants}) or the upper (magenta-dotted line in Figures\,\ref{hd210839p06} and \ref{figOSupergiants}) SCUBA value are respectively matched. \cloutmax\,=\,5\,--\,12  is still consistent with the 160\,\micron\ and with both SCUBA fluxes, respectively. Only extra flux observations at sub-mm and mm spectral ranges can help derive definite estimates for wind region 4. For the new distance estimate from \textit{GAIA}, $d$\,=\,0.62\,kpc, the corresponding scaled maximum mass-loss rate is \Mdotmax\,=\,1.3\,\massloss, and the same clumping properties described above (Table\,\ref{tableclfactors}) provide the best-fit solutions (Figure\,\ref{figOSupergiants}). \\

\noindent\textit{HD\,152408}

The flux observations at 2 and 6\,cm  are well determined, thus \Mdotmax\ is definite. In the corresponding panel in Figure\,\ref{figOSupergiants}, we present our best-solution fitting all the data (magenta-dotted line) and the best possible one in the fixed-regions approach (solid line), respectively. The maximum mass-loss rate is 9.5\,\massloss\, one of the largest values in the sample. For the fixed-regions approach the best solution is obtained by setting \clin\,=\,\clout\,=\,\clfar\,=\,1, and 
\clmid\,=\,3.25 (Table\,\ref{tableclfactors}). However, the best-fit solution for HD\,152408 is achieved for the adopted-regions approach by slightly increasing to \clmid\,=\,3.8, provided that the inner wind region is extended  to \rmid\,=\,3.5\,\Rstar\ (Table\,\ref{tableclbestfit}). The SCUBA measurement (1.3\,mm) further allows us to constrain the maximum possible value of the clumping degree in Region\,4 to \cloutmax\,=\,3. 
\\

\noindent\textit{HD\,151804 (V973 Sco)}

Discussed in detail in the main text. \\ 

\noindent\textit{HD\,30614 ($\alpha$ Cam)}

This source was analysed by \citetalias{Puls2006}, and 
we therefore show \citetalias{Puls2006}'s best solution together with our best-fits models.
It can be observed in the fit-diagram that \citetalias{Puls2006}'s solution underestimates fluxes at MIR and FIR wavelengths, and that although their \Mdotmax\ is still consistent with the additional measured fluxes at 6\,cm \footnote{The difference between the flux observations at 6\,cm  is not large enough to consider HD\,30614 a variable source (\citealt{Scuderi1998}).}, the additional radio flux at 21\,cm  points to a slightly larger \Mdotmax. 
We found that a value of \Mdotmax= 1.75\massloss\, with \clin= 6, \clmid  = 3.3 and \clout=\clfar= 1, provide the best possible solution in the fixed-regions approach (solid line; Table\,\ref{tableclfactors}). However, the best-fit solution (magenta-dotted line) for the estimated \Mdotmax\ is achieved by extending to \rmid= 4\,\Rstar, so that \clin= 6 and \clmid= \clout= \clfar= 1 (Table\,\ref{tableclbestfit}). In all simulations the maximum value of \clout
is constrained by \cloutmax=2.  \\

\noindent\subsection{B\,Supergiants}\label{apxBSg}


\textit{HD\,37128 ($\epsilon$\,Ori)}

There are well determined available radio fluxes for this source, therefore \Mdotmax\ is definite. For \Mdotmax\,=\,1.25\,\massloss, we present the two fit-solutions consistent with all the data, corresponding to the fixed- (solid line) and the adapted-regions (magenta-dotted line) approach, respectively (Figure\,\ref{figBSupergiants1}). For the fixed-regions approach, the best possible solution corresponds to a \,1 (Table\,\ref{tableclfactors}). The best-fit solution, however, is obtained in the adapted-regions approach by increasing a factor\,2 the clumping degree in the intermediate wind region, \clin\,=\,5, \clmid\,=\,8 and \clout\,=\,\clfar\,=\,1, provided that \rout\,=\,8 \Rstar\ (Table\,\ref{tableclbestfit}).\\

\noindent\textit{HD\,38771 ($\kappa$\,Ori)}

Only one radio observation, an upper limit at 6\,cm, is available for this source. Thus, only an upper limit for \Mdotmax\ is provided. The panel corresponding to this star in Figure\,\ref{figBSupergiants1} shows the fit-solutions for both the fixed- (solid line) and the adapted-regions (magenta-dotted line) approach.  
We estimated \Mdotmax\,$\lesssim$\,0.7\,\massloss. For this upper limit, the best possible solution in the fixed-regions approach corresponds to a significantly clumped inner wind, \clin\,=\,8, and \clmid\,=\,\clout\,=\,\clfar\,=\,1\,(Table\,\ref{tableclfactors}). 
The best-fit solution is achieved in the adapted-regions approach, with a considerable large clumping degree in the inner wind region, \clin\,=\,20, and again \clmid\,=\,\clout\,=\,\clfar\,=\,1, provided that \rmid\,=\,1.55\,\Rstar\ (Table\,\ref{tableclbestfit}). 
Despite of the lack of flux observations from 100\,\micron\ to radio wavelengths, \cloutmax\,=\,8 is still consistent with all data. Only mm and well-determined radio flux measurements can provide definite \Mdotmax\ and the clumping structure of the outer wind for this source.\\

\noindent\textit{HD\,154090 ($\kappa$\,Sco)}

Only one radio measurement is available at 3.5\,cm, but it is well determined and thus \Mdotmax\ is definite. Figure\,\ref{figBSupergiants1} displays fit-solutions for both the fixed- (solid line) and the adapted-regions approach (magenta-dotted line). 
For \Mdotmax\,=\,1.\,\massloss, a moderate clumping degree in the inner and the intermediate wind regions, \clin\,=\,2 and \clmid\,=\,4.5, and \clout\,=\,\clfar\,=\,1, provide the best possible fit solution in the fixed-regions approach (Table\,\ref{tableclfactors}). 
The best-fit solution for this source is achieved in the adapted-regions approach, with a moderate, and almost constant, clumped wind restricted to r\,$\lesssim$\,15\,\Rstar, \clin\,=\,4, \clmid\,=\,3.5 and \clout\,=\,\clfar\,=\,1, provided that \rin\,=\,1.4\,\Rstar\ instead of \rinc (Table\,\ref{tableclbestfit}).  In the absence of sub- and mm observations, \cloutmax\,=\,8 is still consistent with the 70 and 100\,\micron\ and 3.5\,cm  fluxes. Only additional flux observations at mm can better constrain wind region 4. \\

\noindent\textit{HD\,193237 (P\,Cyg)}

Discussed in detail in the main text. \\ 

\noindent\textit{HD\,24398 ($\zeta$\,Per)}

Discussed in detail in the main text. \\




\noindent\textit{HD\,169454}

This source is one of the four eBHGs in our sample. The available radio fluxes at 2 and 6\,cm  are well determined, and thus \Mdotmax\ is definite. In Figure\,\ref{figBSupergiants1} we present fit-solutions for the fixed- (solid line) and the adapted-regions (magenta-dotted line) approach. For \Mdotmax\,=\,10.4\,\massloss, the best possible solution in the fixed-regions approach is provided for a weak intermediate clumped wind, \clmid\,=\,2.5, and an unclumped inner and outer wind, \clin\,=\,\clout\,=\,\clfar\,=\,1 (Table\,\ref{tableclfactors}). However, a slightly better solution is obtained in the adapted-regions approach, by increasing the clumping degree only in the intermediate wind region, \clmid\,=\,4.5, provided it begins at \rmid\,=\,4 \Rstar\ (Table\,\ref{tableclbestfit}). In all simulations, the single mm flux estimate (at 1.3\,mm) constrains the maximum clumping degree in Region\,4, still consistent with all data, \cloutmax\,=\,2.5 \\

\noindent\textit{HD\,152236 ($\zeta^1$\,Sco)}

This star is another of the four eBHGs in our sample. For the well determined radio measurements, we obtain \Mdotmax\,=\,6.2\,\massloss. Figure\,\ref{figBSupergiants2} displays the derived solutions for the fixed- (solid line) and the adapted-regions approach (magenta-dotted line). For the fixed-regions approach, we obtain a weak clumping degree in the intermediate wind region, \clmid\,=\,2.8, and \clin\,=\,\clout\,=\,\clfar\,=\,1 (Table\,\ref{tableclfactors}). However, a slightly better solution is obtained in the adapted-regions approach with a marginally larger clumping factor in the intermediate wind region, provided that it is restricted by \rout\,=\,10\,\Rstar\ 
(Table\,\ref{tableclbestfit}). \\

\noindent\textit{HD\,41117 ($\chi^2$\,Ori)}

All available radio measurements are well determined, and therefore \Mdotmax\ is definite. For \Mdotmax\,=\,1.8\,\massloss, the best-fit solution is achieved in the fixed-regions approach for \clin\,=\,7.3, \clmid\,=\,1.35 and \clout\,=\,\clfar\,=\,1 (solid line in the corresponding panel in the Figure\,\ref{figBSupergiants2}). The lack of observations in the sub- and mm regime lead to \cloutmax\,=\,6 consistent with FIR and radio observations at 2\,cm. \\

\noindent\textit{HD\,194279}

Our measured fluxes at 70 and 100\,\micron\ are lower than theoretical predictions for an homogeneous, unclumped wind, whereas the available, well-determined radio fluxes at 3.5 and 6\,cm are consistent with those. We derived two solutions intending to fit consistently the FIR and/or radio fluxes (corresponding fit-diagram in Figure\,\ref{figBSupergiants2}). The described clumping structures below correspond to the fixed-regions approach, which provides the best-fit solution for this source.    

A first solution (solid line), matching radio observations, is provided for \Mdotmax\,=\,2.12\,\massloss\ and for a constant \clin\,=\,\clmid\,=\, \,1 (first entry Table\,\ref{tableclfactors}). However, it can be observed that although this solution reproduces the radio fluxes, it only hardly matches the flux measurement at 70\,\micron, and overestimates the one at 100\,\micron. 

A second, and better solution (magenta-dotted line), is achieved for a much lower \Mdotmax\,=\,0.505\,\massloss. For this value, the minimum clumping degree at r\,$\lesssim$\,15\,\Rstar\ and a strong one at r\,$\gtrsim$\,15\,\Rstar, perfectly match all observations, i.e. \clin\,=\,\clmid\,=\,1 and \clout\,=\,\clfar\,=\,20 (second entry in Table\,\ref{tableclfactors}).

The second solution fits all flux observations throughout the wind and provides the best-fit for this source. However, if this is the case, HD\,194279 is the only single star and radio thermal emitter in our sample with such a high clumping degree in the outer wind regions (r\,$\gtrsim$15\,\Rstar). On the other hand, if HD\,194279 turned out to be a non-thermal emitter, then this clumping structure would follow the same trend found for the other non-thermal emitters in our sample as HD\,37043 (SB) or CygOB2\#12. Due to the scarcity of flux observations of HD\,194279 at large wavelengths, the high uncertainties derived in the estimated FIR fluxes and the fact that the two radio measurements are consistent with thermal emission, we consider the constant clumping degree solution to be more likely (first entry in Table\,\ref{tableclfactors}). Only re-observing at FIR and radio wavelengths can help discriminating between these scenarios.\\

\noindent\textit{HD\,198478} 

\begin{table*}[htp]
\begin{center}
\begin{tabular}{c|c|c|c|c|c|c|c|c|c c|c|c|c}\hline\hline
\teff	&\!log $g$\!	&\vinf	&$\beta$	& \xhe		&\Rstar	& \Mdotmax 	& \clin\	& \clmid	& \clout 	& \cloutmax	& \clfar	& log $Q'$	& \!Sol.\!	\\
	\hline
17.5\!	& \!2.12\!	& 470	& 1.3		& 0.2	& 35.8	&$\leq$ 0.38	& 40.0		&	6.5		& 1.0		& 10.0	& 1.0		& -8.75		& 1a		\\
		&			&		&			& 0.1	&	      &$\leq$ 0.322 & 40.0	&	6.5		&	1.0		& 10.0	&	1.0		& -8.83		& 1b		\\
		&			& 200	& 1.3		& 0.2	&		  &$\leq$ 0.165 & 45.0	&	6.0		&	1.0 	& 10.0	&	1.0		& -9.11		& 2a		\\			
		&			&		&			& 0.1	&		  &$\leq$ 0.14  & 45.0	&	6.0		&	1.0		& 10.0	&	1.0		& -9.18		& 2b		\\
	\hline			
\end{tabular}
\caption{Minimum clumping structure and \Mdotmax\ of HD\,198474 derived in this work for the interval of stellar parameters provided by \cite{Markova2008} (see text and Figure\,\ref{fighd198478}). \teff\ is in kK; log g is the gravity without centrifugal correction; YHe denotes the helium fraction by number used in the simulations; and \vinf, \Rstar\ and \Mdotmax\ are in units of km\,s$^{-1}$, \Rsun\ and 10$^{-6}$ \Msunyr, respectively. \cloutmax\ gives the maximum possible clumping factor in Region 4 still consistent with the data. log $Q'$ = log (\Mdotmax/\Rstar$^{1.5}$) is in units of \Msunyr and \Rsun, and, for clarity, the last column indicates the label for the solution displayed in Figures \ref{fighd198478} and \ref{figBSupergiants2}. }
\label{tablehd198478}	
\end{center}
\end{table*} 

\begin{figure*}[t]
\begin{center}
\includegraphics[width=9cm]{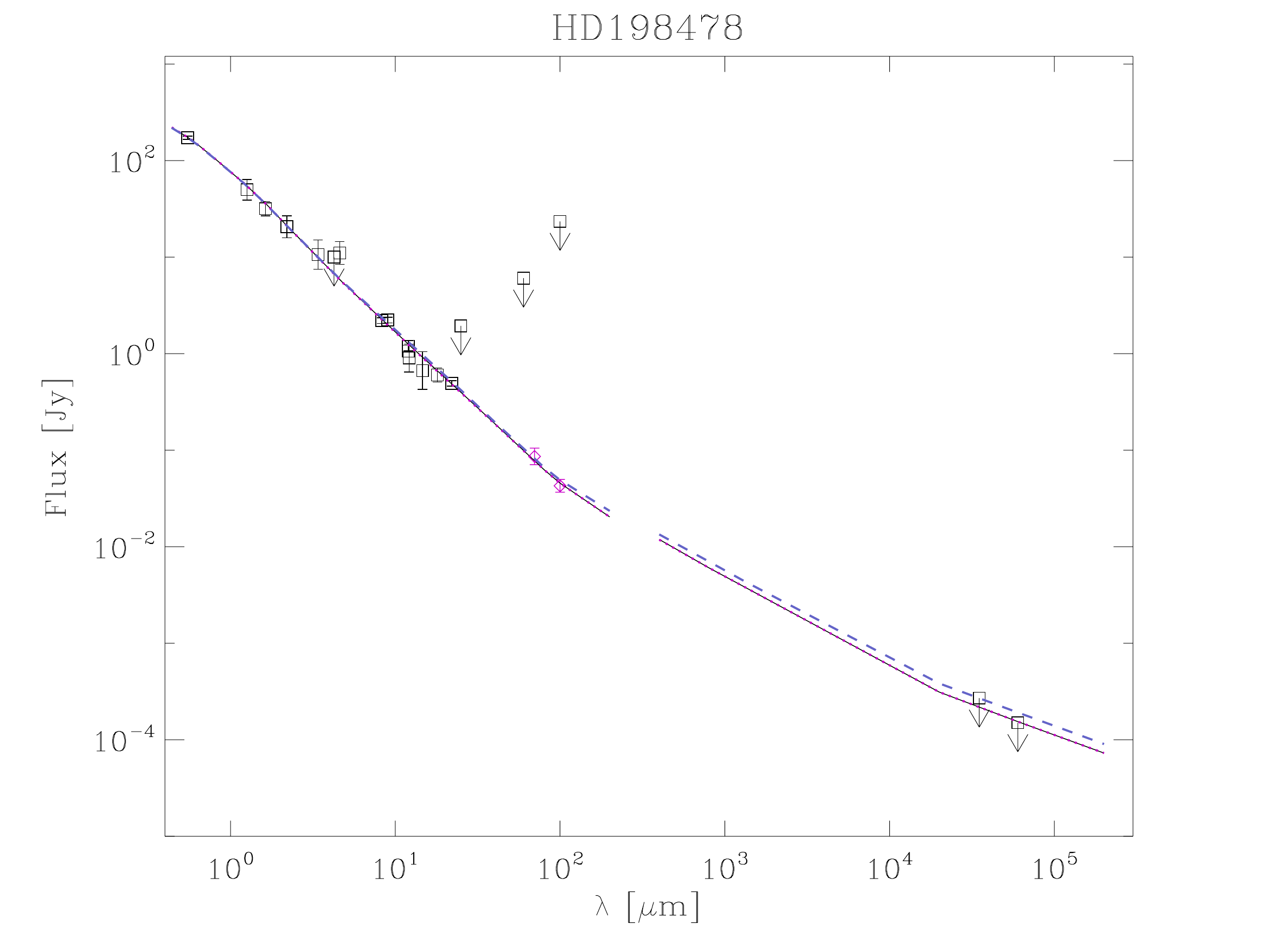}
\includegraphics[width=9cm]{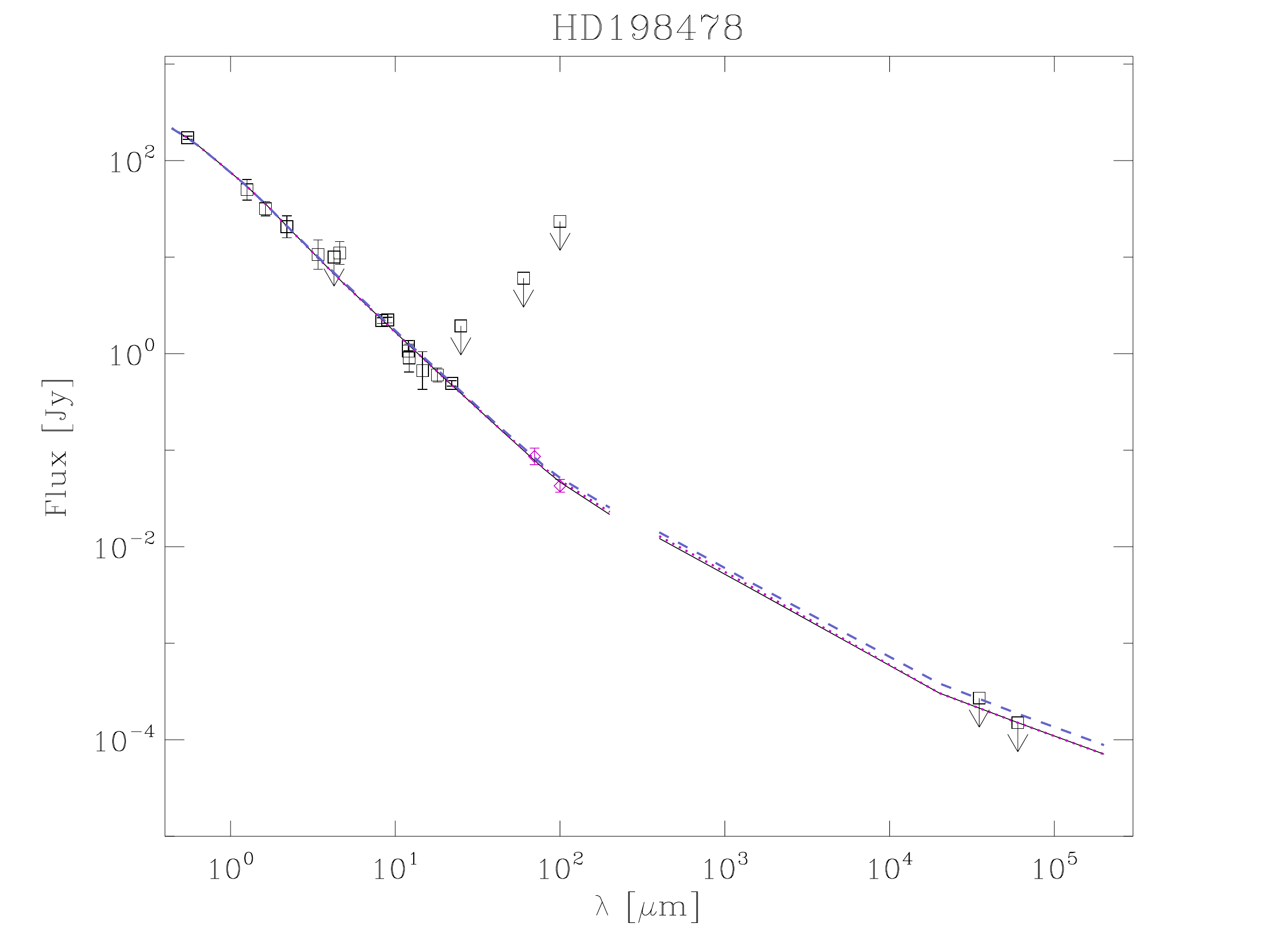}
\caption{Observed and best-fit fluxes vs. wavelength for HD\,198478 depending on the input parameters used in the simulations (see Table\,\ref{tablehd198478} and text in Appendix\,\ref{apxBSg}). Colours and symbols as in Figure\,\ref{hd210839p06}. \textit{Left:} Best-fit models \,470\,kms$^{-1}$ for YHe\,=\,0.2 (solid line; Sol. 1a) and YHe\,=\,0.1 (magenta-dotted line; Sol. 1b). \textit{Right:} Best-fit models for \vinf\,=\,200\,kms$^{-1}$ for YHe\,=\,0.2 (solid line; Sol. 2a) and YHe\,=\,0.1 (magenta-dotted line; Sol. 2b). The blue-dashed line in both plots shows how changes in YHe are reflected in the emission flux model.}
\label{fighd198478}
\end{center}
\end{figure*}

 
Due to its stellar pulsation activity 
this star has variable spectral profiles, so that the stellar and wind parameters obtained via optical spectroscopic analysis show substantial scatter (\citealt{Crowther2006}, \citealt{Markova2008}, \citealt{Kraus2015}). 
This is not due to the different approaches to derive stellar parameters, but to the phase of activity of the source when spectra where taken.

In order to study how different input values affect the estimated \Mdotmax\ and clumping structure, we use the range of input parameters as provided by \cite{Markova2008}. In Table\,\ref{tablehd198478} we present our fit solutions in the fixed-regions approach for each combination of \vinf, $\beta$ and \xhe\ by these authors. The two diagrams in Figure\,\ref{fighd198478} display the corresponding solutions present in Table\,\ref{tableclfactors}. For clarity, in Table\,\ref{tableclfactors} we indicate the full range for derived \Mdotmax\ and clumping factors, and Figure\,\ref{figBSupergiants2} presents fit-solutions corresponding to the lowest (magenta-dotted line) and the largest (solid line) \Mdotmax.

Only two radio flux measurements at 3.5 and 6\,cm  are available (\citealt{Scuderi1998}), and both are upper limits. Therefore, the resulting \Mdotmax\ is an upper limit. The minimum $\chi^2$ obtained for the different solutions in Table\,\ref{tablehd198478} are only marginally different (solid and magenta-dashed lines in Figures \ref{fighd198478} and \ref{figBSupergiants2}). Note that although the absolute values of clumping factors change, the overall clumping properties throughout the entire wind are conserved, and the largest difference in \clin\ and \clmid\ is a factor of 1.33. 

Regarding \Mdotmax, the largest difference, a factor 2.7, arises from simultaneously changing \vinf\ and \xhe\ (first and last entry in Table\,\ref{tablehd198478}). However, for a given velocity field (\vinf\ and $\beta$) and a factor of 2 different helium-fraction, \Mdotmax\ and clumping factors vary only by $\sim$\,15\% and less than 12\%, respectively. Moreover, in all simulations the clumping degree in the outermost wind remained constant, i.e. \clout\,=\,\clfar\,=\,1 for r\,$\gtrsim$\,15\,\Rstar, with \cloutmax\,=\,10 (blue-dashed line in Figure\,\ref{figBSupergiants2}). Therefore, we are confident that the presented \Mdotmax\ ranges and clumping properties are reliable. 

In summary, qualitatively the observed SED for HD\,198478 can be perfectly reproduced with a strong clumping degree in the innermost wind region, moderate clumping in the intermediate one, and the minimum clumping degree in the outermost ones, with a range of mass-loss rates (maybe a reflex of mass-loss episodes due to stellar pulsations). \\

\noindent\textit{HD\,80077}

 The available radio measurements at 3.5 and 6\,cm  for this eBHG are well determined, and thus \Mdotmax\ is definite. The best-fit model is obtained in the fixed-regions approach (solid line in corresponding panel in Figure\,\ref{figBSupergiants2}). For \Mdotmax\,=\,3.45\,\massloss a weak clumped inner and intermediate wind, \clin\,=\,2.5 and \clmid\,=\,1.8, and an unclumped outer one,\clout\,=\,\clfar\,=\,1 (Table\,\ref{tableclfactors}) perfectly match the SED of this source. Due to the lack of observations at sub- and mm-wavelengths we estimated that \cloutmax\,=\,6 is still consistent with the observed 70 and 100\,\micron, and 3.5\,cm fluxes (magenta-dotted line). Only flux observations at mm regime will better constrain this wind region.\\[2pt]\\

\noindent\textit{HD\,53138 ($o^{2}$ CMa)}

There are no mm or radio observations available for this source. Therefore, our analysis only provides an upper limit for the maximum mass-loss rate consistent with the observed fluxes at the largest possible wavelengths (at 70 and 100\,\micron). In the corresponding panel in Figure\,\ref{figBSupergiants2}, we present the solutions for both the fixed- and adapted-regions approaches (solid and magenta-dotted line, respectively).

For \Mdotmax\,=\,1.8\,\massloss, the fixed-regions approach yields \clin\,=\,2, and \clmid\,=\,\clout\,=\,\clfar\,=\,1 (Table\,\ref{tableclfactors}). The best-fit solution, however, is achieved in the adapted-regions approach by slightly increasing the clumping degree in the inner wind region, \clin\,=\,3 and \clmid\,=\,\clout\,=\,\clfar\,=\,1, provided that \rmid\,=\,1.3\,\Rstar\ (Table\,\ref{tableclbestfit}). Due to the lack of observations  of this source at $\lambda > 100\,\micron$, \cloutmax\,=\,18 at r\,$\gtrsim$\,15\,\Rstar, and $f^{\rm far}_{\rm max}$\,=\,50. Only mm and radio observations will definitively constrain the \Mdotmax\ and the clumping structure of the outer wind of this source. \\


\noindent\textit{CygOB2\#12}


This is an exotic eBHG\footnote{Although CygOB2\#12 has been classified as eBHG due to its similarities to other eBHGs, the combination of its extremely high \Lstar\ and low \teff\ cannot be consistently reproduced by any theoretical isochrones (\citealt{Clark2012}). Thus, its evolutionary stage remains unclear.} which, in addition, is suspected to be a non-thermal or variable thermal source; several authors have reported short term variations in the measured radio fluxes (\citealt{Bieging1989}, \citealt{Scuderi1998}), and more recently \cite{Morford2016} detected flux variations of $\sim$\,14 days at 21\,cm. However, the nature of this variability, whether it is due to changes in the state of ionisation of the wind like in P\,Cyg, or by the object being a non-thermal source (maybe a binary), still remains to be unveiled. 

In the corresponding panel in Figure\,\ref{figBSupergiants2}, we present the best possible solutions in the fixed-regions (solid and magenta-dotted lines) and the adapted-regions approaches (blue-dashed and orange-dashed-dotted lines). 
CygOB2\#12 is one of the sources in our sample for which \Mdotmax\ is constrained by infrared fluxes instead of by radio ones. We noted that the mass-loss rate required to match the estimated low flux at 21\,cm, \Mdotmax\,$\sim$\,2.4\,\massloss, overestimates by large the observed fluxes at MIR. From MIR and FIR fluxes, we derived \Mdotmax\,=\,1.02\,\massloss.

For this \Mdotmax, the fixed-regions approach yields \clin\,=\,1, \clmid\,=\,\clout\,=\,10 and \clfar\ ranges from 5 to 15 (Table\,\ref{tableclfactors}) to consistently reproduce the variable radio flux values (solid and magenta-dotted lines, respectively). The best-fit solution, however, is achieved in the adapted-regions approach for a larger clumping degree in the intermediate wind region, \clin\,=\,1, \clmid\,=\,12, \clout\,=\,10 and \clfar\,=\,5\,--\,15 (blue-dashed and orange-dashed-dotted lines, respectively), provided that \rmid\,=\,2.5\,\Rstar\ and \rout\,=\,8\,\Rstar\ (Table\,\ref{tableclbestfit}). In all simulations the flux at 1.3\,mm constrains \cloutmax\,=\,12.\\

\subsection{O\,Giants}\label{OGiants}

\textit{HD\,24912 ($\xi$\,Per)}

\begin{figure}[!ht]
\begin{center}
\includegraphics[width=9cm]{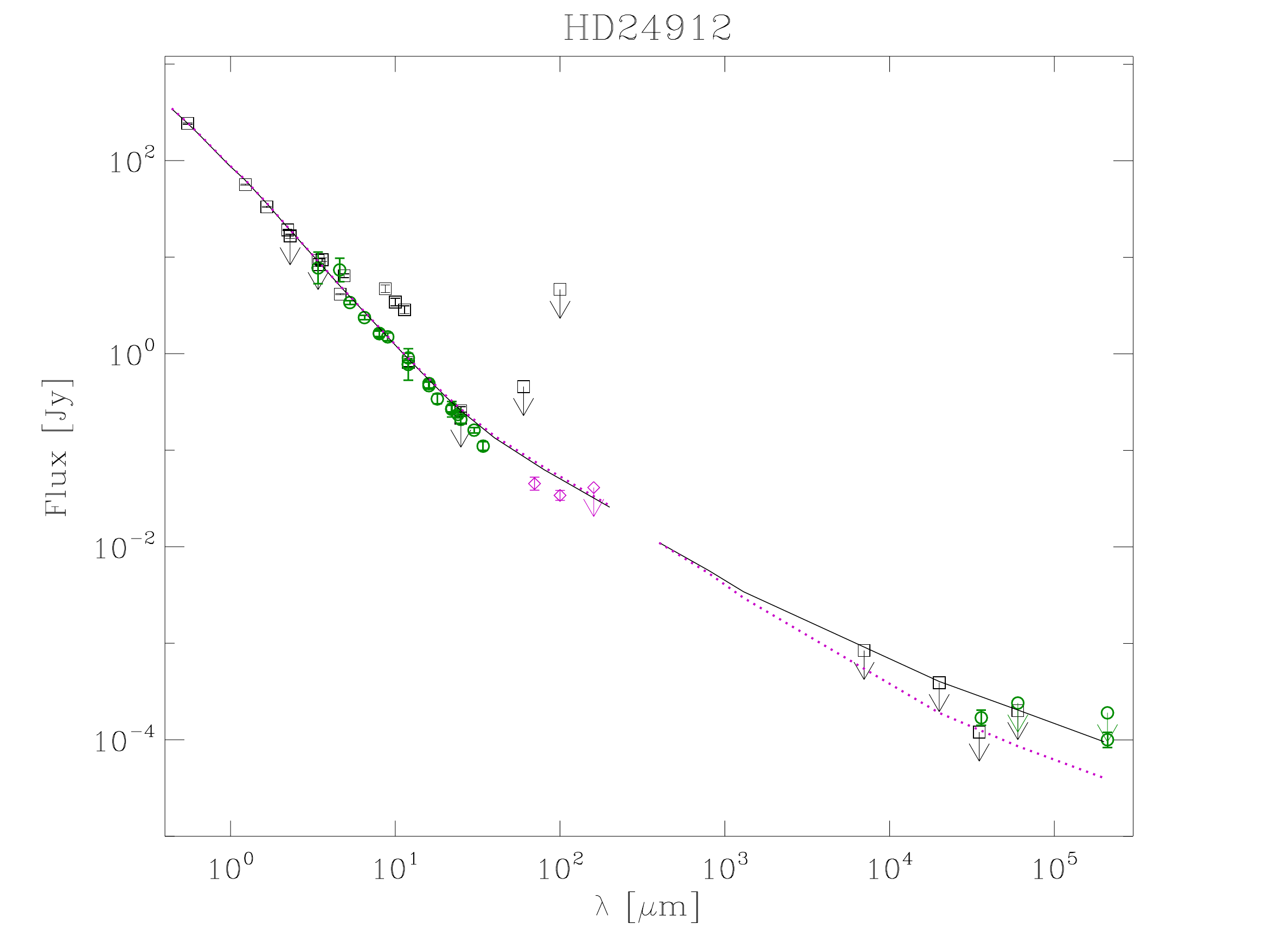}
\caption{Comparison between the best-fit solutions for HD\,24912\,($\xi$\,Per) by \citetalias{Puls2006} (solid and magenta lines, see text) and our flux estimates at 70, 100 and 160\,\micron\ (magenta diamonds). Colours and symbols as in Figure\,\ref{hd210839p06}.}
\label{fighd24912}
\end{center}
\end{figure}

\begin{figure*}[htp]
\begin{center}
\includegraphics[width=9cm]{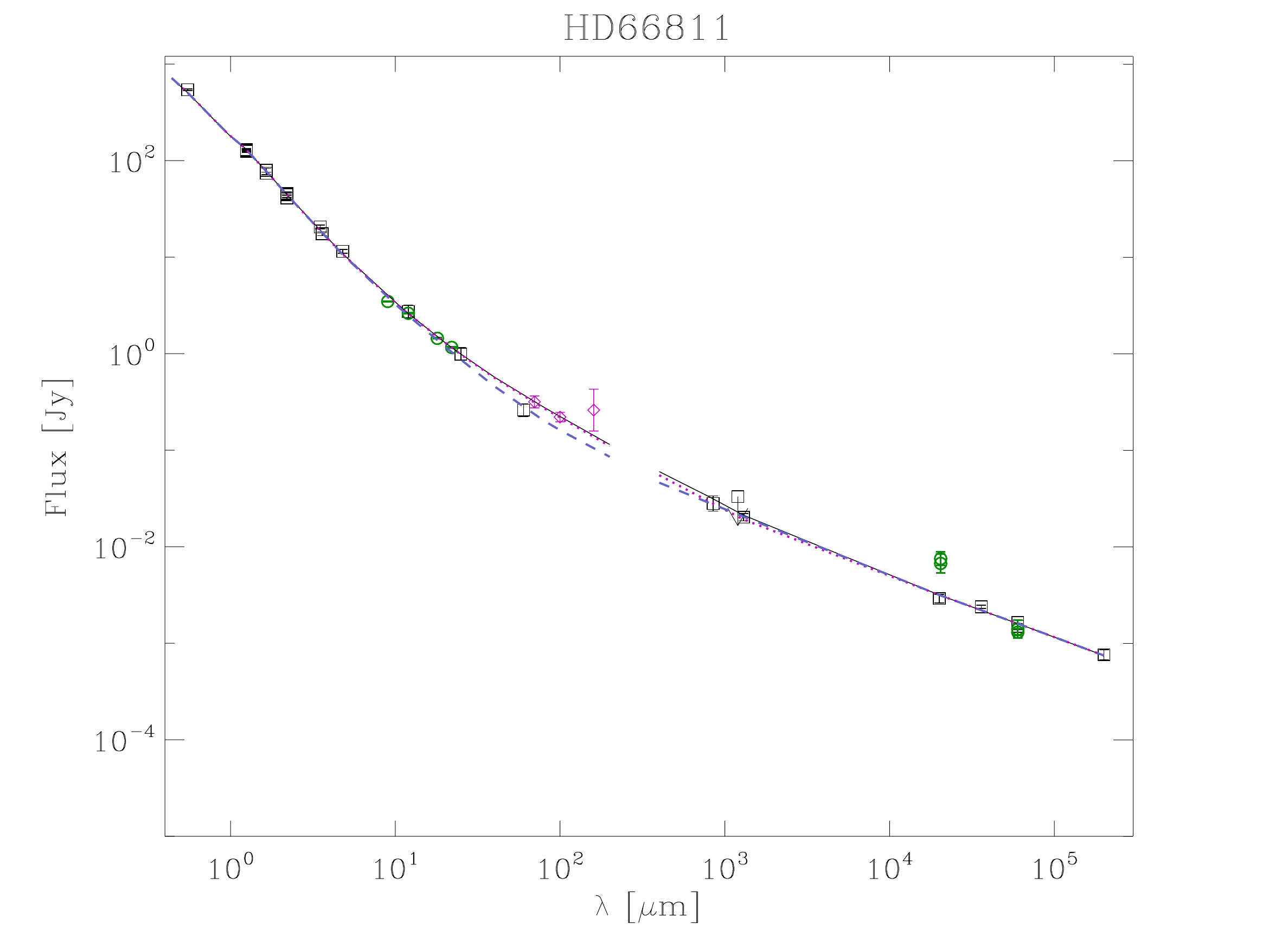}
\includegraphics[width=9cm]{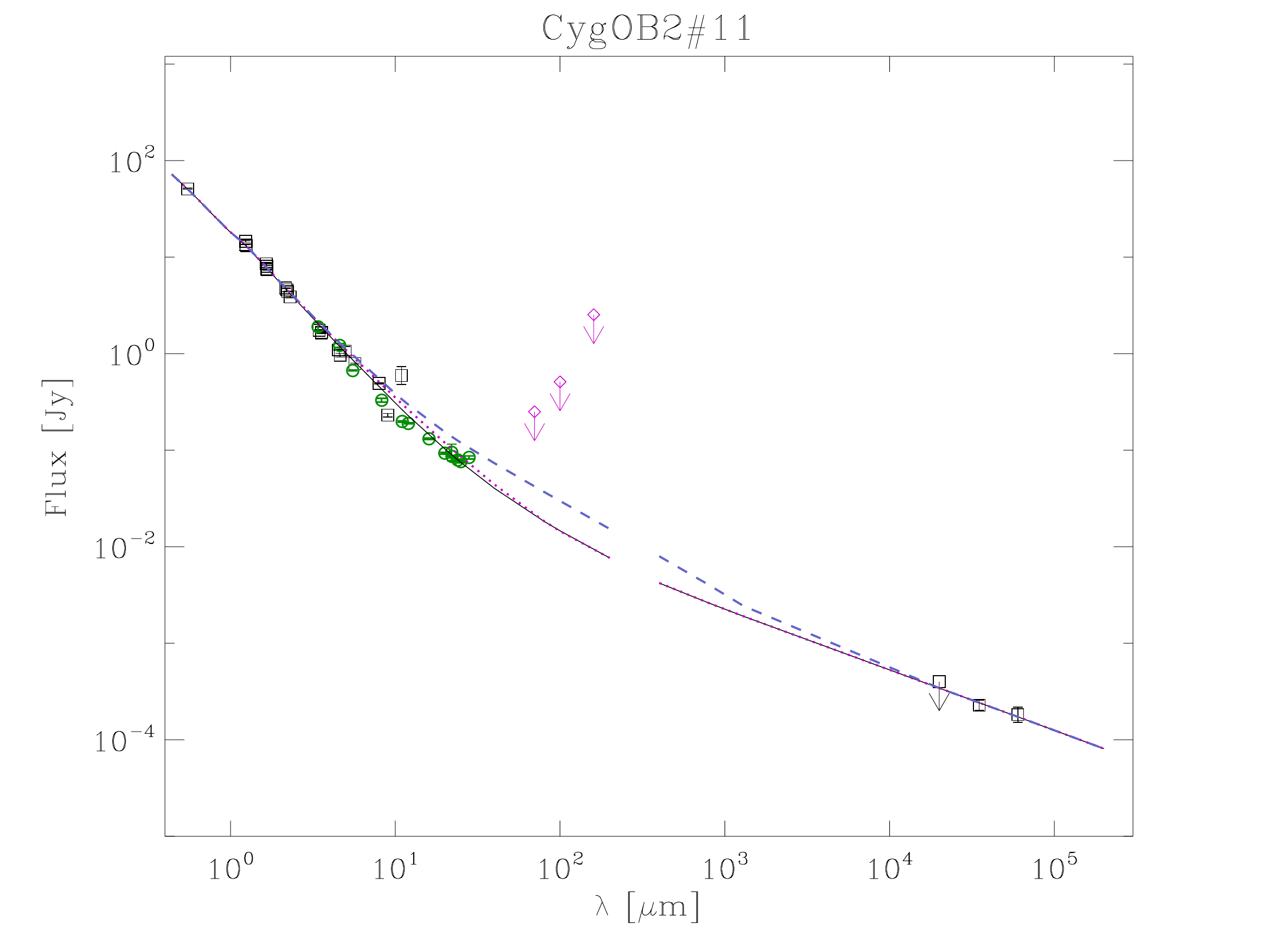}
\includegraphics[width=9cm]{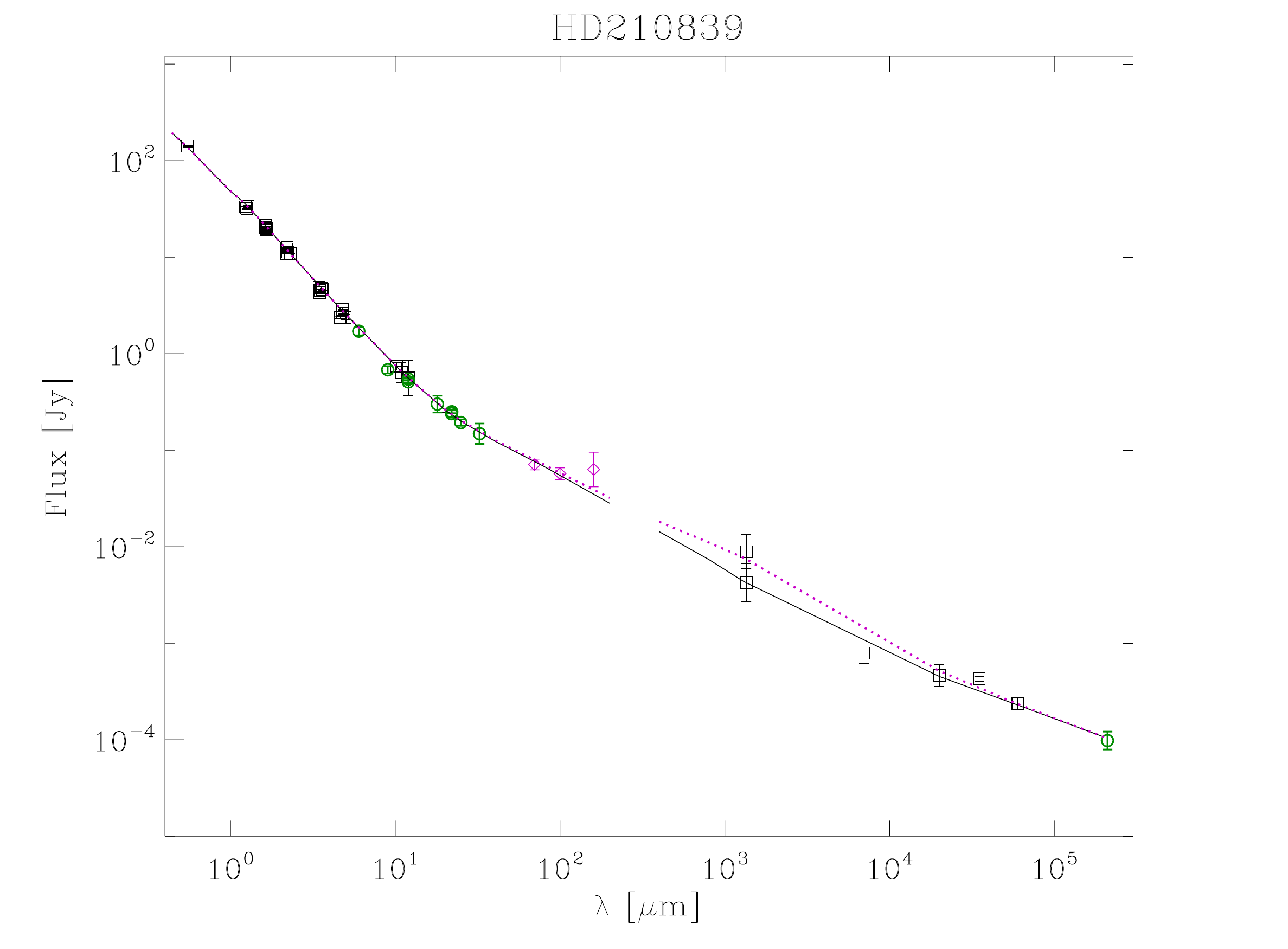}
\includegraphics[width=9cm]{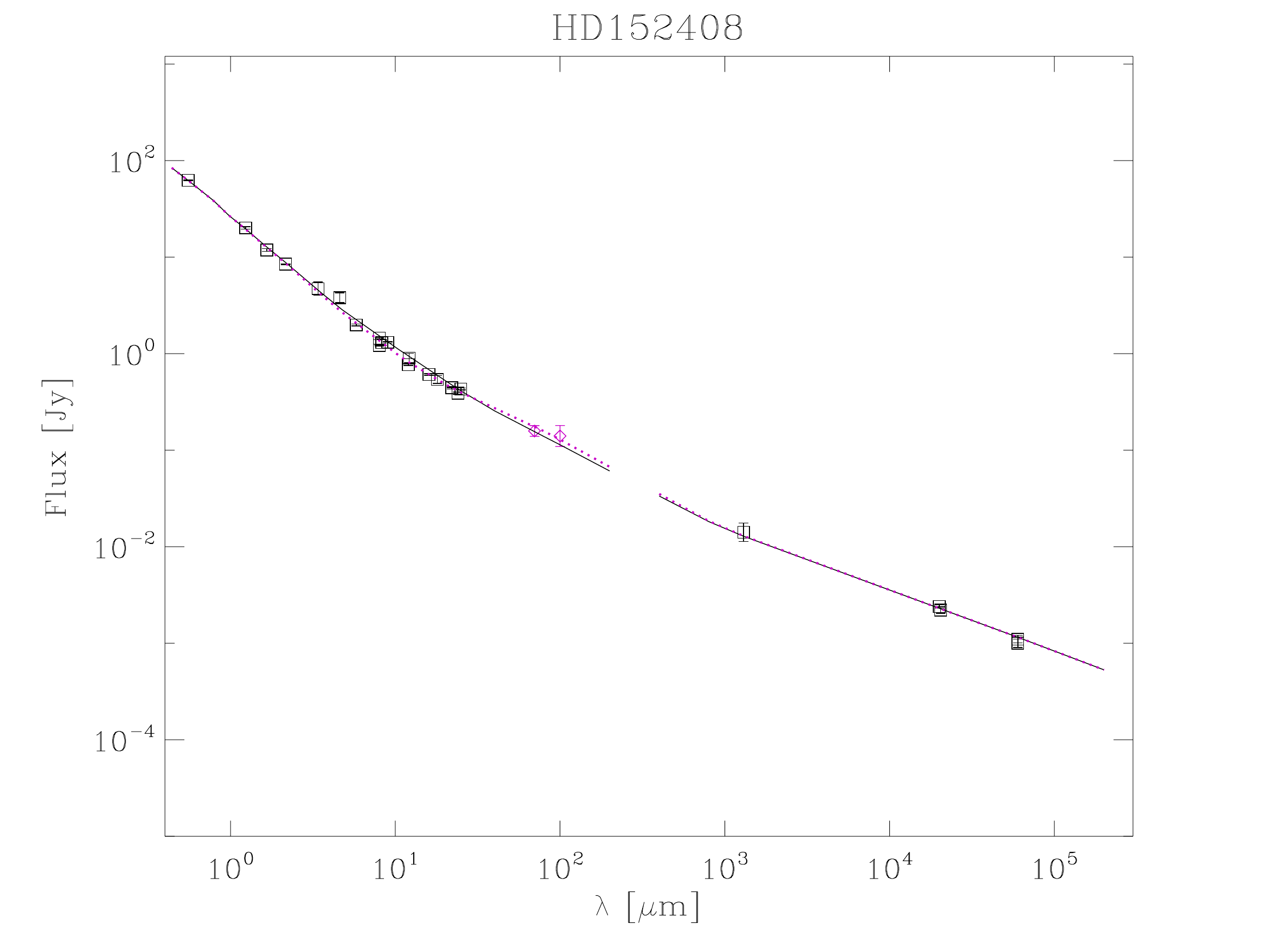}
\includegraphics[width=9cm]{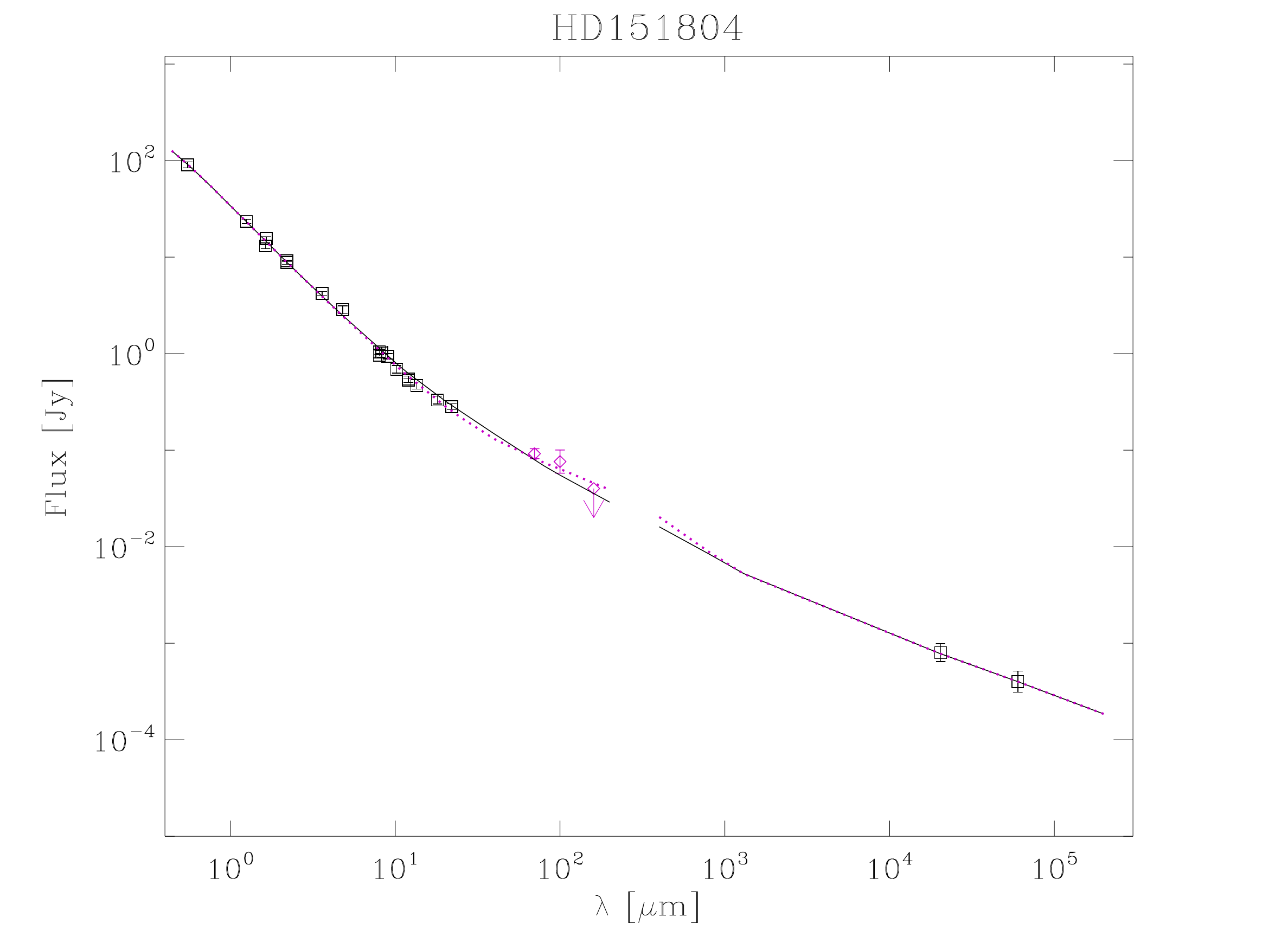}
\includegraphics[width=9cm]{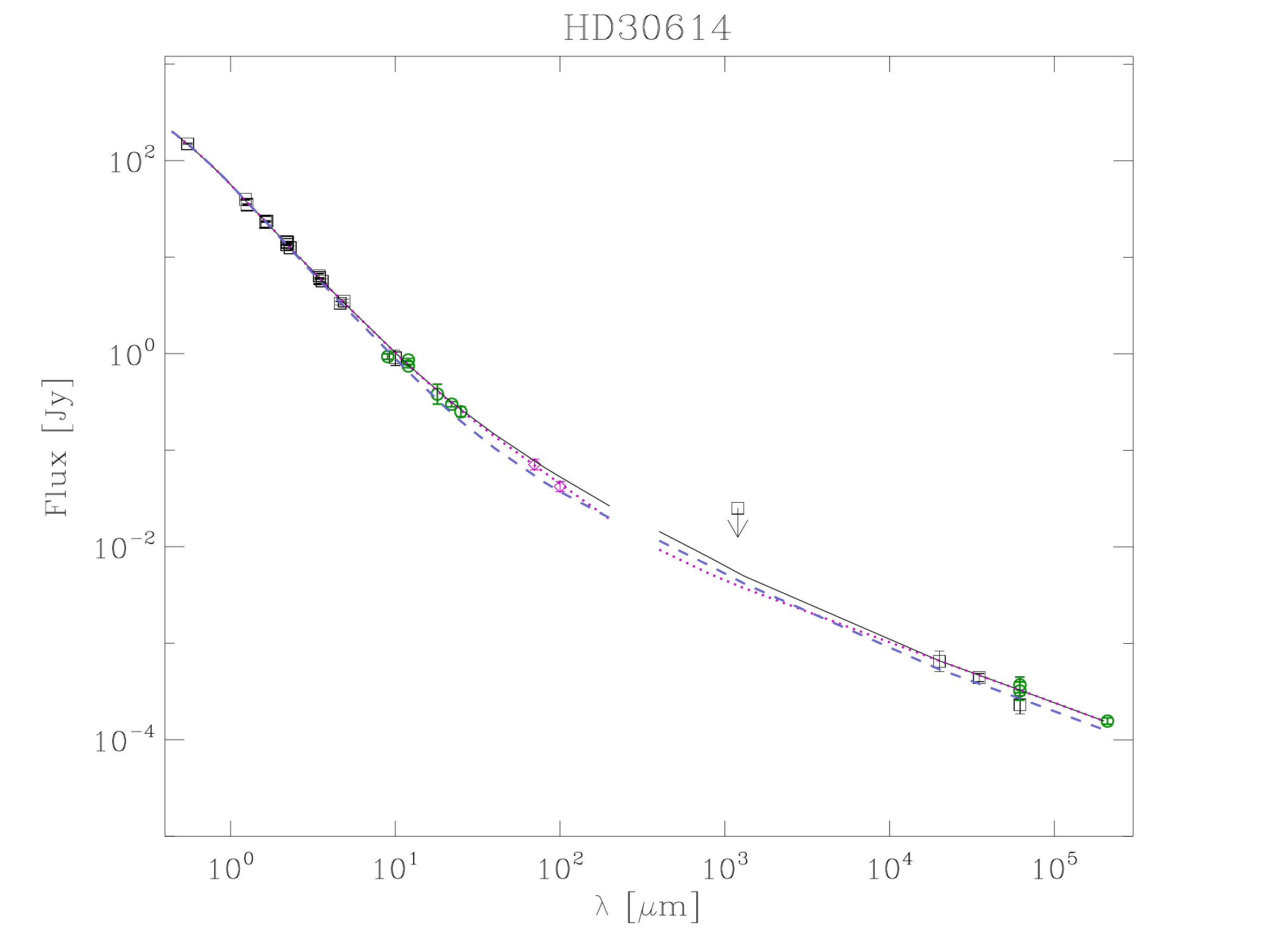}
\caption{Observed and best-fit fluxes vs. wavelength for the O\,Supergiants in our sample. Solid lines represent the best-fit model in the fixed-regions approach derived in this work (see Table\,\ref{tableclfactors} for parameters); magenta-dotted lines are either the best-fit solution (Adapted-regions approach; see Table\,\ref{tableclbestfit} for parameters) or an alternate solution (see comments on individual objects in Section \ref{apxOSg}); and blue-dashed lines correspond to existing previous best-fit models from \citetalias{Puls2006}. Magenta diamonds are our measured FIR fluxes at 70, 100 and 160\,\micron. Black squares and green circles indicate flux values from the literature. For those sources in common with \citetalias{Puls2006}, green circles indicate new available data added to the analysis. Arrows indicate upper limits.}
\label{figOSupergiants}
\end{center}
\end{figure*}

\begin{figure*}[htp]
\begin{center}
\includegraphics[width=9cm]{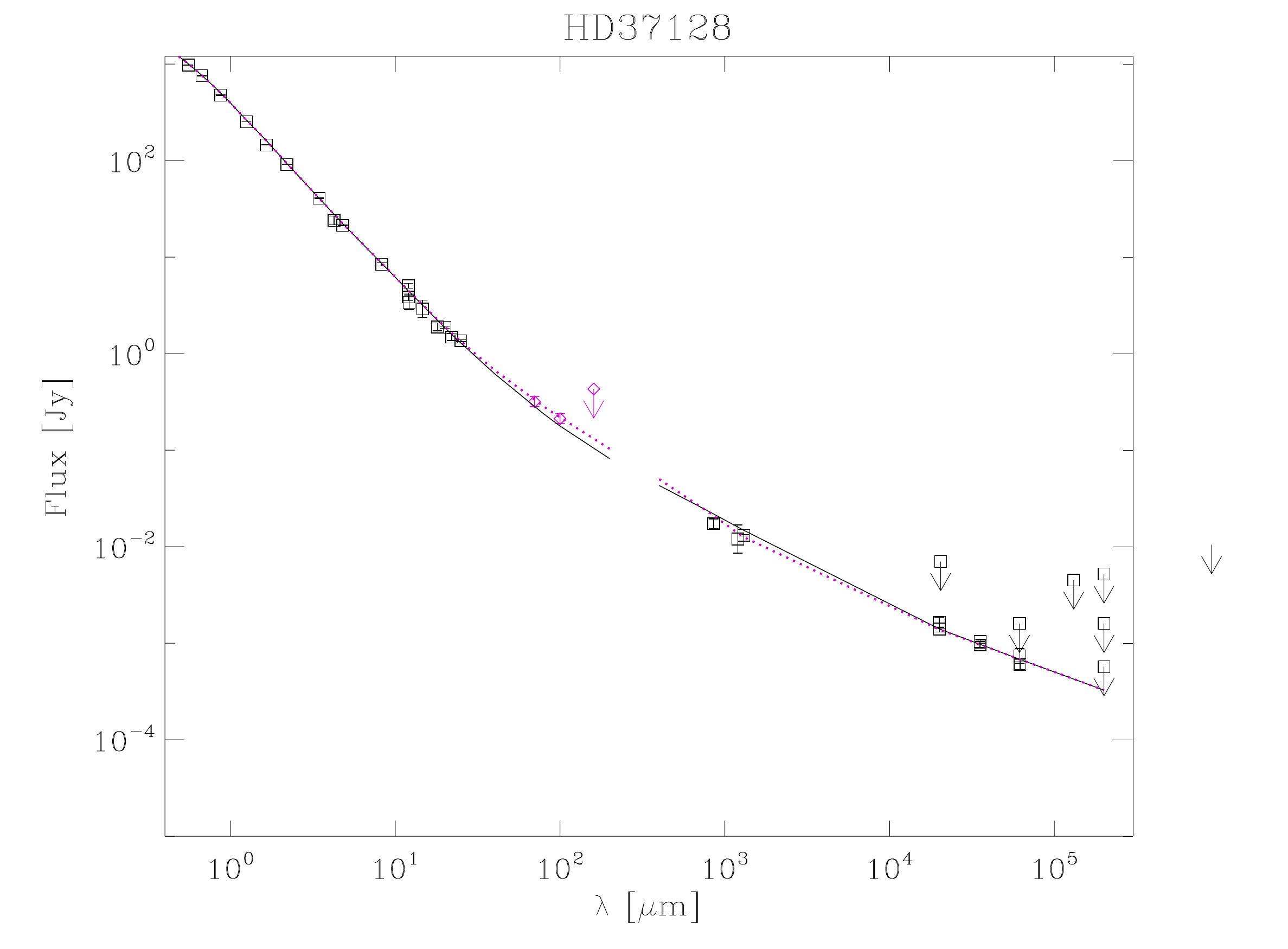}
\includegraphics[width=9cm]{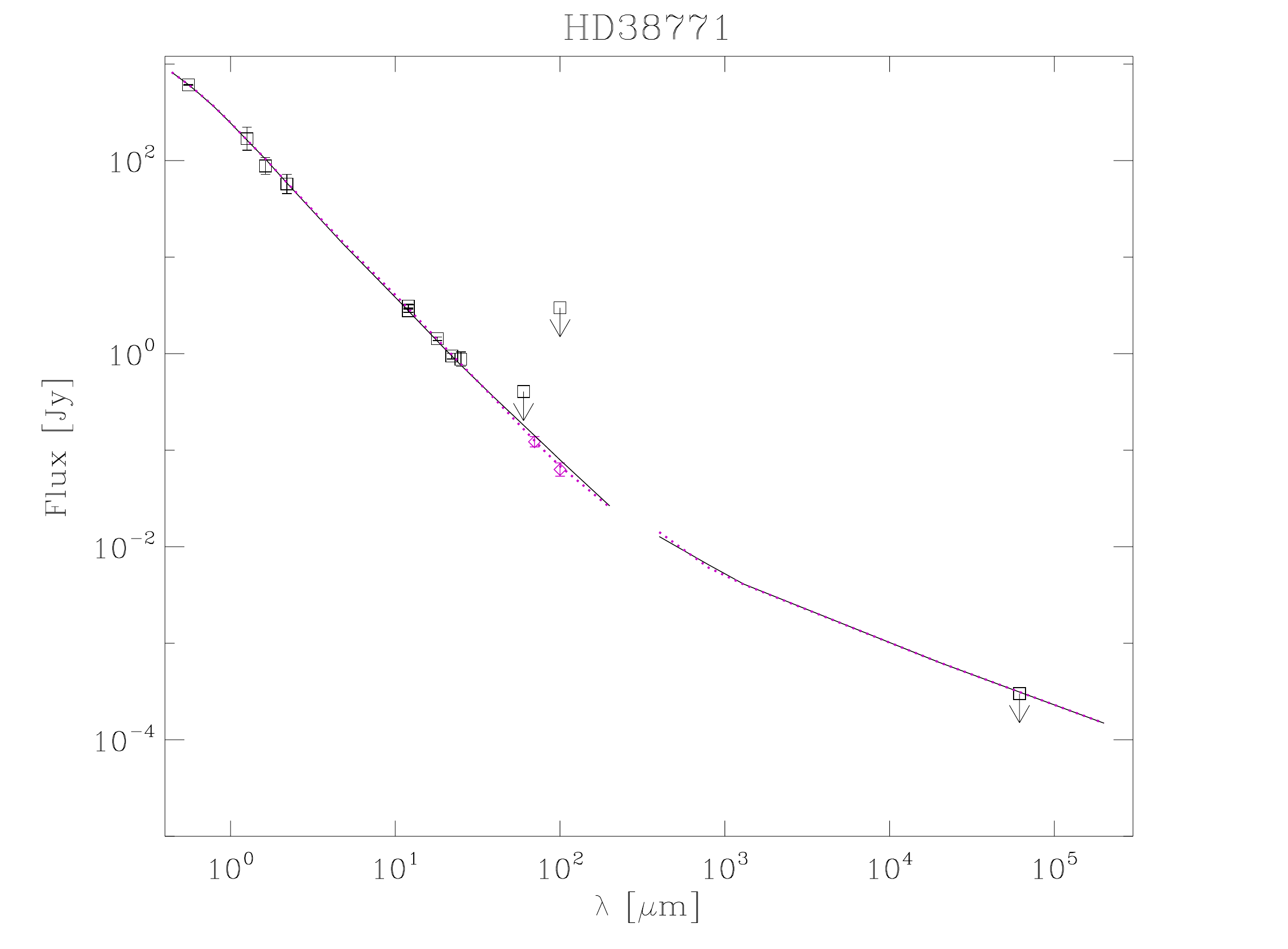}
\includegraphics[width=9cm]{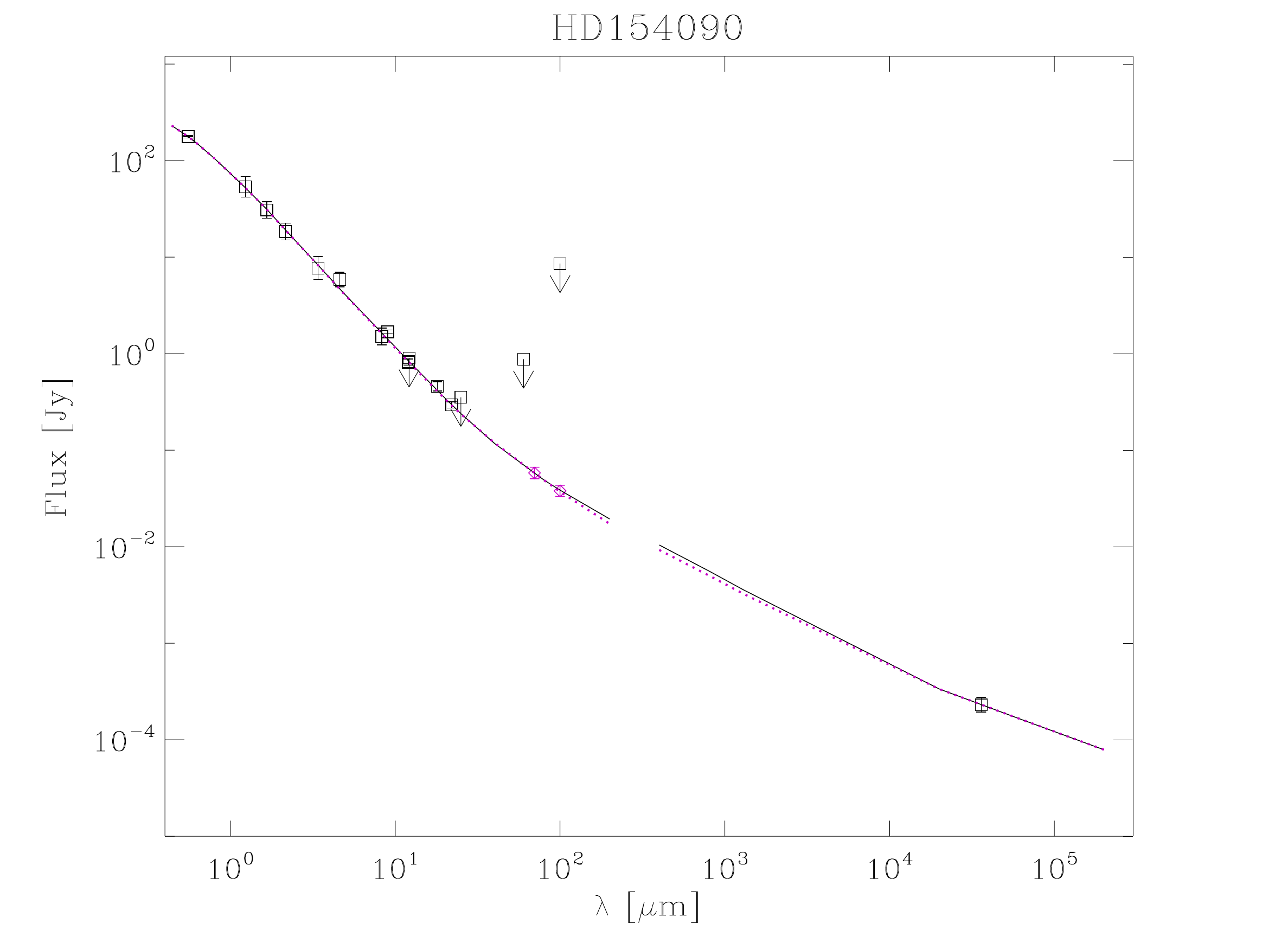}
\includegraphics[width=9cm]{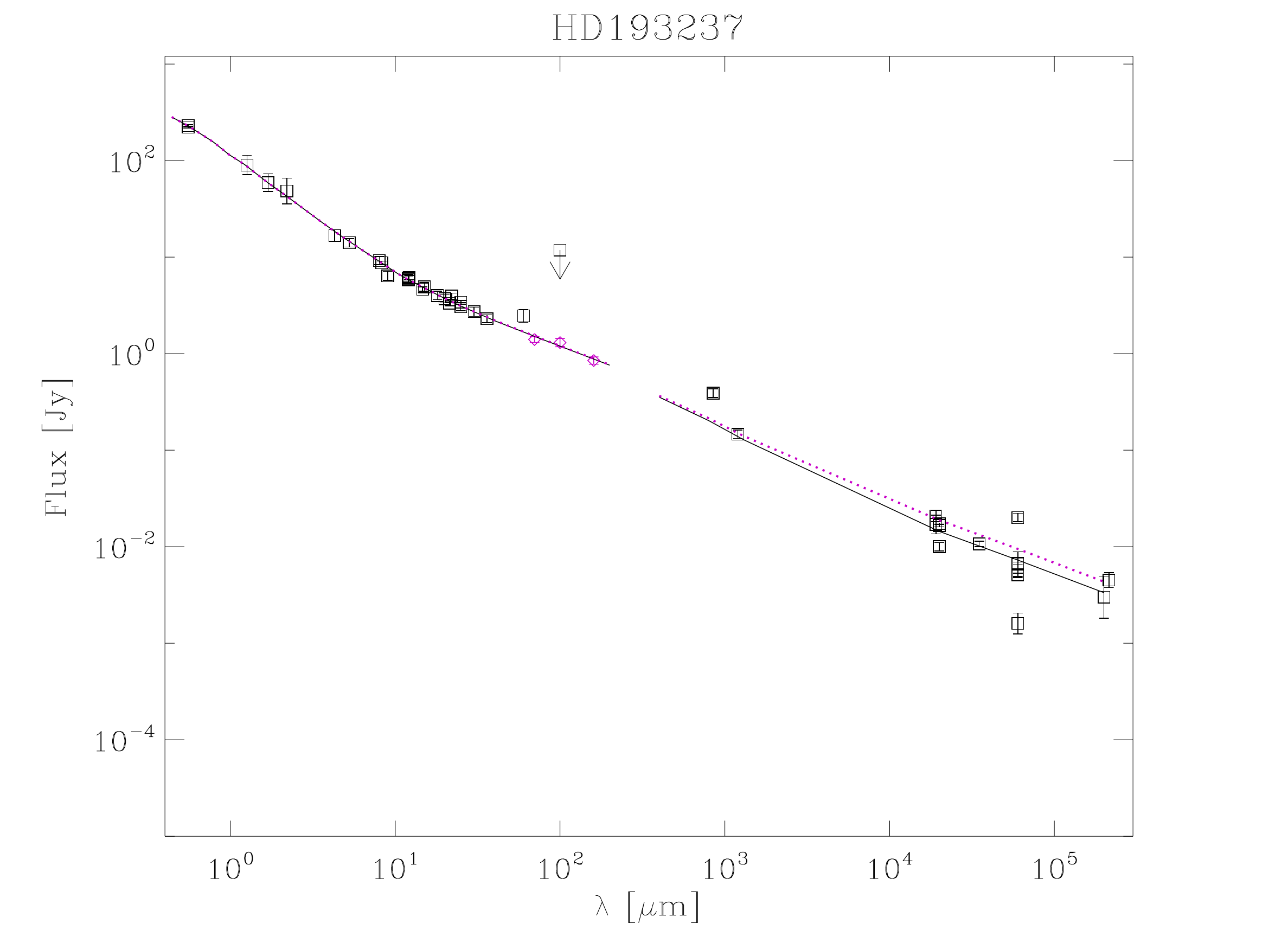}
\includegraphics[width=9cm]{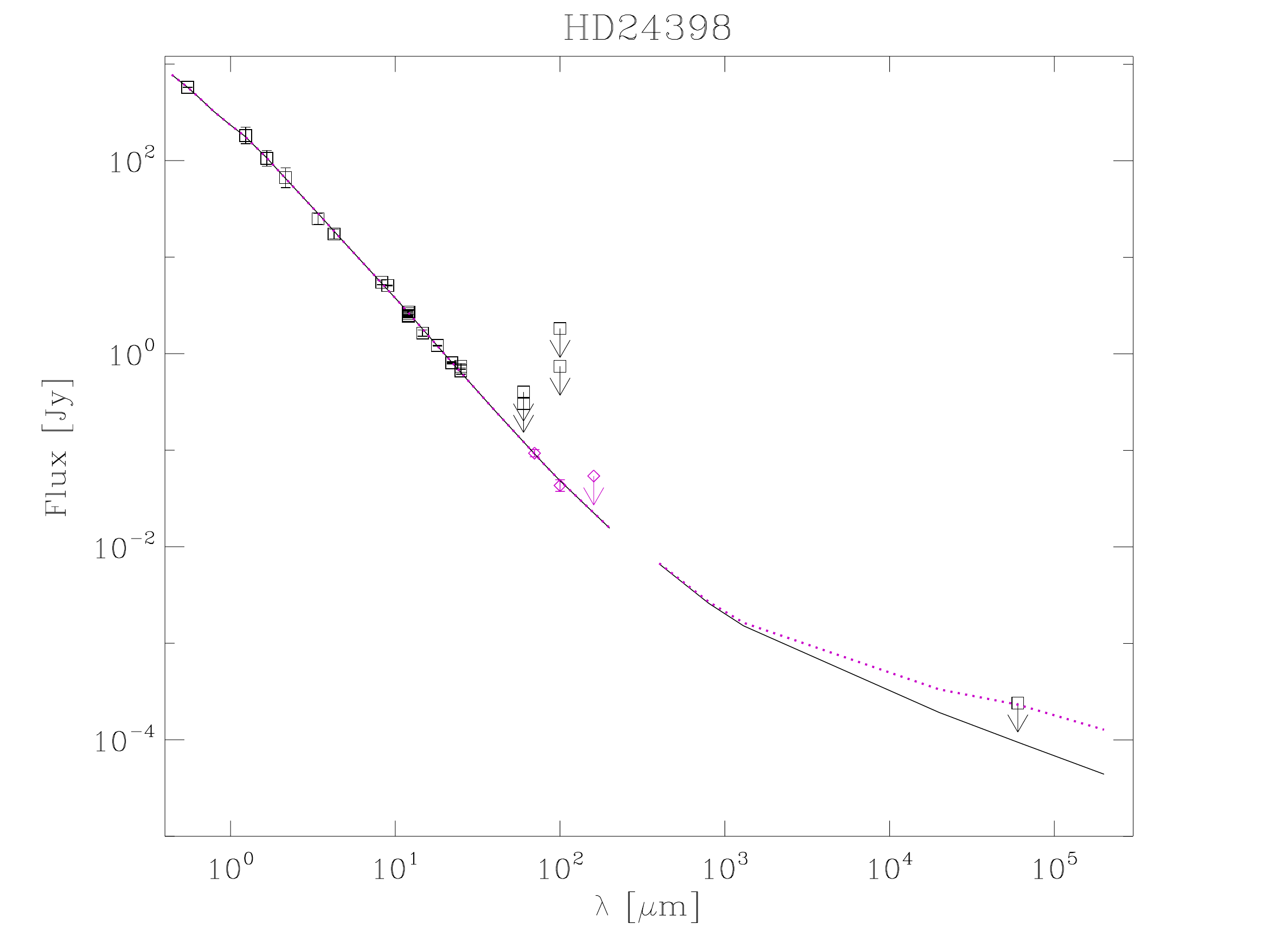}
\includegraphics[width=9cm]{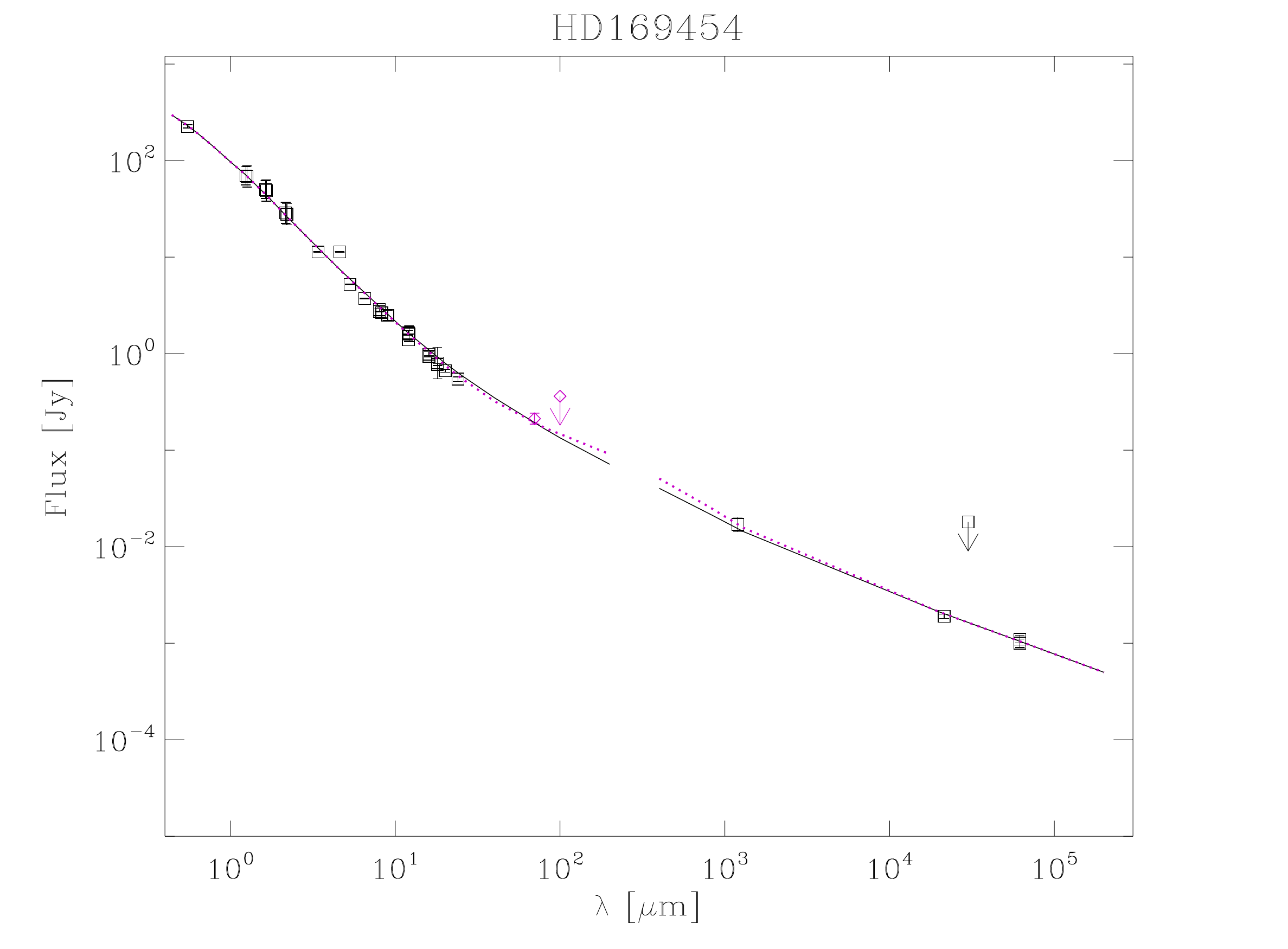}
\caption{Observed and best-fit fluxes vs. wavelength for the B\,Supergiants in our sample. Colours, symbols and line types as in Figure\,\ref{figOSupergiants}. See comments on individual objects in Section \ref{apxBSg}.}
\label{figBSupergiants1}
\end{center}
\end{figure*}

\begin{figure*}[htp]
\begin{center}
\includegraphics[width=9cm]{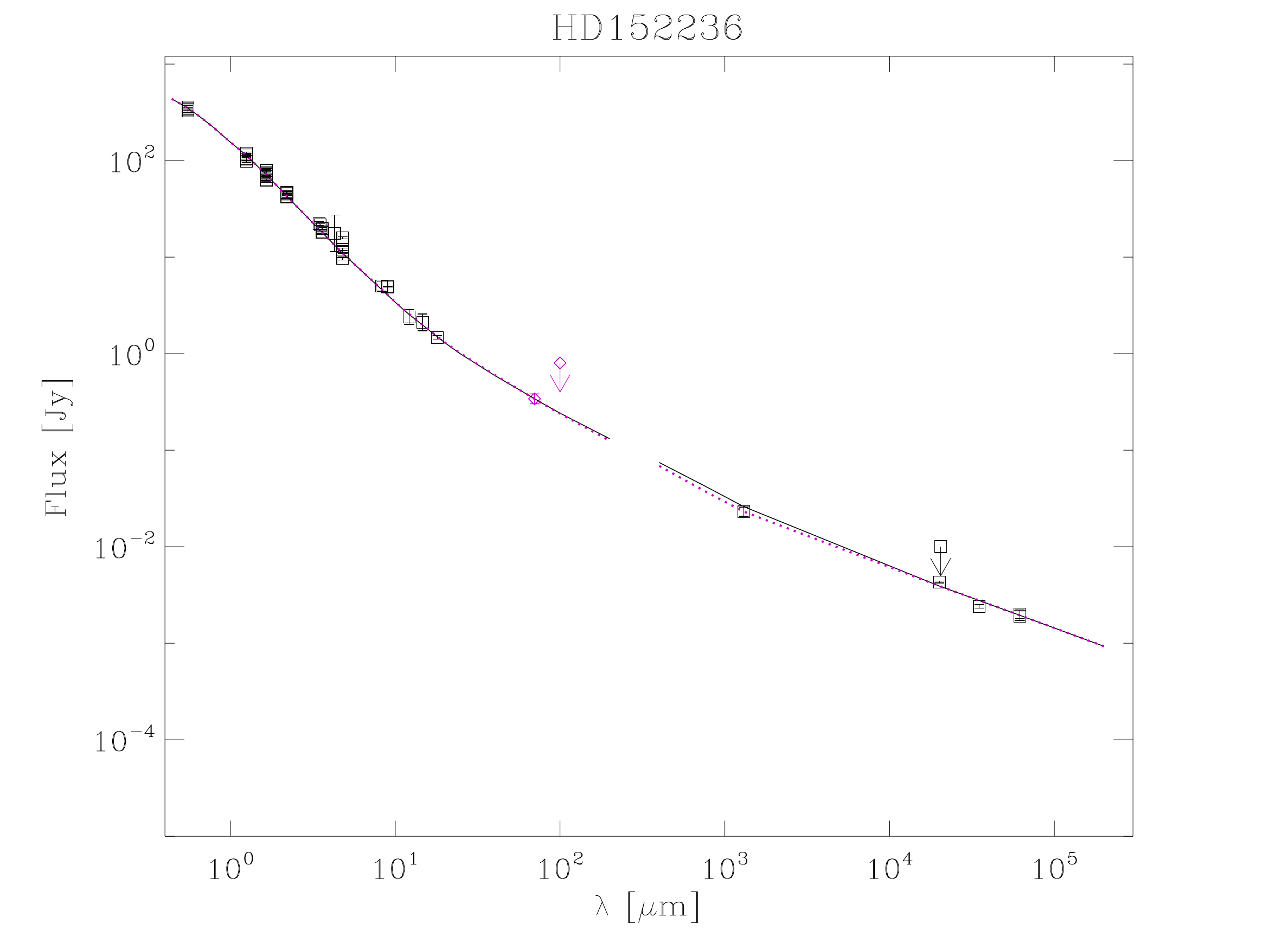}
\includegraphics[width=9cm]{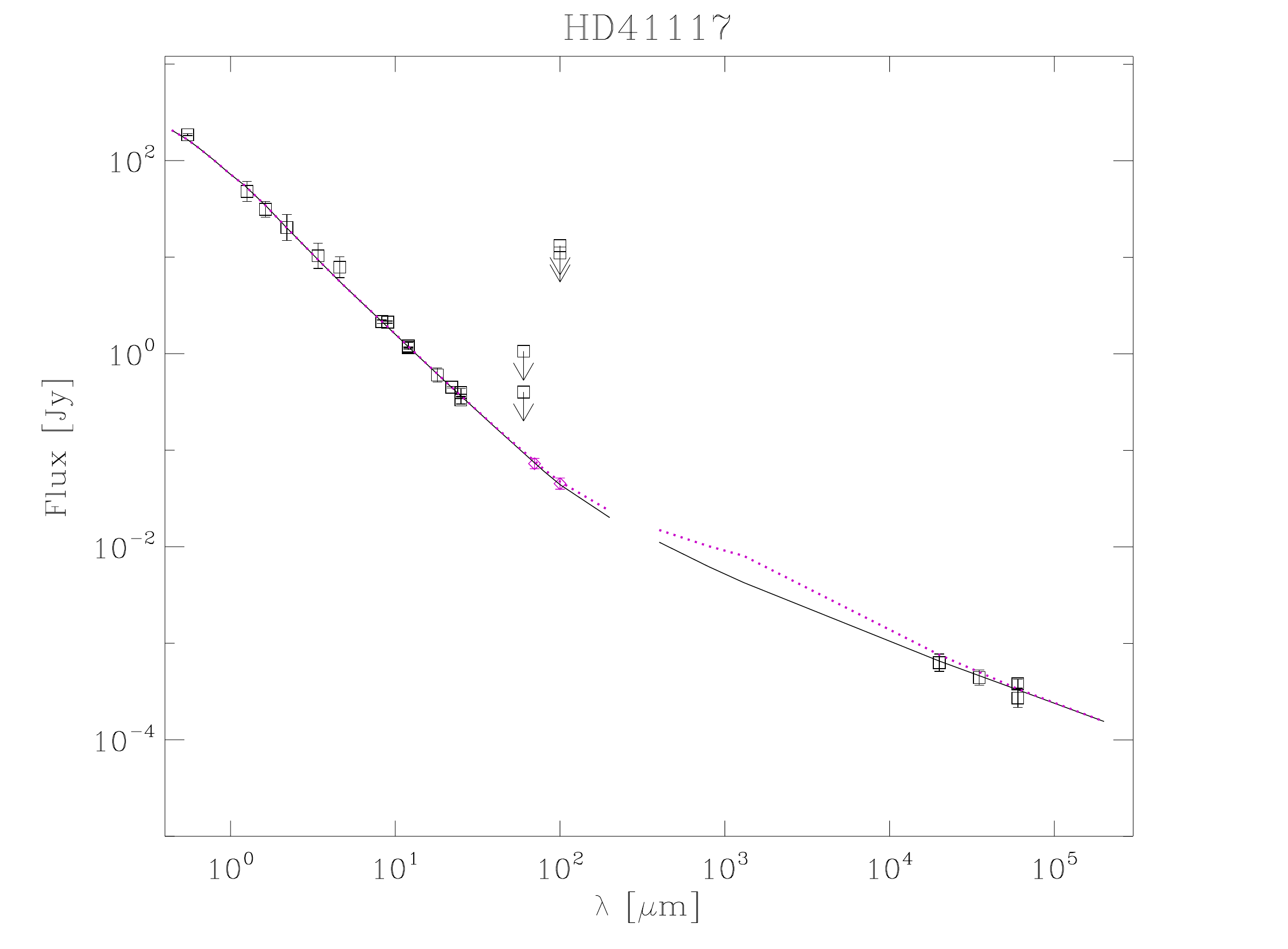}
\includegraphics[width=9cm]{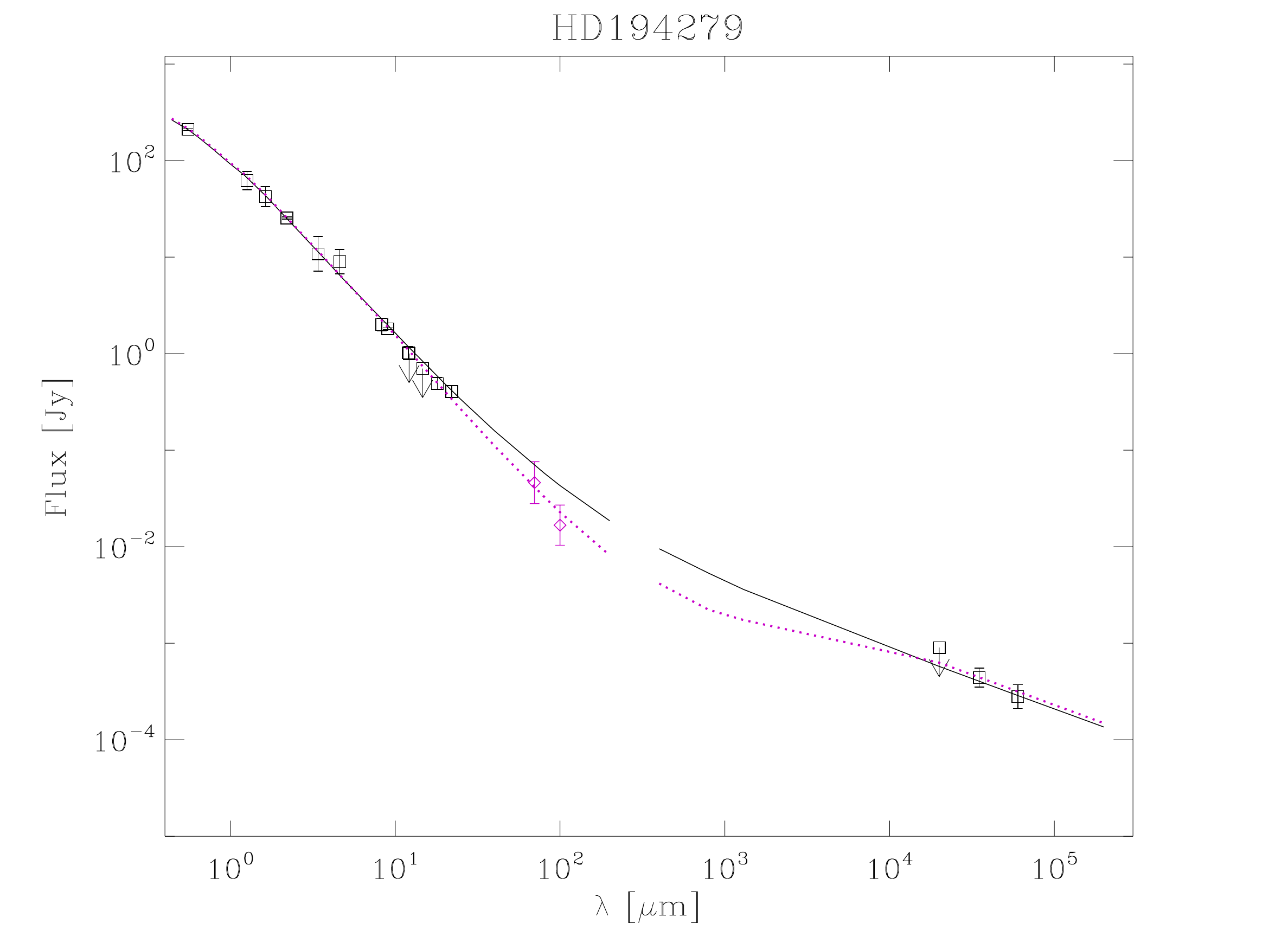}
\includegraphics[width=9cm]{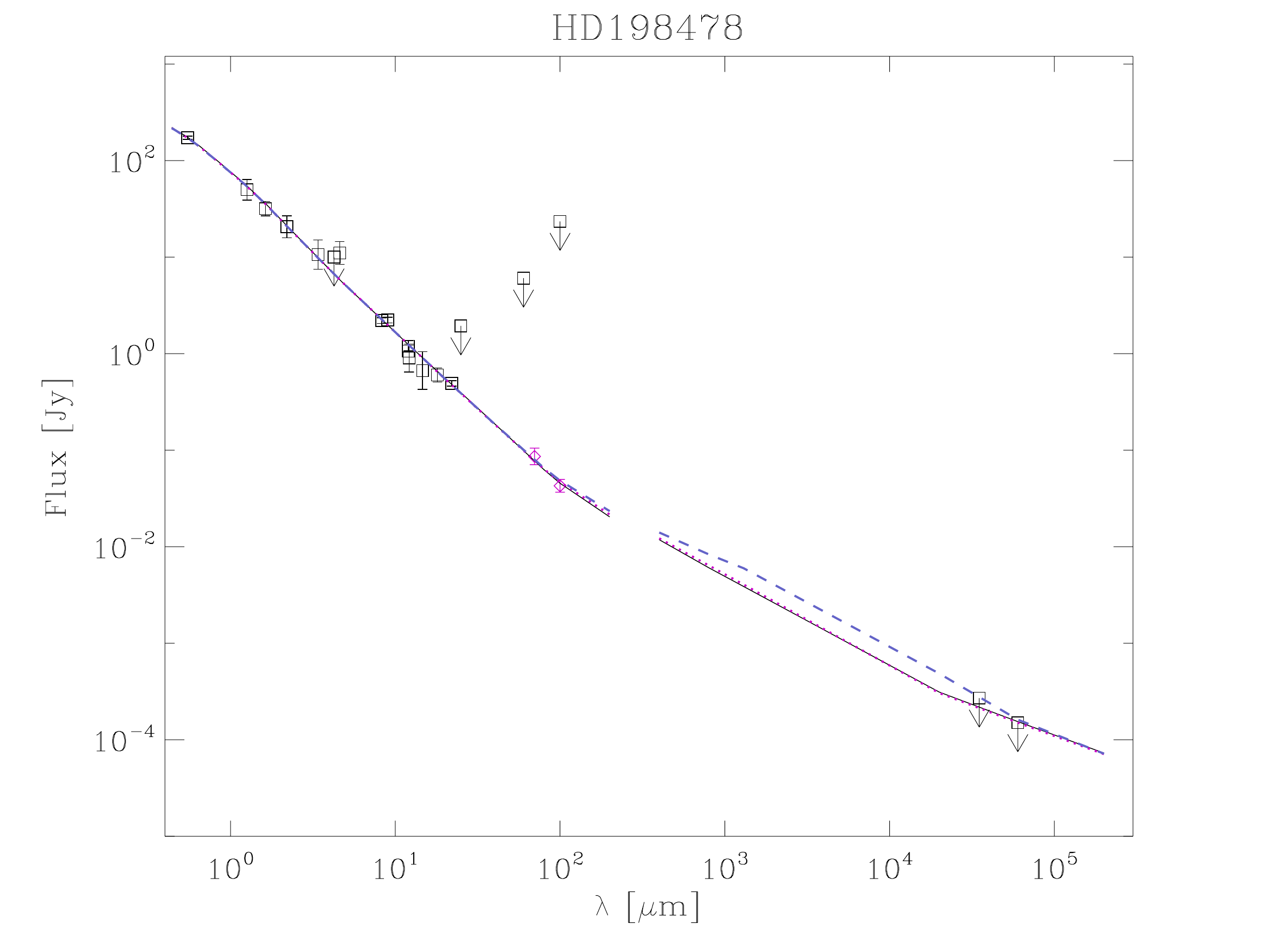}
\includegraphics[width=9cm]{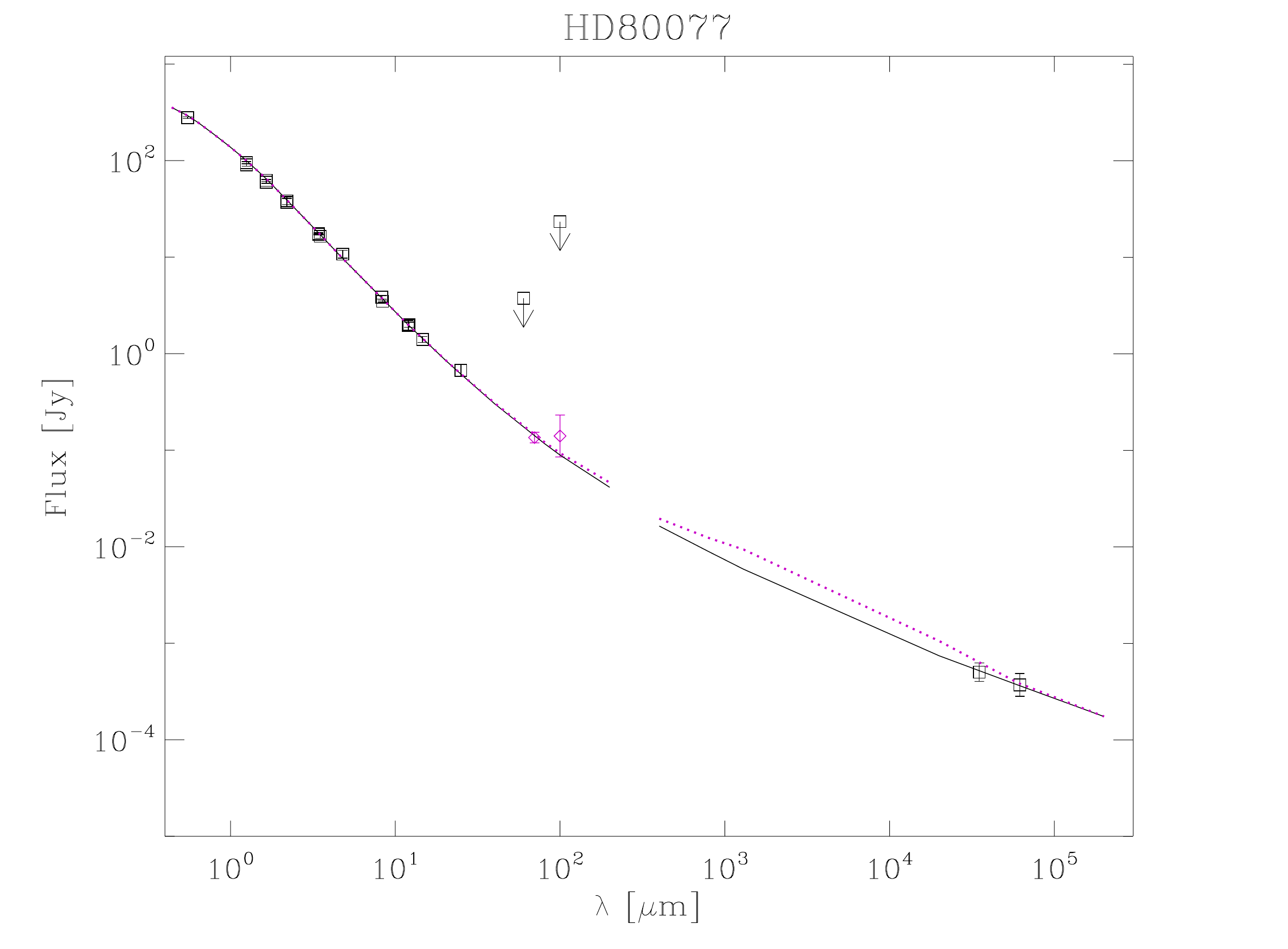}
\includegraphics[width=9cm]{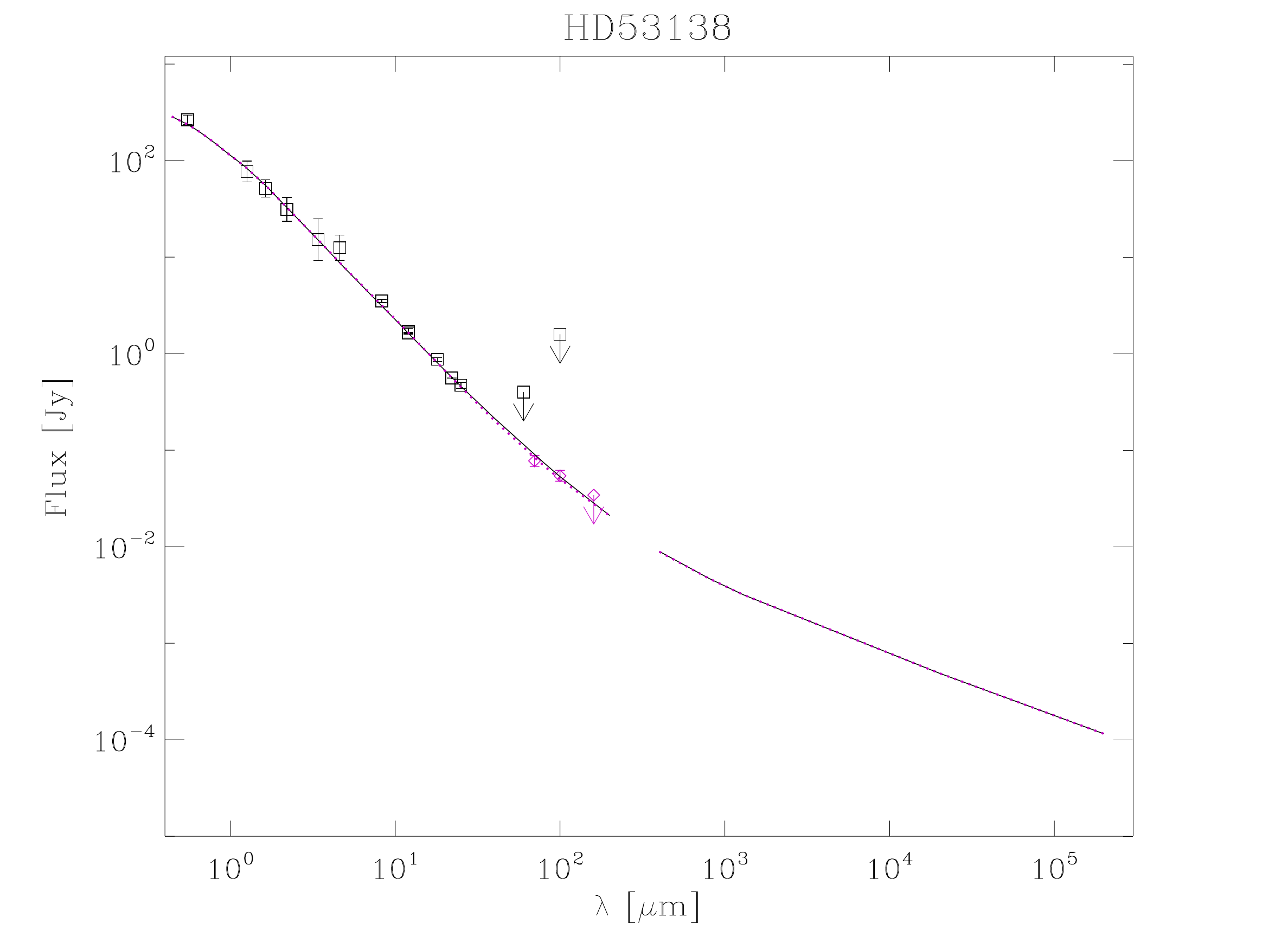}
\includegraphics[width=9cm]{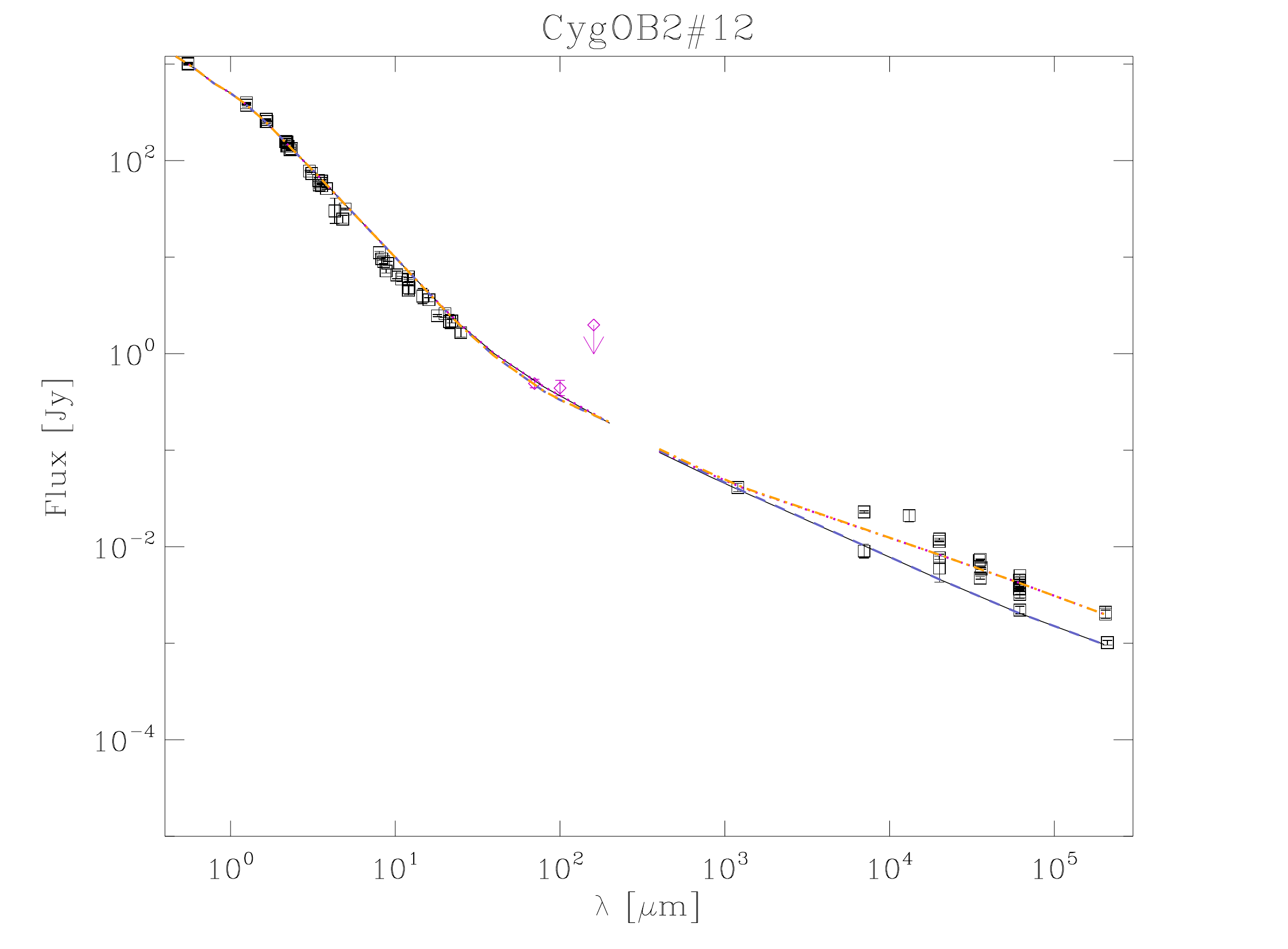}
\caption{As Figure\,\ref{figBSupergiants1}, displaying more B\,Supergiants. See comments on individual objects in Section\,\ref{apxBSg}.}
\label{figBSupergiants2}
\end{center}
\end{figure*}

The clumping stratification of this source was also analysed by \citetalias{Puls2006}. 
These authors derived two possible solutions from fitting the variable upper limits of the radio flux estimates at 0.7, 2, 3.5 and 6\,cm. In Figure\,\ref{fighd24912} we show these best-fit solutions from PO6. The first solution (solid line) is consistent with the larger radio upper flux limits, 
and the second one (magenta-dashed line) with all radio upper flux limits. 
It can be observed that both solutions by \citetalias{Puls2006} overestimate our new flux estimates at 70 and 100\,\micron\ and, moreover, that the additional and well determined fluxes at 3.6\,cm  and 21\,cm  used in our analysis seem to confirm variable radio emission from this source. Note that although these new radio observations are slightly larger (which can be due to either calibration effects or real radio variability of the source), they are also consistent with the upper flux limits previously detected at the radio regime. Moreover, since the radio fluxes estimated for other sources in our sample ($\lambda$\,Cep and $\alpha$\,Cam) by \cite{Schnerr2007} are consistent with the existing radio observations by different authors at different wavelengths, we are also confident in the estimated values provided by the authors for this source. 

\begin{figure*}[htp]
\begin{center}
\includegraphics[width=9cm]{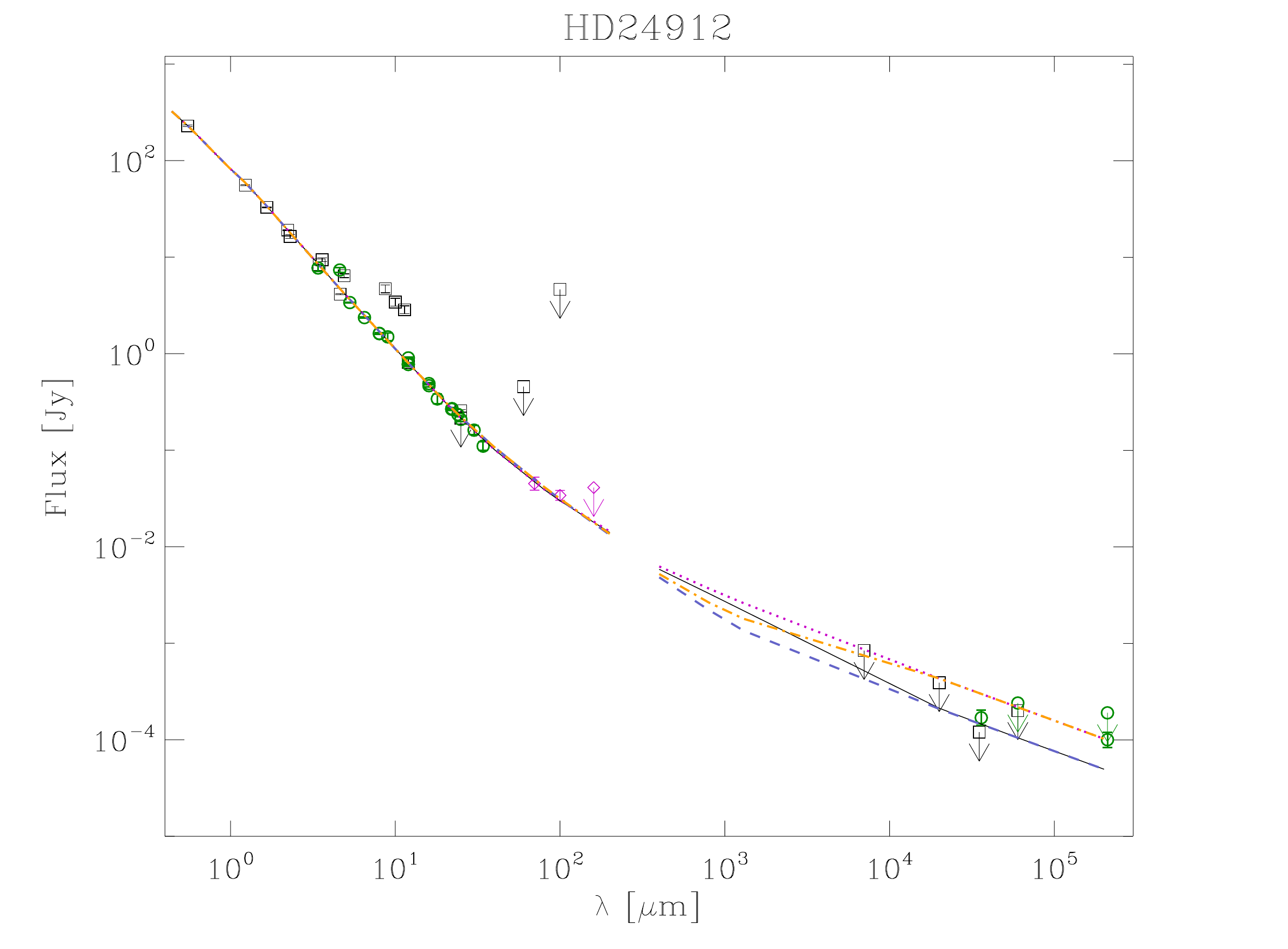}
\includegraphics[width=9cm]{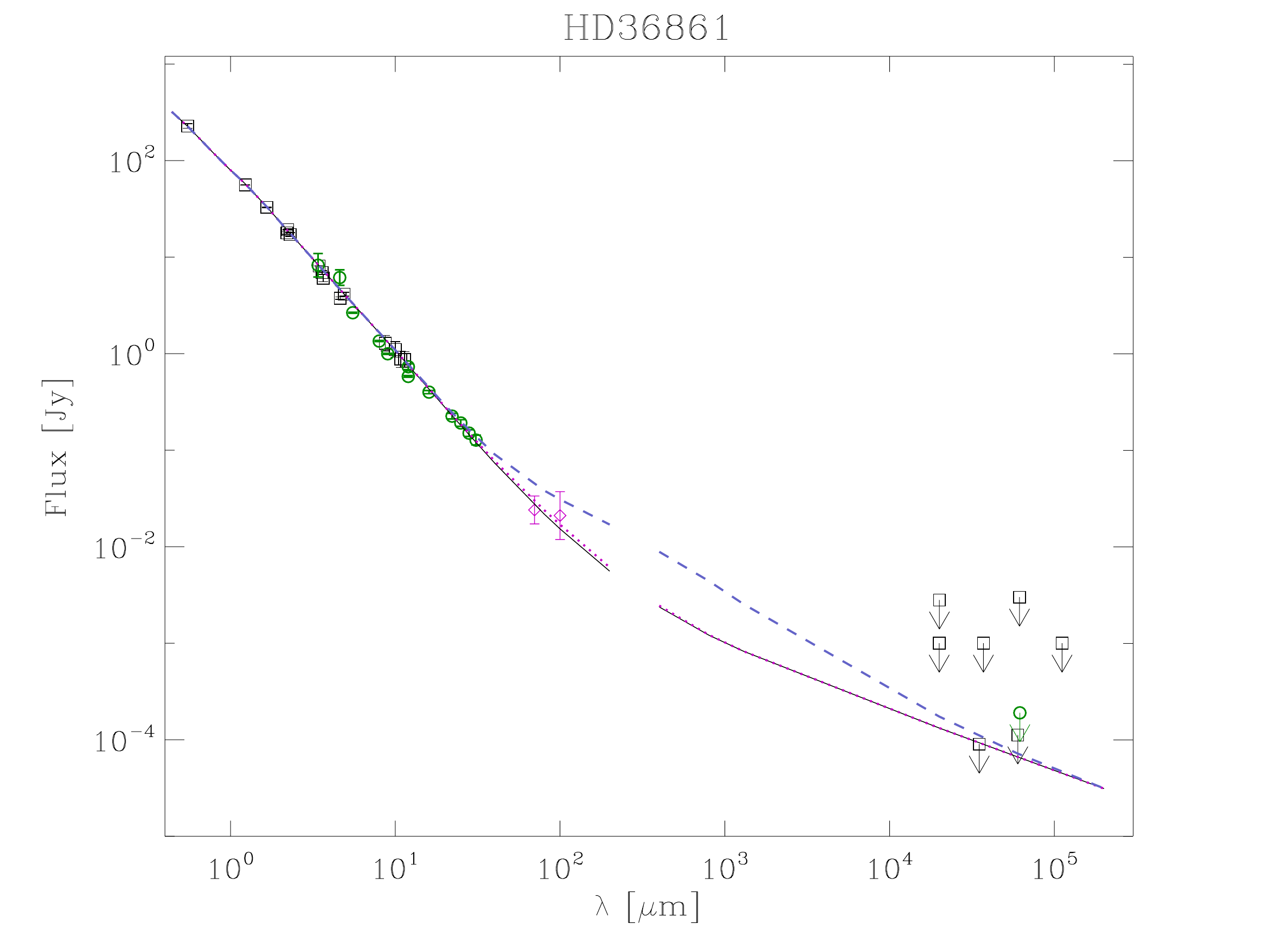}
\caption{Observed and best-fit fluxes vs. wavelength for the two O\,Giants in our sample, HD\,24912 ($\xi$ Per) and HD\,36861 ($\lambda$\,Ori A). Colours, symbols and line types as in Figure\,\ref{figOSupergiants}, except for orange dashed-dotted lines, which indicate, for HD\,24912, previous best-fit models from \citetalias{Puls2006} (for HD\,36861, the results from \citetalias{Puls2006} are still indicated as magenta-dotted). See comments on individual objects in Section \ref{OGiants}.}
\label{figOGiants}
\end{center}
\end{figure*}

We obtain \Mdotmax\,=\,1.4\,\massloss. In order to fit the aforementioned variable radio fluxes we derived several solutions for the clumping structure. The corresponding fit-diagram in Figure\,\ref{figOGiants} displays our fit-solutions in the fixed-regions (solid and magenta-dotted lines) and adapted-regions approaches (blue-dashed and orange dashed-dotted lines). In the fixed-regions approach an almost constant and moderate clumping degree for the entire wind, \clin\,=\,\clmid\,=\,3.5 and \clout\,=\,\clfar\,=\,3 (second value in Table\,\ref{tableclfactors}), fits perfectly the upper flux limits and the well determined measurement at 21\,cm  (magenta-dotted line in Figure\,\ref{figOGiants}), whereas a moderate clumped wind restricted to r\,$\lesssim$\,15\,\Rstar\ and an unclumped outer one, \clin\,=\,\clmid\,=\,3.5 and \clout\,=\,\clfar\,=\,1 (first value in Table\,\ref{tableclfactors}), are consistent with all radio fluxes but the flux value at 21\,cm  (solid line in Figure\,\ref{figOGiants}).    

The best-fit solution for this source, however, is achieved in the adapted-regions approach for a much more structured clumped wind. For the estimated \Mdotmax, a significantly clumped inner wind, \clin\,=\,5, and a constant \clmid\,=\,\clout\,=\,\clfar\,=\,1 (first value in Table\,\ref{tableclbestfit}) provide the best-fit solution consistent with all radio data but the flux value at 21\,cm  (blue-dashed line in Figure\,\ref{figOGiants}), with \rin\,=\,1.1\,\Rstar\ and \rmid\,=\,4\,\Rstar. To fit the larger upper radio flux limits and the flux at 21\,cm  (orange dashed-dotted lines in Figure\,\ref{figOGiants}), the clumping degree for the outer wind regions have to be increased to \clout\,=\,\clfar\,=\,3 (second value in Table\,\ref{tableclbestfit}). The lack of observations in the mm regime gives \cloutmax\,=\,5.   

Finally, the new measured FIR fluxes for this source enforce a certain clumping degree within 1.05\,--\,1.1\,$\lesssim$\,r/\Rstar\,$\lesssim$\,2\,--\,4. The observed variability at the MIR (from 4.6 to 11.4\,\micron) and radio regimes may be explained by means of either clumping or some other physical process such as co-rotational interaction zones or a wind compressed equatorial region, as suggested by \citetalias{Puls2006}. In particular, a potential magnetic field interacting with the stellar wind leads to wind confinement and channelling (\citealt{ud-Doula2002}, \citealt{Owocki2008}), and in presence of relativistic electrons, non-thermal synchrotron emission arises and may dominate the radio spectrum under certain conditions (\citealt{Trigilio2004}). Another possibility is that $\xi$\,Per is a binary system, or both magnetic field and colliding winds are present as some numerical simulations suggest (\citealt{vanLoo2005}). However, positive magnetic field detections for $\xi$\,Per have not been reported so far, and there is no observational evidence of a close companion, although several sources in its surroundings have been detected. Therefore, the nature of this variability remains unanswered, and only mm and additional radio observations can help ascertain it. \\

\noindent\textit{HD\,36816 ($\lambda$ Ori A)}

This source was previously analysed by \citetalias{Puls2006}, and there is a new distance estimate available from \textit{GAIA}. 
All radio flux observations are upper limits, and consequently the estimated \Mdotmax\ is an upper limit. In the right panel in Figure\,\ref{figOGiants} we present \citetalias{Puls2006}'s best-fit solution (magenta-dotted line) and the best-fit model derived in this work (solid line), corresponding to the fixed-regions approach. 

\citetalias{Puls2006} derived \Mdotmax\,$\lesssim$\,0.4\,\massloss\ (for $d$\,=\,0.5\,kpc) and \clin\,=\,2, \clmid\,=\,\clout\,=\,\clfar\,=\,1 (with \rinc, \rmidc,\rout\,=\,10\,\Rstar\ and \rfarc). Moreover \clmidmax\,=\,20 and \cloutmax\,=\,2 (blue-dashed line), respectively. The clumping stratification derived by \citetalias{Puls2006} for the upper limit \Mdotmax\ is consistent with the additional MIR observations available, but slightly overestimates the obtained fluxes at 70 and 100\,\micron. The best-fit solution, however, is found by decreasing the clumping degree in the inner wind region to \clin\,=\,\clmid\,=\,\clout\,=\,\clfar\,=\,1. 

There are no well-defined flux observations available at wavelengths larger than 100\,\micron. Despite that, we can lower the maximum clumping degree for the intermediate and outer wind regions derived by \citetalias{Puls2006} to \clmidmax\,=\,5 and \cloutmax\,=\,2. Only additional, and well determined observations at mm and radio regimes can provide definite  \Mdotmax\ and clumping factors in Regions\,4 and 5, \clout\ and \clfar, respectively. For the \textit{GAIA} distance used here, d=\,0.27\,kpc, the scaled upper limit maximum mass-loss rate becomes \Mdotmax\,$\lesssim$\,0.16\,\massloss, and the derived clumping structure described above (Table\,\ref{tableclfactors}) provide the best-fit solution.\\

\subsection{OB\,Dwarfs}\label{OBDwarfs}

\textit{HD\,149757 ($\zeta$ Oph)}

\begin{figure*}[htp]
\begin{center}
\includegraphics[width=9cm]{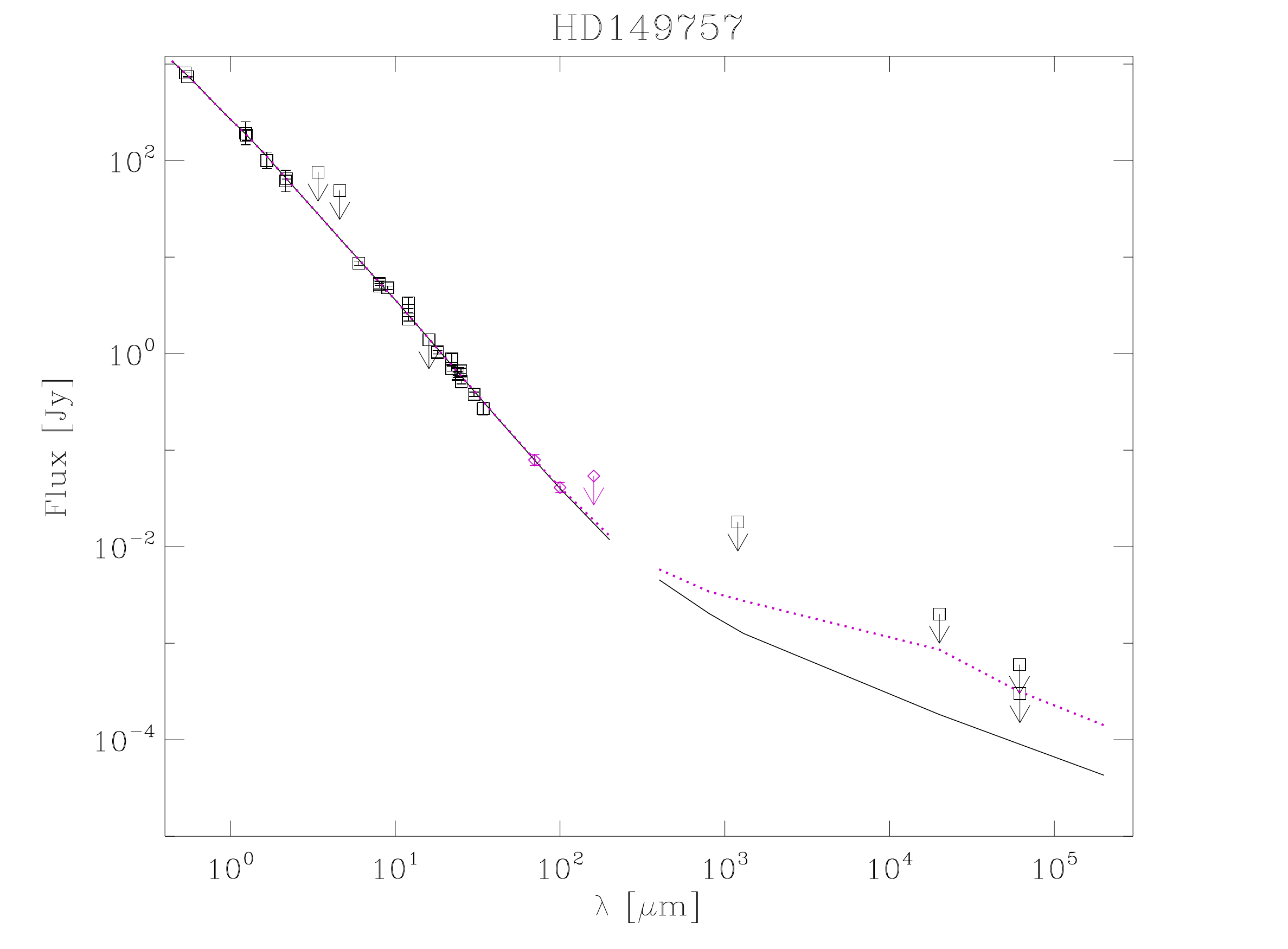}
\includegraphics[width=9cm]{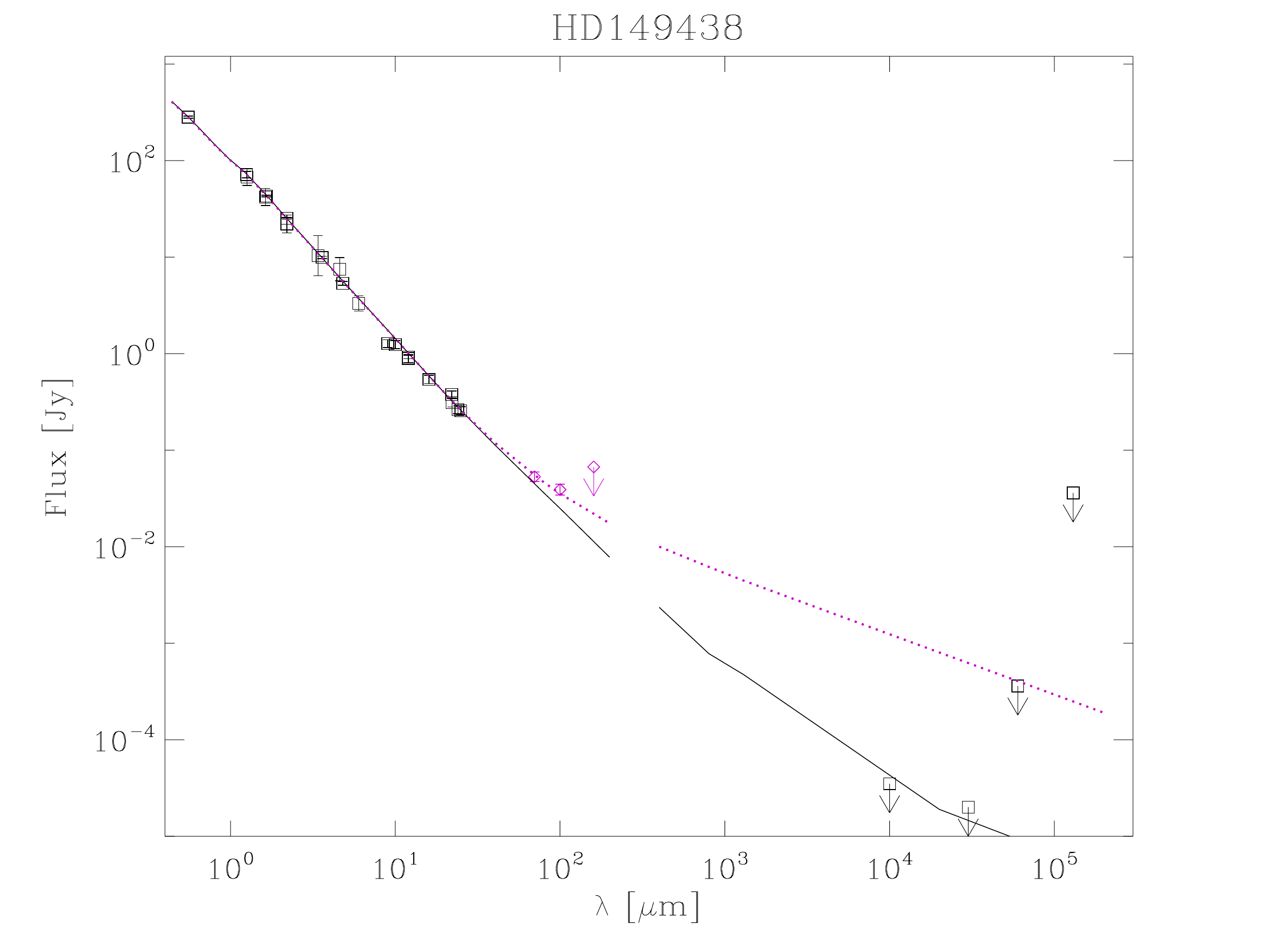}
\caption{Observed and best-fit fluxes vs. wavelength for the OB\,Dwarfs in our sample. Colours, symbols and line types as in Figure\,\ref{figOSupergiants}. See comments on individual objects in Section \ref{OBDwarfs}.}
\label{figOBDwarfs}
\end{center}
\end{figure*}

All available mm and radio observations for this rapidly rotating source are upper limits, therefore \Mdotmax\ is also an upper limit. We noted that if we derive \Mdotmax\ to fit the radio upper limits at 6\,cm, the flux-model emission is consistent with the observations from V- to NIR-band but overestimates fluxes at MIR and FIR wavelengths. Therefore, the upper limit \Mdotmax\ is constrained by the fluxes at 70 and 100\,\micron. The left panel in Figure\,\ref{figOBDwarfs} displays our best-fit solution, which corresponds to the fixed-regions approach. This best-fit solution (solid line) is obtained by \Mdotmax\,$\lesssim$\,0.07\,\massloss\ and a constant \clin\,=\,\clmid\,=\,\clout\,=\,\clfar\,=\,1 ( Table\,\ref{tableclfactors}). This model reproduces all well-defined flux observations and is consistent with all upper limits in the radio regime.We also obtain \cloutmax\,=\,20 and $f^{\rm far}_{\rm max}$\,=\,6 (magenta-dotted line).  Additional well defined mm and radio fluxes are needed to derive definite \Mdotmax\ and clumping properties in the outermost wind region.\\

\noindent\textit{HD\,149438 ($\tau$ Sco)}

$\tau$\,Sco is a confirmed magnetic B-star (\citealt{Petit2013}), and a very slow rotator ($\sim$\,5\,\kms; \citealt{Donati2006}). Mass-loss rates found in the literature range from \Mdot\,=\,0.0013\,\massloss\ (from UV resonance lines; \citealt{Hamann1981}), to \Mdot\,=\,0.0614\,\massloss\ (from optical spectroscopy; \citealt{Mokiem2005}), with a value in-between \Mdot\,=\,0.02\massloss\ (from NIR spectroscopy; \citealt{Repolust2005}). We note that all these derivations used non-magnetic wind models to derive the mass-loss rate, and thus did not take into account the considerable effect the magnetic field has upon the wind structure. 

The right panel in Figure\,\ref{figOBDwarfs} displays our best-fit models for this source, corresponding to the fixed-regions approach.  
Whereas the upper limits at 1 and 3\,cm  (\citealt{Kurapati2017}) are consistent with thermal emission, those at 6\,cm  (\citealt{Bieging1989}) and 13\,cm  (\citealt{Kurapati2017}) point to non-thermal emission. In view of the SED trend up to 160\,\micron, we suspect that the likely flux value at 13\,cm  is much lower than the upper limit provided by \cite{Kurapati2017}, and therefore we do not attempt to fit it. Despite that, the upper flux limits from 1 to 6\,cm  still suggest $\tau$\,Sco is a non-thermal emitter. Since it is a magnetic star, the observed radio variability might be explained by non-thermal synchrotron emission. However, non-thermal radio emission seems to favour centrifugal magnetospheres instead of dynamical ones (\citealt{Kurapati2017}), and $\tau$\,Sco is classified as having a dynamical magnetosphere (\citealt{Petit2013}). Therefore, the detected variability and its nature still needs to be confirmed by deeper radio observations. 

Nevertheless, we derived two possible upper limits for \Mdotmax, attending to variable radio upper flux limits, and being consistent with the well determined fluxes at 70 and 100\,\micron. A first solution (magenta-dotted line) is obtained for a \Mdotmax\,$\lesssim$\,0.315\,\massloss\ and a constant \clin\,=\,\clmid\,=\,\clout\,=\,\clfar\,=\,1 (first entry in Table\,\ref{tableclfactors}). Although this model perfectly matches the observations at $\lambda \lesssim$\,100\,\micron\ and the upper limit at 6\,cm, it overestimates by large the radio upper flux limits at shorter wavelengths. 
A second solution (solid line) is achieved for a much lower \Mdotmax\,$\lesssim$\,0.018\,\massloss, an extremely large clumping degree in the inner wind region, \clin\,=\,300, and \clmid\,=\,\clout\,=\,\clfar\,=\,1 (second entry in Table\,\ref{tableclfactors}). In this case, the model consistently reproduces all observations and the radio upper limits, but the flux at 70\,\micron\ only marginally, and slightly underestimates the one at 100\,\micron. The FIR fluxes and the radio upper limit at 6\,cm could be perfectly matched with this value of \Mdotmax\ by increasing the clumping degree of the wind at r\,$\gtrsim$\,2 \Rstar\ as \clmid\,=\,\clout\,=\,\clfar\,=\,300. However, the radio upper limits at 1 and 3\,cm  still remains overestimated by large. We caution, however, that none of the above takes into account any effects of the magnetic field upon the wind structure. Additional and well determine flux observations at mm and radio regime, along with proper magnetic modelling, may help reaching conclusive results regarding this source. \\

\subsection{Binary OB stars}\label{biOB}

\noindent\textit{HD\,149404} 

\begin{figure*}[htp]
\begin{center}
\includegraphics[width=9cm]{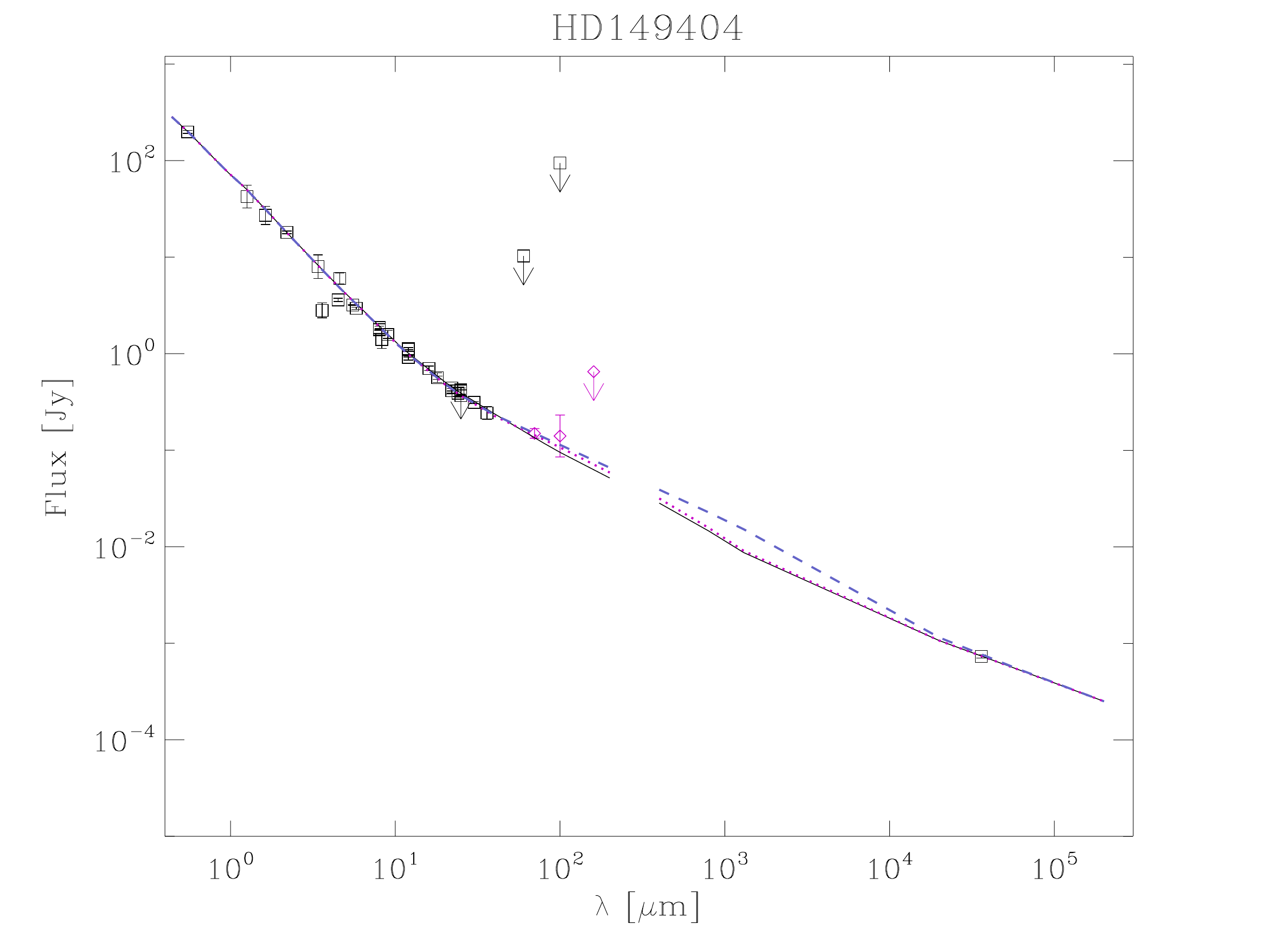}
\includegraphics[width=9cm]{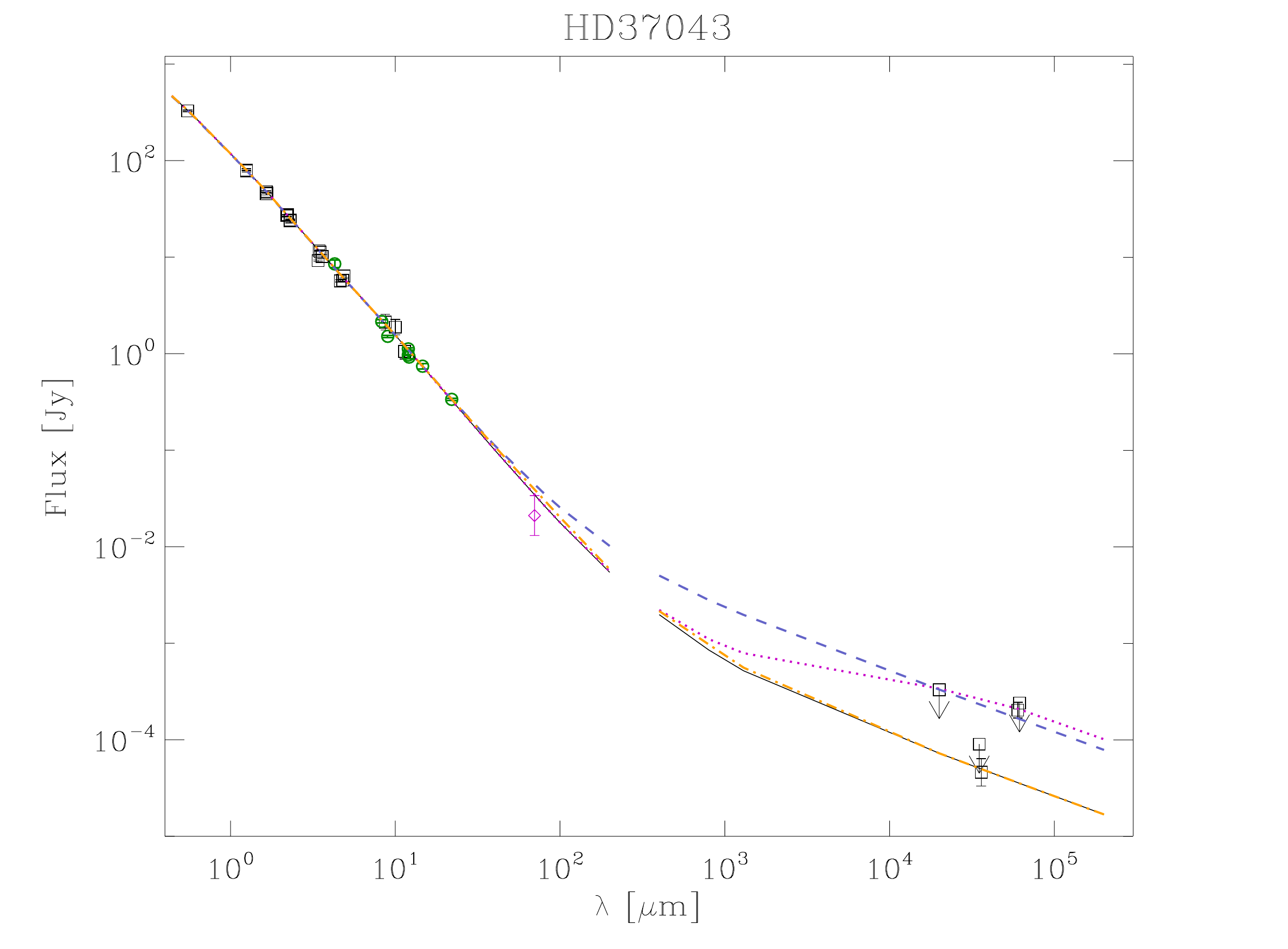}
\caption{Observed and best-fit fluxes vs. wavelength for the two binary systems in our sample, HD\,140494 (V973\,Sco) and HD\,37043 ($\iota$ Ori). Colours, symbols and line types as in Figure\,\ref{figOSupergiants}, except that in the HD\,149404 diagram, the blue-dashed line represents also an alternative solution, whereas for HD\,37043, together with the orange dashed-dotted line, they represent the best-fit models by \citetalias{Puls2006}. See comments on individual objects in Section \ref{biOB}.}
\label{figOBbinaries}
\end{center}
\end{figure*}

This source is a known massive spectral binary (\citealt{Rauw2001}, \citealt{Raucq2016}). From the only reliable radio observation at 3.6\,cm  we obtain \Mdotmax\,=\,8.27\,\massloss. The corresponding panel in Figure\,\ref{figOBbinaries} displays the best possible solution in the fixed-regions (solid line) adapted-regions approaches (magenta-dotted line). The best possible solution in the fixed-regions approach is obtained for \clin\,=\,\clout\,=\,\clfar\,=\,1 and \clmid\,=\,5.4 (Table\,\ref{tableclfactors}). The best-fit solution, however, is achieved in the adapted-regions approach (magenta-dotted line) by increasing the clumping degree in the intermediate wind region, \clin\,=\,\clout\,=\,\clfar\,=\,1 and \clmid\,=\,7, provided that it starts at \rmid\,=\,2.5\,\Rstar (Table\,\ref{tableclbestfit}). For both approaches, \cloutmax\,=\,6 is still consistent with the data (blue-dashed line).\\

\noindent\textit{HD\,37043 ($\iota$\,Ori)}

$\iota$\,Ori is a known binary (\citealt{Pittard2000}, \citealt{Pablo2017}), with consistent observed variable flux at radio wavelengths. This source was previously analysed by \citetalias{Puls2006} as a thermal emitter. 
In the corresponding panel in Figure\,\ref{figOBbinaries}, we present the two solutions derived by \citetalias{Puls2006}, together with our best-fit model, which corresponds to the fixed-regions approach. 

\citetalias{Puls2006} derived a first solution (blue-dashed line) consistent with both the upper limit at 2\,cm  and the well determined radio flux at 6\,cm: \Mdotmax\,=\,0.8\,\massloss, \clin\,=\,\clmid\,=\,\clout\,=\,\clfar\,=\,1 with \rinc, \rmidc, \rout\,=\,10 \Rstar\ and \rfarc; and a second one (orange dashed-dotted line) fitting only the lower flux value at 3.5\,cm: \Mdotmax\,=\,0.25\,\massloss, \clin\,=\,12, \clmid\,=\,\clout\,=\,\clfar\,=\,1 with \rinc, \rmid\,=\,1.3\,\Rstar, \rout\,=\,10\,\Rstar, \rfarc. 

As shown in the fit-diagram, both models provide a marginal fit of the flux value at 70\,\micron. However, if HD\,37043 is a non-thermal source, the lower \Mdotmax\ solution by \citetalias{Puls2006} represents the upper and lower limits of the mass-loss rate and clumping structure for this source, respectively, even in the case that our estimated flux value for this source at 70\,\micron\ is wrong. Note that this solution is also consistent with the upper measured radio fluxes by increasing the clumping in the outer wind regions. We found that the best-fit for all the data (solid line) is provided by the lowest value of \Mdotmax\ derived by \citetalias{Puls2006} (\Mdotmax\,=\,0.25\,\,1 (first values in Table\,\ref{tableclfactors}). To reproduce the upper radio fluxes (magenta-dotted line), large clumping factors at the outer wind regions are required, \clout\,=\,5 and \clfar\,=\,15 (second values in Table\,\ref{tableclfactors}). In all simulations \cloutmax\,=\,10. 
In this case, our best-fit solution is only slightly better than that obtained for the lower \Mdotmax\ and the strong inner clumped wind solution derived by \citetalias{Puls2006}. \\

\end{document}